\documentclass[aps,prd,onecolumn,showpacs,floats,floatfix,letterpaper,nofootinbib,superscriptaddress,preprintnumbers]{revtex4}
\usepackage{amssymb,amsmath,latexsym,mathrsfs}
\usepackage{graphicx}
\usepackage{epsfig}
\usepackage{multirow}
\usepackage{array}
\usepackage{xspace}
\usepackage{color}
\usepackage{url}
\usepackage{eufrak}
\usepackage[bookmarks=true, bookmarksnumbered=true, colorlinks=true,
urlcolor=darkblue, citecolor=darkpowderblue, linkcolor=darkred,
breaklinks=true]{hyperref}   
  
\usepackage{soul}  

\definecolor{darkblue}{rgb}{0,0,0.5}
\definecolor{darkpowderblue}{rgb}{0,0.2,0.6}
\definecolor{darkred}{rgb}{0.55,0,0}

\newcommand{\bea}{\begin{eqnarray}}
\newcommand{\eea}{\end{eqnarray}}

\begin{document}
\title{The 21~cm signal and the interplay between dark matter \\ annihilations and astrophysical processes}
\author{Laura Lopez-Honorez}
\affiliation{Theoretische Natuurkunde,
Vrije Universiteit Brussel and The International Solvay Institutes,\\
Pleinlaan 2, B-1050 Brussels, Belgium.}
\author{Olga Mena}
\affiliation{Instituto de F\'{\i}sica Corpuscular (IFIC)$,$
 CSIC-Universitat de Val\`encia$,$ \\  
 Apartado de Correos 22085$,$ E-46071 Valencia$,$ Spain}
 \author{\'Angeles Molin\'e}
  \affiliation{Instituto de F\'{\i}sica Corpuscular (IFIC)$,$
  	CSIC-Universitat de Val\`encia$,$ \\  
  	Apartado de Correos 22085$,$ E-46071 Valencia$,$ Spain}
\affiliation{CFTP, Instituto Superior Tecnico, Universidade Tecnica de Lisboa$,$\\
Av. Rovisco Pais 1, 1049-001 Lisboa, Portugal}
\author{Sergio Palomares-Ruiz}
\affiliation{Instituto de F\'{\i}sica Corpuscular (IFIC)$,$
 CSIC-Universitat de Val\`encia$,$ \\  
 Apartado de Correos 22085$,$ E-46071 Valencia$,$ Spain}
\author{Aaron C. Vincent}
\affiliation{Institute for Particle Physics Phenomenology (IPPP),\\ Department of Physics, Durham University, Durham DH1 3LE, UK.}

\preprint{CFTP/16-007, IFIC/16-16, IPPP/16/20}

\begin{abstract}
Future dedicated radio interferometers, including HERA and SKA, are very promising tools that aim to study the epoch of reionization and beyond via measurements of the 21~cm signal from neutral hydrogen. Dark matter (DM) annihilations into charged particles change the thermal history of the Universe and, as a consequence, affect the 21~cm signal. Accurately predicting the effect of DM strongly relies on the modeling of annihilations inside halos. In this work, we use up-to-date computations of the energy deposition rates by the products from DM annihilations, a proper treatment of the contribution from DM annihilations in halos, as well as values of the annihilation cross section allowed by the most recent cosmological measurements from the \emph{Planck} satellite. Given current uncertainties on the description of the astrophysical processes driving the epochs of reionization, X-ray heating and Lyman-$\alpha$ pumping, we find that disentangling DM signatures from purely astrophysical effects, related to early-time star formation processes or late-time galaxy X-ray emissions, will be a challenging task. We conclude that only annihilations of DM particles with masses of $\sim100$~MeV, could leave an unambiguous imprint on the 21~cm signal and, in particular, on the 21~cm power spectrum. This is in contrast to previous, more optimistic results in the literature, which have claimed that strong signatures might also be present even for much higher DM masses. Additional measurements of the 21~cm signal at different cosmic epochs will be crucial in order to break the strong parameter degeneracies between DM annihilations and astrophysical effects and undoubtedly single out a DM imprint for masses different from $\sim100$~MeV.
\end{abstract}
\maketitle

\section{Introduction}
\label{sec:introduction}

The redshifted 21~cm line arising from the transition between the singlet and triplet hyperfine levels of neutral hydrogen provides a  \emph{unique} probe of the high redshift cosmic window at $z\gtrsim 6$, including the dark ages (before the first stars have formed) and the Epoch of Reionization (EoR). The first measurements of the 21~cm brightness temperature power spectrum at the EoR could be obtained by the Giant Metrewave Radio Telescope (GMRT)~\cite{Ananthakrishnan:1995, Paciga:2011}, the Low Frequency Array (LOFAR)~\cite{vanHaarlem:2013dsa}, the Murchison Widefield Array (MWA)~\cite{Tingay:2012ps} and the Precision Array for Probing the Epoch of Reionization (PAPER)~\cite{Parsons:2009in}. Currently, only bounds on this signal at redshifts $z \simeq 6-12$ have been published (see, e.g., Ref.~\cite{Paciga:2013fj, Dillon:2013rfa, Pober:2014aca, Ali:2015uua}). The next generation of radio interferometers, such as the Hydrogen Epoch of Reionization Array (HERA)~\cite{reiondotorg} and the Square Kilometer Array (SKA)~\cite{Mellema:2012ht}, should have the sensitivity to fully reconstruct the reionization mechanism and to explore the Universe at even earlier epochs, within the dark ages. Alternatively, a complementary way to detect the 21~cm signal is to measure the global signal, averaged over all directions in the sky. The current Experiment to Detect the Reionization Step (EDGES)~\cite{Bowman:2012hf} and the future Large-Aperture Experiment to Detect the Dark Ages (LEDA)~\cite{Greenhill:2012mn} or the Moon orbiting space observatory Dark Ages Radio Experiment (DARE)~\cite{Burns:2011wf} aim to detect this global signal. In particular, EDGES has already ruled out extremely rapid reionization models~\cite{Bowman:2012hf}. A three-dimensional map of the 21~cm signal could also be obtained using the so-called intensity mapping technique, which measures the collective emission from large regions without resolving individual galaxies. This technique has been considered to study damped Lyman-alpha systems at lower redshifts, in the post-reionization era ($z\lesssim 3$)~\cite{Villaescusa-Navarro:2014cma}, and will be complementary to galaxy redshift surveys~\cite{Wyithe:2007rq, Chang:2007xk, Loeb:2008hg}. The planned experimental set-ups in this field include the GBT-HIM project with the Green Bank Telescope (GBT)~\cite{Chang:2016}, the Canadian Hydrogen Intensity Mapping Experiment (CHIME)~\cite{CHIME}, the Tianlai project~\cite{Chen:2015} and the SKA-mid frequency~\cite{SKA} experiments (see, e.g., Ref.~\cite{Bull:2014rha} and references therein).

Although the primary task of near future 21~cm experiments is to improve our current knowledge of the Universe's reionization history, they also represent an additional tool for fundamental cosmology~\cite{Scott:1990,Tozzi:1999zh, Iliev:2002gj, Barkana:2005xu, Barkana:2004zy, Bowman:2005hj, McQuinn:2005hk, Santos:2006fp, Mao:2008ug, Visbal:2008rg, Clesse:2012th, Liu:2015txa, Liu:2015gaa}. The global 21~cm signal and the fluctuations of the 21~cm brightness temperature power spectrum can improve the cosmological constraints on neutrino physics~\cite{Loeb:2003ya, McQuinn:2005hk, Mao:2008ug, Pritchard:2008wy, Shimabukuro:2014ava, Oyama:2012tq,  Villaescusa-Navarro:2015cca, Oyama:2015gma}, provide a signal of particle dark matter  (DM)~\cite{Loeb:2003ya, Zurek:2007gn, Mesinger:2013nua, Sitwell:2013fpa, Sekiguchi:2014wfa, Shimabukuro:2014ava, Carucci:2015bra}, measure dark energy properties~\cite{Wyithe:2007rq, Chang:2007xk, Morales:2009gs, Archidiacono:2014msa, Chen:2015vdi},  explore signatures of non-gaussianities~\cite{Loeb:2003ya, Cooray:2006km, Pillepich:2006fj, Joudaki:2011sv, Tashiro:2012wr, Chongchitnan:2012we, Chongchitnan:2013oxa, Chen:2015vdi} or isocurvature perturbations~\cite{Takeuchi:2013hza, Sekiguchi:2013lma},  primordial black holes~\cite{Mack:2008nv, Tashiro:2012qe} and alternatives to general relativity~\cite{Loeb:2008hg, Hall:2012wd,Brax:2012cr}. 

Annihilations and decays of DM particles are well known to affect the temperature and ionization history of the early Universe~\cite{Chen:2003gz, Hansen:2003yj, Pierpaoli:2003rz, Padmanabhan:2005es, Shchekinov:2006eb, Furlanetto:2006wp}. The effects of energy injection from DM annihilations and decays have been thoroughly studied to put stringent constraints on DM properties using Cosmic Microwave Background (CMB) data~\cite{Zhang:2006fr, Mapelli:2006ej, Zhang:2007zzh, Natarajan:2008pk, Natarajan:2009bm, Belikov:2009qx, Galli:2009zc, Slatyer:2009yq, Huetsi:2009ex, Cirelli:2009bb, Kanzaki:2009hf, Hisano:2011dc, Galli:2011rz, Finkbeiner:2011dx, Giesen:2012rp, Slatyer:2012yq, Lopez-Honorez:2013lcm, Galli:2013dna, Diamanti:2013bia, Madhavacheril:2013cna, Ade:2015xua, Slatyer:2015jla, Kawasaki:2015peu}. Energy injection from DM annihilations and decays may also affect the 21~cm signal from regions of neutral hydrogen in the intergalactic medium (IGM), leaving potentially detectable signatures~\cite{Shchekinov:2006eb, Furlanetto:2006wp, Valdes:2007cu, Chuzhoy:2007fg, Cumberbatch:2008rh, Natarajan:2009bm, Yuan:2009xq, Valdes:2012zv, Evoli:2014pva}. Thus, both the 21~cm and the CMB power spectra are complementary in the study of the impact of DM annihilations and decays during the dark ages. At high redshifts ($z \gtrsim 30$), the effects of DM annihilations and decays are easy to model, as they occur in the linear density perturbation regime. Unfortunately, the large galactic synchrotron foreground at the relevant frequencies represents an enormous experimental challenge~\cite{Burns:2011wf}, which makes the detection of these effects an arduous task in the near future, beyond the reach of current and planned ground-based radio interferometers. In this work, we discuss the impact of DM annihilations on the 21~cm signal by focusing on the epoch between reionization and the formation of the first galaxies ($10 \lesssim z \lesssim 30$). This region is expected to be visible by future experimental facilities such as HERA and SKA. However, in this redshift range, the astrophysical background associated to the formation of the first luminous sources also affects the 21~cm signal, so the effects from DM annihilations have to be disentangled from it. We therefore discuss some of the dependences on the astrophysical parameters (including the halo mass function, the threshold for halos to host star-forming galaxies, the X-ray efficiency and the stellar Lyman-$\alpha$ flux) before turning to the DM-related parameters (including the DM mass, the annihilation cross section, the minimum halo mass and the effect of substructure) that are relevant to the evaluation of the impact of DM annihilations on the 21~cm signal. 

Finally, we contrast our results with some existing studies in the literature. Recent works~\cite{Valdes:2012zv, Evoli:2014pva} have claimed that DM annihilations at $z \lesssim 50$, which occur mostly in halos, could lead to distinctive features on the 21~cm global signal and on its power spectrum. However, these studies adopted a simplified parametrization of the efficiency of energy injection and deposition  in halos, which leads to a significant overestimation of the effects for DM masses higher than $\sim 1$~GeV. In our work, we critically examine this issue. We additionally stress that, even when neglecting DM annihilation effects, the choice of the parametrization for the halo mass function and of some of the relevant astrophysical parameters yield a prediction for the 21~cm signal that can vary significantly. Combined with a correct integration over the DM energy deposition efficiency, these uncertainties could lead to a mischaracterization of the potential imprint of DM annihilations on the 21~cm signal.  In most cases, such signatures could be mimicked by a different set of astrophysical parameters.
 
The structure of this paper is as follows. In Secs. \ref{sec:21-cm-signal} and \ref{sec:evo}, we describe the general physics behind the cosmological 21~cm signal (both the global signal and the power spectrum), the impact of different parametrizations of the halo mass function and the effect of varying some astrophysical parameters. This is followed in Sec. \ref{sec:dark-matter-impact} by the description of the impact of DM annihilations on the 21~cm signal, where we show our main results, along with a critical comparison with previous results. We conclude in Sec. \ref{sec:conclusions}.

\section{The 21 cm signal and its power spectrum}
\label{sec:21-cm-signal}

The 21~cm line is produced by spin-flip transitions in neutral hydrogen between the more energetic triplet state and the ground singlet state. The intensity of the redshifted line depends on the ratio of the populations of these two hyperfine levels. This ratio is quantified by the spin temperature, $T_S$, which is determined by three competing effects: 1) absorption and stimulated emission of CMB photons; 2) atomic collisions, which are important at high redshifts, well before the EoR; and 3) resonant scattering of Lyman-$\alpha$ (Ly$\alpha$) photons that turn on with the first sources. The latter process is known as the Wouthuysen-Field effect~\cite{Wouthuysen:1952, Field:1958} (see Ref.~\cite{Hirata:2005mz} and references therein).  

The redshifted 21~cm signal from the EoR and the dark ages can be measured either in emission or in absorption against the CMB blackbody spectrum. Therefore, at a given observed frequency $\nu$, the redshifted signal is usually represented by the contrast between the temperature of the high-redshift hydrogen clouds and that of the CMB, i.e., the differential brightness temperature $\delta T_b$. For these small frequencies (Rayleigh-Jeans limit), from the equation of radiative transfer along the line of sight, $\delta T_b$ yields
\begin{equation}
\delta T_b(\nu)= \frac{T_S-T_{\rm{CMB}}}{1+z} (1-\exp^{-\tau_{\nu_0}})~,
\label{eq:Tb}
\end{equation}
where $\tau_{\nu_0}$ is the optical depth of the IGM for the 21~cm frequency $\nu_0 = 1420.4$~MHz and all quantities are evaluated at redshift $z=\nu_0/\nu-1$. Expanding Eq.~(\ref{eq:Tb}) up to first order in perturbation theory, the differential brightness temperature can be written as~\cite{Madau:1996cs, Furlanetto:2006jb, Pritchard:2011xb, Furlanetto:2015apc}
\begin{equation}
\delta T_b(\nu) \simeq 27 \, x_\textrm{HI} \, (1 + \delta_b) \left( 1 - \frac{T_\textrm{CMB}}{T_S}\right) \left( \frac{1}{1+H^{-1} \partial v_r / \partial r} \right) \, \left( \frac{1+z}{10}\right)^{1/2} \left(\frac{0.15}{\Omega_m h^2} \right)^{1/2} \left( \frac{\Omega_b h^2}{0.023}\right)\,\textrm{mK} ~,
\label{eq:Tbdev}
\end{equation}
where $x_\textrm{HI}$  represents the fraction of neutral hydrogen, $\delta_b$ is the baryon overdensity, $\Omega_b h^2$ and $\Omega_m h^2$ refer to the current baryon and matter contribution to the Universe's mass-energy content, $H(z)$ is the Hubble parameter and $\partial v_r / \partial r$ is the comoving gradient of the peculiar velocity along the line of sight. The above expression is exact if $\partial v_r/ \partial r$ is constant over the width of the 21~cm line and $\partial v_r / \partial r\ll H$.  

After galaxy formation starts, $T_S$ is expected to be much smaller than $T_{\rm{CMB}}$ so that $\delta T_b(\nu)$ is negative and the 21~cm signal is in absorption (plus fluctuations due to the pattern of overdensities). This absorption regime continues until the X-ray emission from galaxies heats the gas so that $T_S > T_{\rm CMB}$ and the 21~cm signal is in emission~\cite{Pritchard:2011xb}. Note that when the IGM is observed in emission and $T_S\gg T_{\rm CMB}$, the signal saturates to a level which is independent of $T_S$, whereas when it is in absorption and $T_S \ll T_{\rm CMB}$, the absolute value of the differential brightness temperature is larger by a factor of $T_{\rm CMB}/T_S$. In more detail, the evolution of the IGM temperature follows different stages:
\begin{enumerate}
\item At very high redshifts ($z\gtrsim300$), the IGM is very dense, and therefore {\it the spin temperature is coupled via collisions} to the kinetic gas temperature $T_K$, which is in turn coupled to the CMB temperature by Compton scatterings between CMB photons and residual free electrons. Thus, $T_S \simeq T_K \simeq T_\textrm{CMB}$ and the all-sky averaged 21~cm differential brightness temperature is  $\overline{\delta T_b}\simeq 0$.
\item  After the CMB decoupling redshift ($300\gtrsim z\gtrsim 100$), $T_S$ and $T_K$ are still coupled but $T_S \simeq T_K < T_\textrm{CMB}$, since the IGM temperature decreases adiabatically ($T_S \propto (1+z)^2$), i.e., faster than the CMB temperature ($T_\textrm{CMB}\propto(1+z)$). In this period, $\overline{\delta T_{b}}$ is negative.
\item As the {\it Universe cools down due to its expansion}, the density of the IGM decreases, $T_S$ decouples from $T_K$ and starts approaching $T_\textrm{CMB}$ again. In the reionization model considered here, $\overline{\delta T_b}\simeq 0$ around $z\sim 30$.
\item  When the {\it first astrophysical sources start to form}, $T_S$ slowly couples to $T_K$ again, via scatterings in the Ly$\alpha$ resonance (the so-called Wouthuysen-Field process or Ly$\alpha$ pumping), and $\overline{\delta T_b}$ decreases again. The 21~cm line is again observable in absorption, with a large dip at $z\sim 20$ and reaches $\overline{\delta T_b} \simeq -160$ mK for our default simulation parameters.  
\item {\it Heating from X-ray sources} at a later time ($z \lesssim 20$) makes $T_S> T_\textrm{CMB}$ and the 21~cm signal changes from absorption to emission.
\item Finally at $z\lesssim 10$, the {\it IGM is fully ionized} and, consequently, $\overline{\delta T_b}\simeq 0$. 
\end{enumerate}

\begin{figure}
\begin{tabular}{c c}
\includegraphics[width=0.53\textwidth]{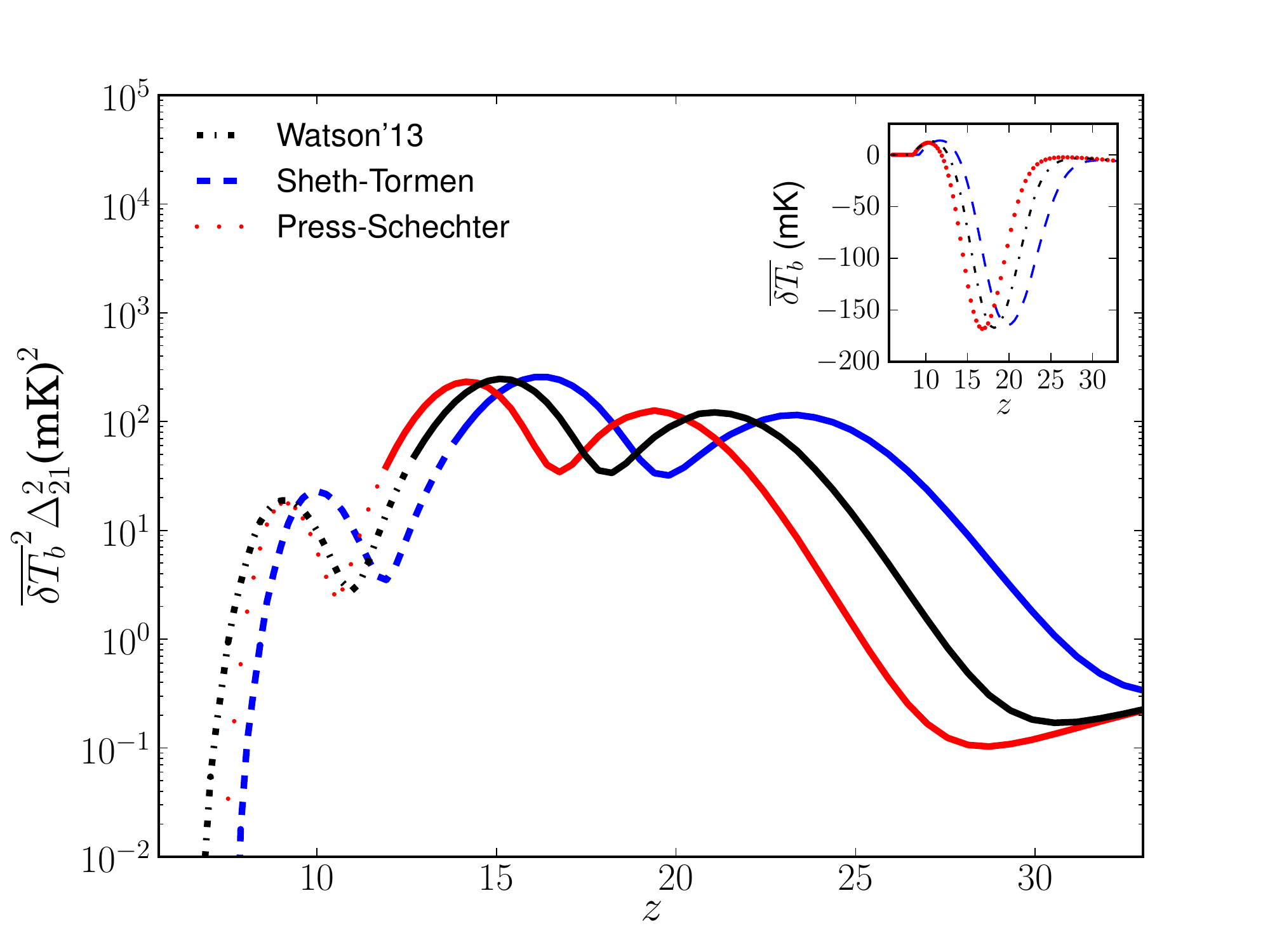} &  \hspace{-7mm}
\includegraphics[width=0.53\textwidth]{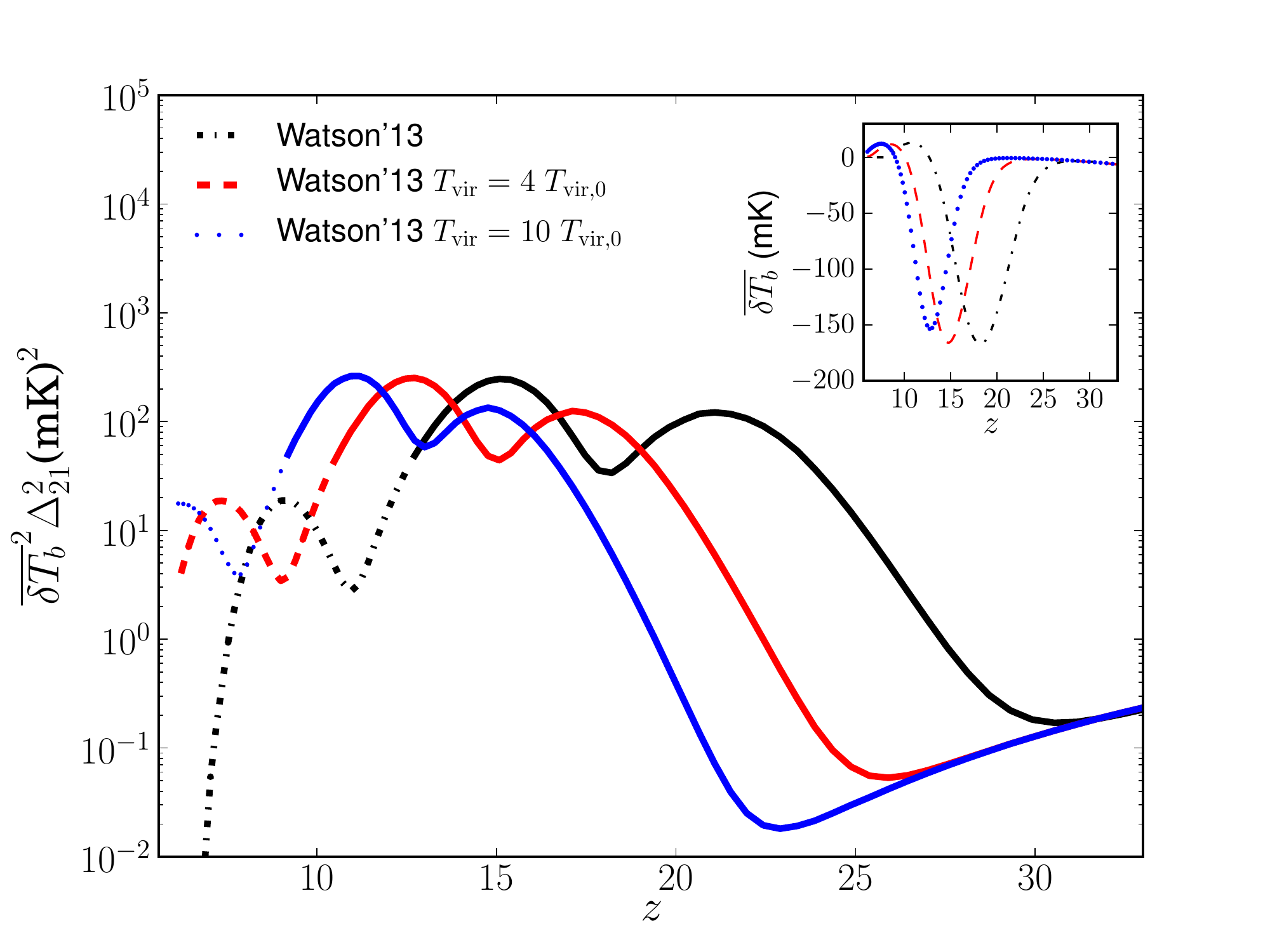} 
\\
\includegraphics[width=0.53\textwidth]{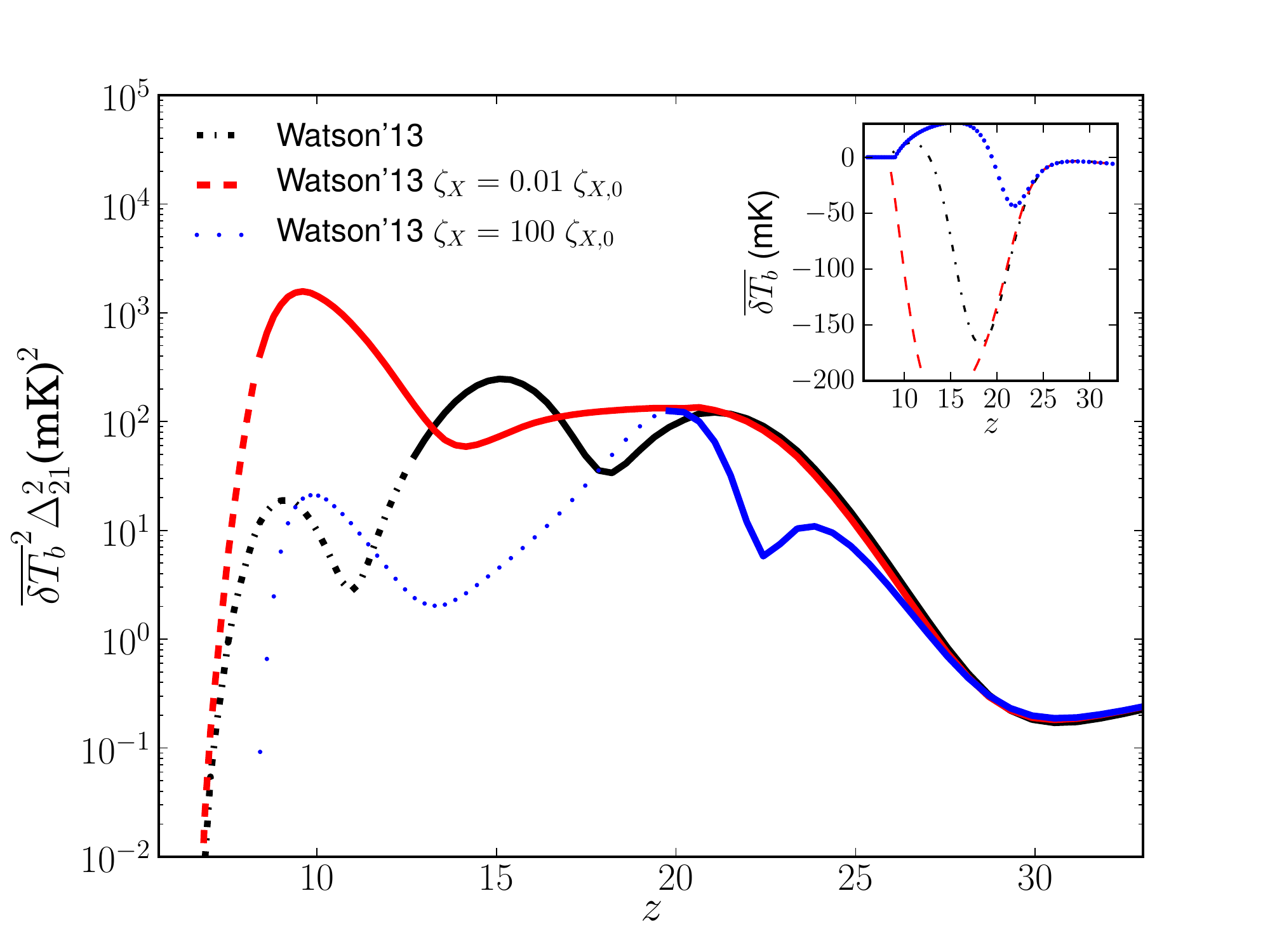} &  \hspace{-7mm} 
\includegraphics[width=0.53\textwidth]{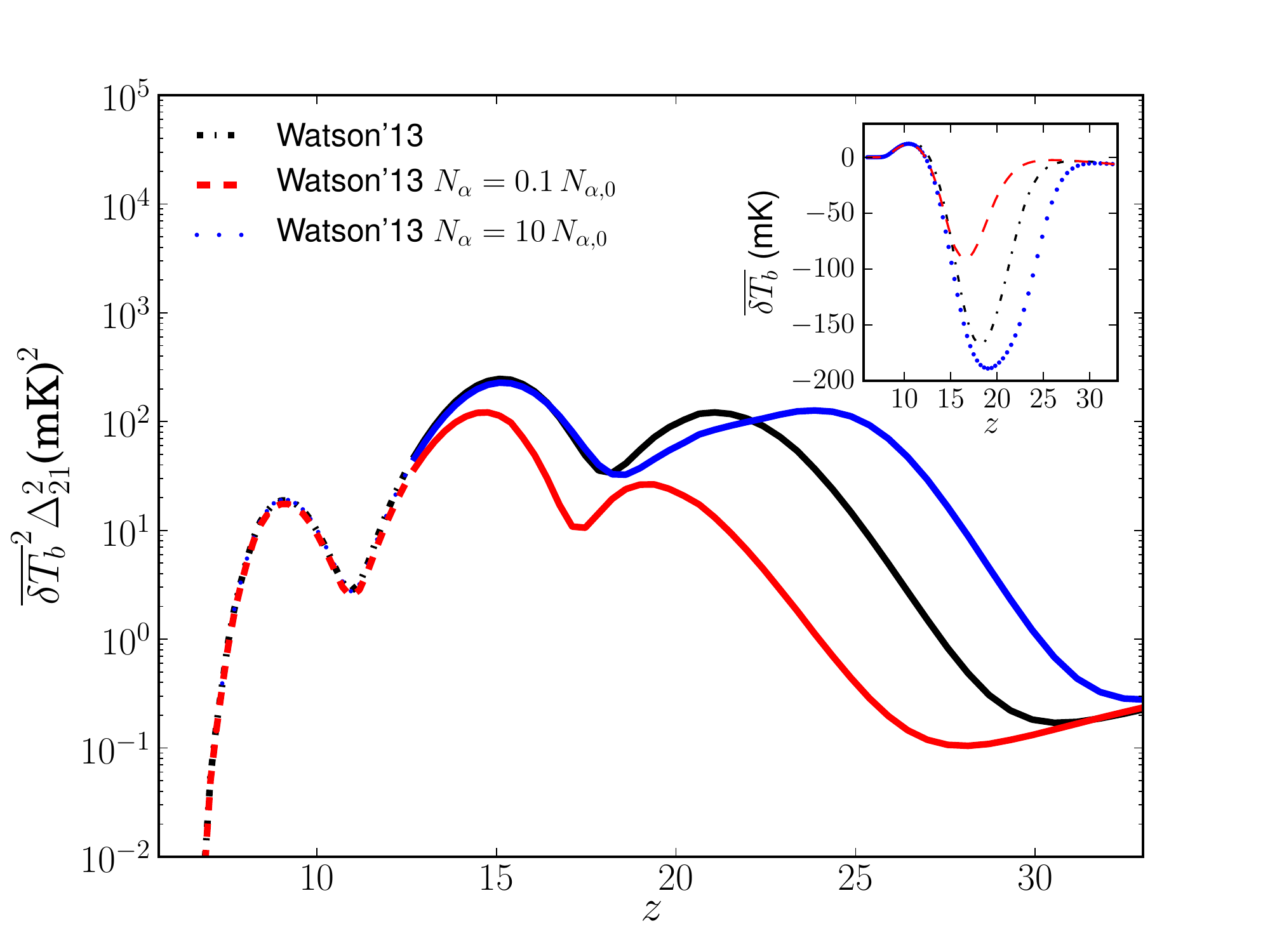}
\end{tabular}
\caption{{\it Global 21~cm signal} represented by the all-sky averaged differential brightness temperature $\overline{\delta T_b}$ as a function of redshift (insets) and the associated \emph{21~cm power spectrum of fluctuations} at a scale $k=0.1 \, {\rm Mpc}^{-1}$ (main panels). In the main panels, we distinguish when the signal is in absorption (solid lines) or emission (non-solid lines). \textit{Top-left panel}: three different halo mass functions (see Sec.~\ref{sec:halo-mass-function}): Press-Schechter~\cite{Press:1973iz, Bond:1990iw} (leftmost red lines); our default, Watson'13~\cite{Watson:2012mt} (middle black lines); and Sheth-Tormen~\cite{Sheth:1999mn, Sheth:1999su} (rightmost blue lines). \textit{Top-right panel}: variations of the minimum virial temperature, $T_{\textrm{vir}}$, with respect to the default value, $T_{\textrm{vir},0} = 10^4$~K (Sec.~\ref{sec:halo-viri-temp}). \textit{Bottom-left panel}: variations of the number of X-ray photons per solar mass, $\zeta_{\rm X}$, with respect to the default value, $\zeta_{\rm X,0} = 10^{56}$ (Sec.~\ref{sec:number-x-ray}). \textit{Bottom-right panel}: variations of the number of photons per stellar baryon between Ly$\alpha$ and the Lyman limit, $N_\alpha$, with respect to the default value obtained by assuming $\sim 4400$ ionizing photons per stellar baryon from Pop II stars, as defined in Ref.~\cite{Barkana:2004vb} (Sec.~\ref{sec:Jalpha}). For the top-right and the bottom panels, we use the W13 halo mass function.}
\label{fig:tbps}
\end{figure}

So far, we have addressed the global 21~cm signal, i.e., the all-sky averaged differential brightness temperature. Although this contains information about the mean contribution from all sources, fluctuations are expected to be significant and of diverse origins (matter distribution, discrete nature of luminous sources, HII regions, galactic halos).  One of the greatest advantages of using an atomic line is that we have access not only to the spatial distribution of the signal, but also to its time dependence via the redshifting of frequency with Hubble expansion.  Noting that the evolution of the signal is much more pronounced at small scales, its separation from the smooth astrophysical foregrounds could in principle be madex easier by studying the spatial fluctuations rather than the global signal. Thus, we turn to the differential brightness temperature power spectrum, defined as
\begin{equation}
\langle  \widetilde{\delta}_{21} (\mathbf{k}, z)  \widetilde{ \delta}_{21}^* (\mathbf{k}^\prime, z) \rangle \equiv (2\pi)^3 \delta^D (\mathbf{k} - \mathbf{k}^\prime) P_{21}(k,z) ~,
\label{eq:P21}
\end{equation}
where $\delta^D$ is the Dirac delta function, the brackets denote an ensemble average, and $\widetilde{\delta}_{21}(\mathbf{k}, z)$ refers to the Fourier transform of ${\delta}_{21}(\mathbf{x}, z)={\delta T}_{b}(\mathbf{x}, z)/ \overline{\delta T_b}(z)-1$. The power spectrum $P_{21}(k,z)$ carries information about the correlations in the spin temperature field and is expected to provide the highest signal-to-noise ratio measurement of the 21~cm line around the EoR in the near future. In what follows, we work with the dimensionless 21~cm differential brightness temperature power spectrum, defined as
\begin{equation}
\Delta^2_{21} (k,z) =\frac{k^3}{2 \pi^2} P_{21}(k,z)~.
\end{equation}

In order to compute the time evolution of both the 21 cm global signal and the power spectrum, we make use of the publicly available package {\tt 21cmFAST}\footnote{Unless indicated otherwise, we use the default inputs of the code.}~\cite{Mesinger:2010ne}, which generates realizations of the evolved density, ionization, peculiar velocity and spin temperature fields from semi-analytic calculations.  The stages 3-6 associated to the global signal evolution mentioned above can clearly be seen in the insets of the four panels of Fig.~\ref{fig:tbps}, for different halo mass function models and astrophysical parameters. In the main panels, we show the 21~cm power spectrum associated to the differential brightness temperature depicted in the insets, for the same redshift range. We show the signal at a scale $k=0.1 \, {\rm Mpc}^{-1}$, which is expected to be reasonably free from foregrounds~\cite{Pober:2013ig}. The 21~cm power spectrum shows three characteristic peaks, associated to the 21~cm absorption/emission stages described above. More precisely, they correspond (from lower to higher redshifts) to the epochs of reionization, heating from X-ray sources and Ly$\alpha$ pumping~\cite{Pritchard:2006sq, Mesinger:2010ne, Baek:2010cm}.

\section{Evolution equations and free parameters}
\label{sec:evo}

As discussed in the previous section, the spin temperature $T_S$ can couple to the gas temperature $T_K$ via collisions and scattering of Ly$\alpha$ photons. Thus, the 21~cm signal depends on the evolution of $T_K$ and the ionized fraction, as well as on the Ly$\alpha$ background. The equations that track the heating history of the gas are obtained from the coupled evolution equations for the ionized fraction in the mostly-neutral IGM, $x_e({\bf x}, z)$, and  the gas kinetic temperature, $T_K({\bf x}, z)$, which are given by~\cite{Mesinger:2010ne}
 \begin{eqnarray}
 \frac{dx_e ({\bf x}, z)}{dz} &=& \frac{dt}{dz} \left(\Lambda_{\rm ion}
 - \alpha_{\rm A} \, C \, x_e^2 \, n_b \, \mathfrak{f}_{\rm H} \right) ~, 
 \label{eq:xe} \\
 \frac{d T_K ({\bf x}, z)}{dz} & = &\frac{2}{3 \, k_B \, (1+x_e)} \, \frac{dt}{dz} \, \sum_\beta \epsilon_\beta 
 + \frac{2 \, T_K}{3 \, n_b} \, \frac{dn_b}{dz}  - \frac{T_K}{1+x_e} \, \frac{dx_e}{dz} ~,
 \label{eq:TK}
\end{eqnarray}
where $n_b=\bar{n}_{b, 0} (1+z)^3 (1+\bar{\delta_b}({\bf x}, z))$ is the total baryon number density, $k_B$ is the Boltzmann constant, $\epsilon_\beta({\bf x}, z)$ is the heating rate per baryon from different sources including Compton scattering (effective at high redshifts, $z \gtrsim 300$) and X-ray heating (the dominant source of energy injection at the epochs of relevance for our study~\cite{Venkatesan:2001cd, Chen:2003gc, Pritchard:2006sq, Zaroubi:2006fx}), $\Lambda_{\rm ion}$ is the ionization rate per baryon, $\alpha_{\rm A}$ is the case-A recombination coefficient, $C\equiv \langle n_e^2 \rangle / \langle n_e \rangle^2$ is the clumping factor, with $n_e$ the electron number density, and $\mathfrak{f}_{\rm H}=N_{\rm H}/N_b$ is the hydrogen number fraction. Notice that we follow Ref.~\cite{Mesinger:2010ne}, assuming that hydrogen and singly ionized helium are ionized to the same degree, i.e., $N_{\rm HII}/N_{\rm HI}= N_{\rm HeII}/N_{\rm HeI}=x_e({\bf x}, z)$.
 
The total Ly$\alpha$ background, which determines the strength of the Wouthuysen-Field effect, has two main components,
\begin{equation}
J_{\alpha} = J_{\alpha,X} + J_{\alpha,\star} ~.
\end{equation}
The contribution from X-ray excitation of HI, $J_{\alpha,X}$, can be related to the X-ray heating rate by balancing the total X-ray energy injection rate to that of redshifted Ly$\alpha$ photons produced from the same X-rays~\cite{Chen:2006zr}. The contribution from stellar photons between Ly$\alpha$ and the Lyman limit, $J_{\alpha,\star}$, arises from the sum of different Lyman-$n$ levels and depends on the type of stellar populations.

Although the number of parameters in the problem is large, for the sake of simplicity and of comparison with other works, we describe the ionization, heating and excitation of the IGM in terms of a reduced number of quantities, which are allowed to vary (one at a time): the halo mass function, $dn/dM$; the minimum virial temperature of galaxy-hosting halos, $T_{\rm vir}$; the number of X-ray photons per solar mass, $\zeta_{\rm X}$; and the number of photons per stellar baryon between Ly$\alpha$ and the Lyman limit, $N_\alpha$. The first two of them are related to the star formation rate, which is in turn proportional to the comoving (X-ray and UV) photon emissivities. The latter two are related to the efficiencies of X-rays and UV photons to heat and excite the IGM. On the other hand, the ionization efficiency of UV photons has been fixed to the default value of the {\tt 21cmFAST} code, $\zeta_{\rm UV} = 31.5$.  The value of $\zeta_{\rm UV}$ is constrained by the redshift at which reionization ends, so we leave $\zeta_{\rm UV}$ unchanged (set so that reionization ends at $z \sim 7$). Moreover, our {\tt 21cmFAST} runs show that, within its allowed range of variation, $\zeta_{\rm UV}$ only affects the 21~cm global signal and power spectrum below $z \simeq 10$ (see Refs.~\cite{Mesinger:2012ys, Christian:2013gma, Mirocha:2015jra} for detailed simulations assuming different values of $\zeta_{\rm UV}$), whereas the effects of DM annihilations show up mainly around the X-ray heating and Ly$\alpha$ pumping epochs, i.e., $10 \lesssim z \lesssim 30$.

The star formation rate is given by
\begin{equation}
R_{\rm SFR} (z) = n_b \, f_\star \, \frac{d f_{\rm coll} (>M_{\rm vir})}{dt} ~,
\end{equation}
where $f_{\rm coll} (>M_{\rm vir})$ is the fraction of mass collapsed in halos more massive than some minimum mass, $M_{\rm vir}$, defined in terms of $T_{\rm vir}$ (cf. Eq.~(\ref{eq:mminT})), and $f_\star$ is the fraction of baryons converted into stars, which we set to $f_{\star, 0} =0.1$. One of the usual underlying assumptions is that the correlation between the star formation rate and the photon luminosities can be extrapolated to high redshifts. Therefore, ionization, heating and excitation critically depend on the fraction of mass collapsed in halos, which is defined in terms of the comoving halo mass function, $dn/dM$, i.e., in terms of the comoving number density of halos per unit mass as a function of redshift,
\begin{equation}
f_{\rm coll} (>M_{\rm vir}) = \int_{M_{\rm vir}} \frac{M}{\rho_0} \, \frac{dn (M,z)}{dM} \, dM ~,
\label{eq:fcoll}
\end{equation}
where $\rho_0$ is the matter density today.

Next, we describe the impact of the variation of these parameters on the 21~cm signal.

\begin{figure}
	\includegraphics[width=0.9\textwidth]{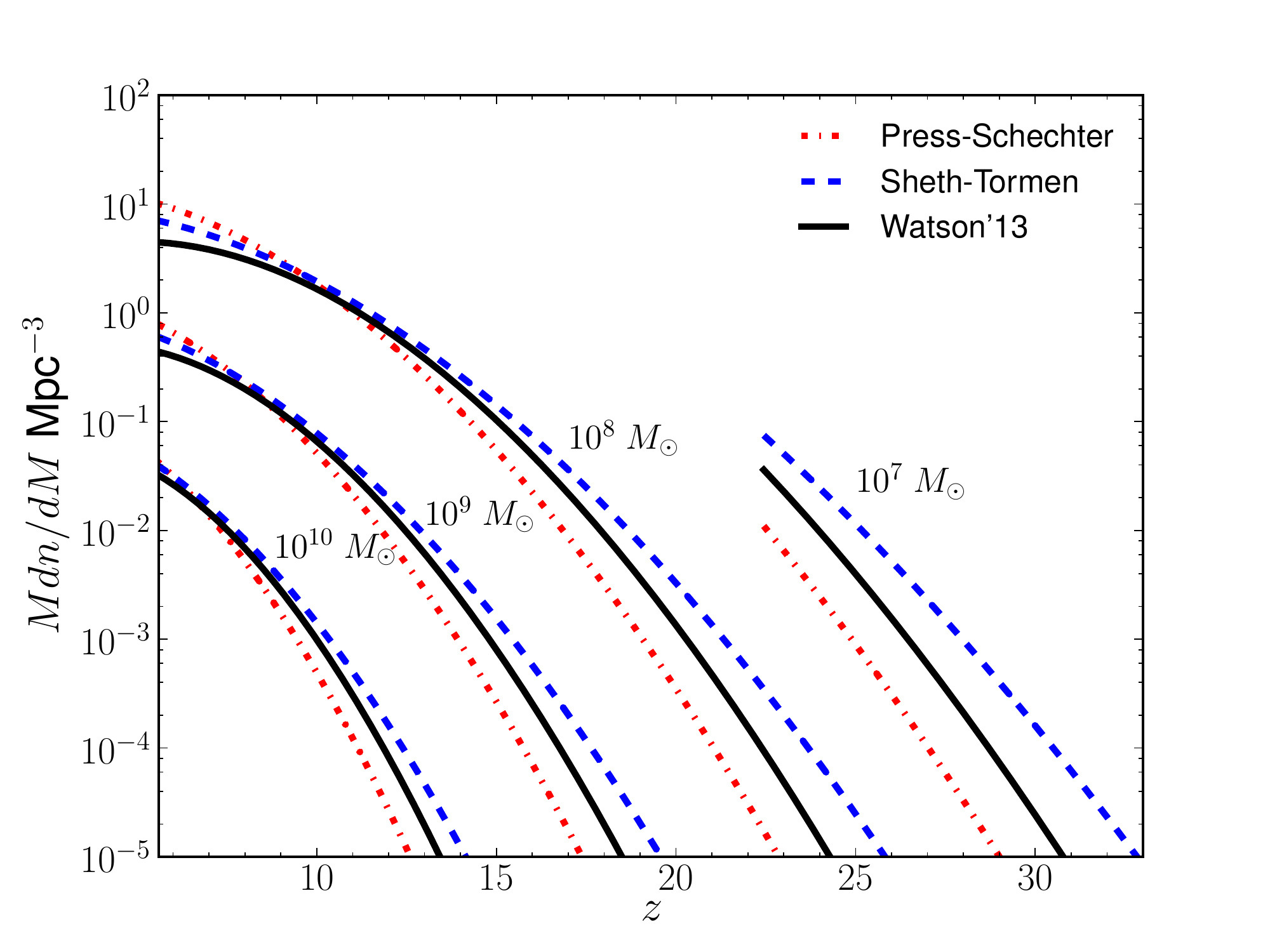} 
	\caption{{\it Comoving halo mass function} as a function of redshift for different values of the halo mass, ranging from $M= 10^{7}M_\odot$ to $10^{10}M_\odot$ (only shown those redshifts for which $M>M_{\rm vir}(z)$, with $T_{\rm vir} = 10^4$~K; see text). We show the results for three parametrizations: Press-Schechter~\cite{Press:1973iz, Bond:1990iw} (dotted red lines); our default halo mass function, Watson'13~\cite{Watson:2012mt} (solid black lines); and Sheth-Tormen~\cite{Sheth:1999su, Sheth:1999mn} (dashed blue lines). }
	\label{fig:hmf}
\end{figure}

\subsection{The halo mass function, $dn/dM$} 
\label{sec:halo-mass-function}

Several approaches have been followed in the literature in order to estimate the density of halos per unit mass (the halo mass function), $dn(M,z)/dM$, mainly from N-body simulations and analytical models, as using observational data is very challenging~\cite{Eke:2005za, Rines:2008tr, Vikhlinin:2008ym, Rozo:2009jj}. Within the context of the 21~cm signal, some previous studies have followed the Barkana-Loeb hybrid prescription~\cite{Barkana:2003qk, Barkana:2007xj}, in which the Press-Schechter (PS) halo mass function~\cite{Press:1973iz, Bond:1990iw} is weighted with the results obtained using the Sheth-Tormen (ST) parametrization~\cite{Sheth:1999mn, Sheth:1999su}, to correct for the underestimation of the halo abundances at high redshifts within the PS formalism. The ST approach matches the numerical simulations better than the PS formalism (at least at low redshifts), but it is also known that ST slightly overestimates the halo abundances at high redshifts~\cite{Iliev:2005sz, Reed:2006rw, Lukic:2007fc, Watson:2012mt}. We therefore use the recent Watson'13 (W13) parametrization (CPMSO + AHF with redshift dependence)~\cite{Watson:2012mt} as our default halo mass function\footnote{The linear matter power spectrum is needed to obtain these halo mass functions, and we compute it using the code provided with Ref.~\cite{Eisenstein:1997jh}, available at \url{http://background.uchicago.edu/~whu/transfer/transferpage.html}.} (as done in Refs.~\cite{Lopez-Honorez:2013lcm, Diamanti:2013bia, Moline:2014xua}) rather than {\tt 21cmFAST}'s default ST parametrization. We illustrate the effects of choosing different halo mass functions on the 21~cm signal in the top-left panel of Fig.~\ref{fig:tbps}.

In Fig.~\ref{fig:hmf}, we show the predictions for the halo mass function of these three models PS, ST and W13, as a function of redshift and for different choices of halo masses $M$ (large enough so that they can host star-forming galaxies, i.e., $M > M_{\rm vir}$). Notice that, with respect to the other two cases, the PS formalism underpredicts the number of halos at high masses and redshifts. However, it overpredicts the halo number density for relatively low masses at low redshifts. The ST halo mass function predicts a slightly higher halo abundance than the W13 parametrization for all masses at all redshifts. Therefore, during the dark ages, the W13 parametrization lies between PS and ST, and is in better agreement with the results of simulations~\cite{Iliev:2005sz, Reed:2006rw, Lukic:2007fc, Watson:2012mt}. As can be seen from the figure, the halo mass function is a decreasing function with the halo mass, so the dominant contribution to $f_{\rm coll}$ comes from halos with masses just above $M_{\rm vir}$ at each redshift, i.e., it comes from halos with masses in the range $10^7 \, M_{\odot} \lesssim M \lesssim 10^8 \, M_{\odot}$ (for $T_{\rm vir} = 10^4$~K).

\subsection{Minimum virial temperature of halos hosting galaxies, $T_{\rm vir}$} 
\label{sec:halo-viri-temp}

The usual approach is to assume that the threshold for hosting star-forming galaxies ($M_{\rm{vir}}$ in Eq.~(\ref{eq:fcoll})) is determined by the cooling of gas at a fixed minimum virial temperature (see, e.g., Ref.~\cite{Barkana:2000fd}), which is assumed to be independent of redshift. This introduces a non-trivial redshift dependence on the minimum virial mass, which is given by~\cite{Barkana:2000fd}
 \begin{equation}
 M_{\rm{vir}} (z) \simeq 10^8 \left(\frac{T_{\textrm{vir}}}{2 \times 10^4 \ \textrm{K}}\right)^{3/2} \left(\frac{1+z}{10}\right)^{-3/2} M_\odot~.
 \label{eq:mminT}
 \end{equation} 
 The default value in the {\tt 21cmFAST} code is $T_{\textrm{vir}, 0}=10^4$~K, which has been identified in the literature with the atomic cooling threshold, as star formation in halos with smaller virial temperatures (and consequently smaller masses) has been argued to be inefficient~\cite{Evrard:1990, Blanchard:1992, Tegmark:1996yt, Haiman:1999mn, Ciardi:1999mx}. This default minimum temperature would thus correspond to a minimum halo mass of $M_{\rm vir} \simeq 3 \times 10^{7} \, M_\odot$ at a redshift $z = 10$.  An upper limit on $T_{\textrm{vir}}$ of $\sim 2 \times 10^5$~K results from the behavior of the emissivities inferred from observations of the Ly$\alpha$ forest (see, e.g., Refs.~\cite{Mesinger:2012ys, Greig:2015qca} and references therein). Therefore, in the top-right panel of Fig.~\ref{fig:tbps}, we consider minimum virial temperatures within the range $[10^4 - 10^5]$~K.

\subsection{Number of X-ray photons per solar mass, $\zeta_{\rm X}$} 
\label{sec:number-x-ray}

In contrast to the ionization efficiency $\zeta_{\rm UV}$, which does not significantly affect the 21~cm signal in the redshift interval of interest here, the uncertainties on the efficiency for ionization, heating and Ly$\alpha$ production by X-ray sources do play an important role. These processes depend on the total X-ray emission rate, which is proportional to the star formation rate and the number of X-ray photons per solar mass in stars, $\zeta_{\rm X}$, i.e., the X-ray efficiency is directly proportional to the product $f_\star \zeta_{\rm X}$. Changing either $f_\star$ or $\zeta_{\rm X}$ leads to similar signatures on the 21~cm observables\footnote{Note that $\zeta_{\rm UV}$ and the Ly$\alpha$ flux are also proportional to $f_\star$. Whereas this dependence could be important for the Ly$\alpha$ flux, it is less so for $\zeta_{\rm UV}$. As mentioned above, within the range allowed by the measured optical depth at reionization, the precise value of $\zeta_{\rm UV}$ turns out not to be so crucial here.}. The default values assumed for these parameters are $f_{\star,0}=0.1$ and $\zeta_{\rm X,0} = 10^{56} \, M_\odot^{-1}$, which would imply, approximately, $N_X \simeq 0.1$ X-ray photons per stellar baryon. In the bottom-left panel of Fig.~\ref{fig:tbps}, we consider $\zeta_{\rm X}$ varying within the range $[10^{54} - 10^{58}] \, M_\odot^{-1}$ (see also, e.g., Refs.~\cite{Valdes:2012zv, Mesinger:2012ys, Christian:2013gma,  Pacucci:2014wwa} for analyses of wide ranges).

\subsection{Number of photons per stellar baryon between Ly$\alpha$ and the Lyman limit, $N_\alpha$} 
\label{sec:Jalpha}

Here, as in {\tt 21cmFast}, we treat the number of photons per stellar baryon between Ly$\alpha$ and the Lyman limit and the ionization rate (driven by $\zeta_{\rm UV}$) as two independent quantities. The default value, $N_{\alpha, 0}$, is obtained by assuming Pop II stars as defined in Ref.~\cite{Barkana:2004vb} and normalizing their emissivity to $\sim 4400$ ionizing photons per stellar baryon. These are the default parameters in {\tt 21cmFast}. In the bottom-right panel of Fig.~\ref{fig:tbps}, we show the 21~cm signal for $0.1 \le N_\alpha/N_{\alpha, 0} \le 10$.

Notice that if the Ly$\alpha$ background flux and the ionization rate are generated by the same stellar population, the number of ionizing photons and the number of photons between Ly$\alpha$ and the Lyman limit are not independent of each other. However, different models for the spectral energy distributions, with different metallicities and stellar initial mass functions, predict different values for the ratio of these two numbers. Whereas in the case of Pop II stars the number of ionizing photons is smaller, Pop III stars, with a much harder spectra, emit more photons contributing to the Ly$\alpha$ background (see, e.g., Refs.~\cite{Ciardi:2003hg, Furlanetto:2006tf, Mirocha:2015jra}). Moreover, only a fraction, $f_{\rm esc}$, of the ionizing photons escape local absorption and thus, this extra parameter also regulates the relative intensity of the ionization rate and the Ly$\alpha$ flux. Thus, in principle, we could treat them as independent quantities.

\subsection{Impact on the 21~cm signal}
\label{sec:impact-21-cm}

Here, we first discuss the impact of the choice of the halo mass function, and then we review how different values for $T_{\rm vir}$, $\zeta_{\rm X}$  and $N_\alpha$ can affect the global 21~cm signal and its power spectrum (see also Refs.~\cite{Mesinger:2012ys, Mesinger:2013nua, Christian:2013gma}). Furthermore, we discuss the possible degeneracies between these parameters. In the following, the remaining parameters in the {\tt 21cmFAST} code are set equal to their default values, and we have integrated from $z=34.5$ to $z=6$. Unless otherwise noted, for all the cases we obtain values for the reionization optical depth of $\tau \sim 0.079$. Note that the \emph{Planck} result combining TT, TE and EE spectra for $\ell \ge 30$ with the low-$\ell$ temperature plus polarization spectra is $\tau_{\rm \it Planck} = 0.079 \pm 0.017$, whereas after adding \emph{Planck} lensing data, the result is $\tau_{\rm {\it Planck}+lensing} = 0.063 \pm 0.014$~\cite{Ade:2015xua}.

In the top-left panel of Fig.~\ref{fig:tbps}, we depict the 21~cm global signal (inset) and the 21~cm differential brightness temperature power spectrum (main panel) for the three halo mass functions previously discussed in Sec.~\ref{sec:halo-mass-function}. Notice that since the PS model (red lines) systematically underpredicts the number of halos at high redshifts and overpredicts it for low masses at low redshifts~\cite{Springel:2005nw, Heitmann:2006hr, Lukic:2007fc, Watson:2012mt}, the peaks in the power spectrum as well as the global absorption dip and the emission peak occur at lower redshifts than for the other two halo mass functions. For the ST model (blue lines), the abundance of halos at high redshifts is larger: the first astrophysical sources switch on at earlier times and therefore, both X-ray heating and Ly$\alpha$ pumping take place at higher redshifts.  The predictions from the more recent W13 halo mass function (black lines) lie in between the ST and the PS ones. However, let us note that, whereas the optical depth at reionization for the PS and W13 models is $\tau_{\rm PS} = 0.077$ and $\tau_{\rm W13} = 0.079$, respectively, for the ST case we obtain $\tau_{\rm ST} = 0.089$. The top-left panel of Fig.~\ref{fig:tbps} clearly illustrates that the uncertainties on the halo mass function have an important impact on the 21~cm signal. In the rest of the panels of Fig.~\ref{fig:tbps}, we show the effects of changing other astrophysical parameters (one at a time and keeping the rest fixed), while using the W13 halo mass function.

In the top-right panel of Fig.~\ref{fig:tbps}, the minimum virial temperature of both X-ray and UV-sources is increased by a factor of 4 (red lines) and by a factor of 10 (blue lines) with respect to our default value (black lines). This range is chosen consistently with the discussion in Sec.~\ref{sec:halo-viri-temp}, although for $T_{\rm vir} = 10^5$~K we obtain a rather small $\tau = 0.047$.  Notice that larger values of the minimum virial temperature induce a shift of the global 21~cm signal to later epochs, as the larger halo mass (or temperature) threshold implies a delay in the formation of X-ray and UV sources. This parameter displays an important correlation with the choice of model for the halo mass function. For instance, the results for the PS halo mass function with $T_{\textrm{vir}}=10^4$~K are very similar to those obtained assuming the ST halo mass function with $T_{\textrm{vir}}=4 \times 10^4$~K (not shown in this work). As we further illustrate in the following sections, these uncertainties limit our ability to discern the effects of DM annihilations from those of astrophysical processes.

In the bottom-left panel of Fig.~\ref{fig:tbps}, the efficiency of X-ray sources is varied by a factor of 0.01 (red lines) and by a factor of 100 (blue lines) with respect to our default value (black lines).  These changes are comparable to some of those explored in the literature (see, e.g., Refs.~\cite{Valdes:2012zv, Mesinger:2012ys, Christian:2013gma,  Pacucci:2014wwa}). The main effect of a higher X-ray efficiency is that the gas kinetic temperature $T_K$ increases at earlier redshifts, shortening the Ly$\alpha$ pumping period and leading to a dip in the brightness temperature at $z \simeq 15-25$, which is less pronounced for larger values of $\zeta_{\rm X}$. Moreover, it leads to an earlier transition from absorption to emission, and consequently shows a higher peak in the differential brightness temperature at $z \simeq 10-15$. This is also translated into the 21~cm power spectrum by a shift of the X-ray heating peak (middle peak), which occurs mainly in emission, to earlier epochs. In addition, we observe a larger difference between the height of the X-ray heating peak and the height of the other two peaks (associated to the epochs of reionization and Ly$\alpha$ pumping). On the other hand, a much lower value of $\zeta_{\rm X}$ would delay the heating of the IGM, shifting the absorption trough to later times. As a result, the X-ray heating peak in the power spectrum would also shift to lower redshifts and could even merge with the reionization peak, which would occur mainly in absorption. This also implies a higher intensity for this peak ($\gtrsim 10^3$~mK$^2$) at $z \sim 10$, as there is a larger contrast of temperatures among different regions.

In the bottom-right panel of Fig.~\ref{fig:tbps}, the number of photons per stellar baryon between Ly$\alpha$ and the Lyman limit is varied by a factor of 0.1 (red lines) and by a factor of 10 (blue lines) with respect to our default value (black lines). The main effect of a higher $N_\alpha$ is to produce a deeper trough in the differential brightness temperature and to shift the Ly$\alpha$ pumping peak of the power spectrum to earlier times. Conversely, a lower value of $N_\alpha$ produces a shallower trough in the differential brightness temperature, shifts the Ly$\alpha$ pumping peak of the power spectrum to later times and can even suppress the intensity of the X-ray heating peak. Note that this parameter is correlated with $\zeta_X$. In principle, the deeper trough in $\overline{\delta T_b}$ for a higher $N_\alpha$ could be compensated by increasing $\zeta_X$. However, this would shift even further the Ly$\alpha$ pumping peak to earlier times. Therefore, changes in $N_\alpha$ cannot be completely compensated by changes in $\zeta_X$ and changes in $T_{\rm vir}$ are also needed.

In summary, even in the absence of signatures from DM annihilations, the large uncertainties of the astrophysical parameters add yet more complexity to the canonical scenario, resulting in important correlations affecting the 21~cm signal. Given these existing degeneracies, extracting information about cosmology or astrophysics from the EoR and beyond will be a challenging task.

\section{The impact of DM annihilations on 21~cm observables}
\label{sec:dark-matter-impact}

The main goal of this paper is to analyze the impact of DM annihilation into Standard Model particles on the 21~cm signal. These annihilations represent an extra source of ionization, heat and excitation of the IGM, which must be added to the astrophysical contributions to the ionization rate, the gas kinetic temperature and the Ly$\alpha$ background flux. In what follows, we discuss the effects of DM annihilations on the $21$~cm signal and for the astrophysical parameters we use our default set of values. Unless otherwise noted, each case we show yields a reionization optical depth of $\tau \sim 0.079$.

\subsection{DM energy deposition}
\label{sec:DMdeposition}

The DM energy deposition is defined as the energy that goes into ionization, heating and excitation of the IGM from the DM annihilation products, whereas the DM energy injection is the total energy liberated by those annihilations. In the case of the smooth DM distribution, before halos start to form, the energy injection rate per unit volume at redshift $z$ is given by
\begin{equation}
\left( \frac{d E}{d V d t}\right)^{\rm smooth}_{\rm injected}=(1+z)^6 \, \rho^2_{\rm DM,0} \, \frac{\left< \sigma v\right>}{ m_{\rm DM}} ~, 
\label{eq:injectedEnergy}
\end{equation}
where $\rho_{\rm DM,0}$ is the DM density today, $\left< \sigma v \right>$ is the thermal average of the DM annihilation cross section times the relative velocity and $m_{\rm DM}$ is the DM particle mass. 

The energy injection and deposition rates can be very different. This is due to the creation of invisible products in DM annihilations and to the free-streaming of electrons and photons, which may travel for a significant time before depositing their energy into the IGM. In previous works (see, e.g., Refs.~\cite{Slatyer:2009yq, Slatyer:2012yq, Galli:2013dna, Lopez-Honorez:2013lcm, Diamanti:2013bia}) the energy deposition rate was split into three parts: the DM energy injection rate, $(dE/dtdV)_{\rm injected}$; the total DM energy deposition efficiency into the IGM, $f(z)$; and the distribution of this energy into different channels, depending on the fraction $\chi_c(x_e(z))$, once this energy has cascaded below a deposition kinetic energy threshold of 3~keV. The index $c = \{ {\rm HI}, {\rm HeI}, {\rm heat}, {\rm Ly}\alpha \}$ corresponds to the hydrogen and helium ionization, heating and excitation (Ly$\alpha$) channels, respectively. We write the dependence of the DM energy deposition rate per baryon for each channel $c$, in terms of the DM energy injection rate, as
\begin{equation}
\epsilon^{\textrm{DM}}_c({\bf x},z)   \equiv  \frac{1}{n_{b}} \left(\frac{dE_c ({\bf x},z)}{dtdV}\right)^{\rm smooth}_{\rm deposited} \equiv  \frac{1}{n_{b}} \, f_c(z) \, \left(\frac{dE ({\bf x},z)}{dtdV}\right)^{\rm smooth}_{\rm injected} ~,
\label{eq:eps} 
\end{equation}
where $f_c(z)$ denotes the DM energy deposition efficiency for a given channel $c$ and $f(z)= \sum_cf_c(z)$. In general, $f(z)$ depends on the DM mass and on the DM annihilation channel, although we omit the explicit referral to these dependences in what follows.

Thus far, the standard approach has been to use $f_c(z) = \chi_cf(z)$ with the simple parametrization of $\chi_c(x_e)$ by Shull and Van Steenberg \cite{Shull:1985} (implemented by Chen and Kamionkowski~\cite{Chen:2003gz}, hereafter SSCK). However, here we follow a more up-to-date approach. Recent work by Slatyer~\cite{Slatyer:2015kla} in the context of DM annihilations and decays and their impact on CMB observables~\cite{Slatyer:2015jla} updated earlier calculations of the energy injection and deposition efficiencies~\cite{Slatyer:2009yq, Slatyer:2012yq} by computing channel-dependent transfer functions $T^c(z,z',E)$, which map the injected energy at redshift $z'$ to the deposited energy at redshift $z$. Integrating this function over $z'$ and over the injected particle spectrum yields the channel-dependent efficiency $f_c(z)$. This is more accurate than the SSCK approach, since it is the result of a full calculation of the behavior of $e^+e^-$ pairs and photons throughout the Universe's history, based on the redshift-dependent properties of the IGM.  Crucially, Ref.~\cite{Slatyer:2015kla} showed that a non-negligible fraction of the energy deposited below 3 keV was in the form of particles that are not energetic enough to ionize, heat or excite the medium, and which instead freely stream, leading only to distortions of the CMB spectrum. Once this effect has been accounted for, a proper computation of $f_c(z)$ can yield a 10\%-50\% difference with respect to using the SSCK approach\footnote{To connect with the notation used here, this would be written as $f_c^{\rm SSCK}(z) = \chi_c^{\rm SSCK}(x_e)f(z)$.}. For our computations, in order to evaluate the channel-dependent efficiency $f_c(z)$, we use the numerical transfer function tables $T^c(z,z',E)$ provided as supplementary material\footnote{\url{http://nebel.rc.fas.harvard.edu/epsilon/}} to Ref.~\cite{Slatyer:2015kla}.
 
The terms to be added to the evolution equations for the ionized fraction and for the kinetic temperature of the gas (see Eqs.~(\ref{eq:xe}) and (\ref{eq:TK})), in the presence of DM annihilations, take the following form:
 \begin{eqnarray}
\left. \Lambda_{\rm ion} \right|_{\textrm{DM}} & = & \mathfrak{f}_{\rm H} \, \frac{\epsilon^{\rm DM}_{\rm HI}}{E_{\rm HI}} +\mathfrak{f}_{\rm He} \, \frac{\epsilon^{\rm DM}_{\rm HeI}}{E_{\rm HeI}} ~, \\
\left. \frac{dT_K}{dz} \right|_{\textrm{DM}} & = & \frac{dt}{dz} \, \frac{2}{3 \, k_B \, (1+x_e)} \, \epsilon_{\rm heat}^{\textrm{DM}} ~, 
 \end{eqnarray}
where $E_{\rm HI, HeI}$ are the ionization energies for hydrogen and helium and $\mathfrak{f}_{\rm He}=N_{\rm He}/N_b$ is the helium number fraction. The Ly$\alpha$ flux resulting from collisional excitations due to energy deposition by DM annihilations in the IGM reads as
 \begin{equation}
J_{\alpha,\mathrm{DM}}=\frac{c\, n_b}{4\pi} \frac{\epsilon_{\mathrm{Ly}\alpha}^{\textrm{DM}}}{h\nu_\alpha}  \frac{1}{H(z)\nu_\alpha} ~,
\label{eq:Jalph}
 \end{equation}
where $\nu_\alpha$ is the emission frequency of a Ly$\alpha$ photon.

\subsection{The impact of halo formation}
\label{sec:structure-formation}

We now turn to the effect of structure formation on the energy deposition into the IGM from DM annihilations. Once halos start to form, the Universe turns into a clumpy structure and the global DM annihilation rate, which depends on the square of the local DM density, is greatly enhanced. Thus, the effects of DM annihilations on the 21~cm signal at $10 \lesssim z \lesssim 30$ are expected to be completely dominated by the contribution from halos~\cite{Chuzhoy:2007fg, Cumberbatch:2008rh}. The energy injection rate at redshift $z$ can be written as~\cite{Cirelli:2009bb}
\begin{equation}
 \left(\frac{dE (z)}{dtdV}\right)_{\rm injected} = \frac{\left< \sigma v \right>}{m_{\rm DM}} \, \rho_{\rm DM,0}^{2} \, (1+z)^{6}  \, [1 + {\cal B}(z)]  ~.
 \label{eq:DMenergy}
 \end{equation}
The function ${\cal B}(z)$ refers to the boost due to structure formation\footnote{The function ${\cal B}(z)$ corresponds to the function $G(z)$ in our previous work~\cite{Lopez-Honorez:2013lcm}.} and it depends on the abundance of halos and their density profiles,
\begin{equation}
{\cal B}(z) = \frac{1}{\rho^2_{\rm DM,0} \, (1+z)^3} \, \int_{M_{\rm min}} \frac{d n (M,z)}{d M} \, dM \, \int_0^{R_{\rm vir}}  \, \rho^2(r) \, 4 \pi r^2 dr  ~,
\label{eq:Bz}
\end{equation}
where $dn/dM$, as in the previous sections, is the comoving halo mass function, $\rho(r)$ is the DM halo density profile, which also depends on the halo mass and redshift, and $R_{\rm vir}$ is the halo virial radius. The value of the minimum halo mass, $M_{\rm min}$, depends on the DM particle properties and on the cosmological model~\cite{Schmid:1998mx, Zybin:1999ic, Boehm:2000gq, Chen:2001jz, Hofmann:2001bi, Berezinsky:2003vn, Boehm:2003xr, Boehm:2004th, Green:2005fa, Loeb:2005pm, Profumo:2006bv, Bertschinger:2006nq, Bringmann:2006mu, Bringmann:2009vf, vandenAarssen:2012ag, Cornell:2012tb, Gondolo:2012vh, Cornell:2013rza, Shoemaker:2013tda, Diamanti:2015kma} and thus, it can lie within a wide range. Here we consider three representative values, $M_{\rm min} =10^{-12} \, M_\odot, \, 10^{-6} \, M_\odot$ and $10^{-3} \, M_\odot$. In order to compute ${\cal B}(z)$ we use our default halo mass function, W13~\cite{Watson:2012mt}, and we assume DM in halos to be distributed with a Navarro-Frenk-White (NFW) profile~\cite{Navarro:1995iw, Navarro:1996gj} with a concentration-mass relation, dependent on redshift, given by Ref.~\cite{Prada:2011jf} (P12). 

Nevertheless, Eq.~(\ref{eq:Bz}) only considers the smooth DM distribution of halos, but it does not include the contribution from substructure inside those halos. We also add this contribution by making the substitution~\cite{Strigari:2006rd, Kuhlen:2008aw}
 \begin{equation}
 \int_0^{R_{\rm vir}}  \, \rho^2(r) \, 4 \pi r^2 \, dr \rightarrow \int_0^{R_{\rm vir}}  \, \rho^2(r) \, 4 \pi r^2 \, dr + \int_{M_{\rm min}}^{M} \frac{dn_{\rm sub}}{dm} \, dm \int_0^{r_{\rm vir}}  \, \rho_{\rm sub}^2(r_{\rm sub}) \, 4 \pi \,  r_{\rm sub}^2 \, dr_{\rm sub} ~, 
 \end{equation}
where $dn_{\rm sub}/dm$ is the comoving subhalo mass function, $m$ is the subhalo mass, $\rho_{\rm sub}(r_{\rm sub})$ is the subhalo density profile, which depends on the subhalo mass and redshift, and $r_{\rm vir}$ is the subhalo virial radius. For the subhalo mass function in a host halo of mass $M$, we use $dn_{\rm sub}/dm = A/M \left(m/M\right)^{-\alpha}$, where values of $\alpha$ in the range $[1.9, \, 2]$ are found in simulations~\cite{Diemand:2006ik, Madau:2008fr, Springel:2008cc}. We have chosen $\alpha=2$, which gives rise to the largest effect due to substructure, and we set $A=0.012$~\cite{Sanchez-Conde:2013yxa}. On the other hand, following previous works~\cite{Strigari:2006rd, Martinez:2009jh, Sanchez-Conde:2013yxa, Ishiyama:2014uoa}, we consider the concentration parameter to be equal to that of halos. We also include the effect of substructure, but unlike these previous studies, only up to the first level, which turns out to be closer to the results obtained once tidal forces inside halos and subhalos are taken into account~\cite{Moline:2016pbm}. However, let us note that subhalos are found to be, on average, more concentrated than host halos of the same mass (see, e.g., Refs.~\cite{Ghigna:1999sn, Bullock:1999he, Ullio:2002pj, Diemand:2007qr, Diemand:2008in, Diemand:2009bm}) and larger concentrations in subhalos can lead to a factor of two increase in the annihilation rate~\cite{Moline:2016pbm}. Given all the uncertainties, we do not expect this effect to have an important impact on our results and we do not take it into account.

\begin{figure}
	\includegraphics[width=.9\textwidth]{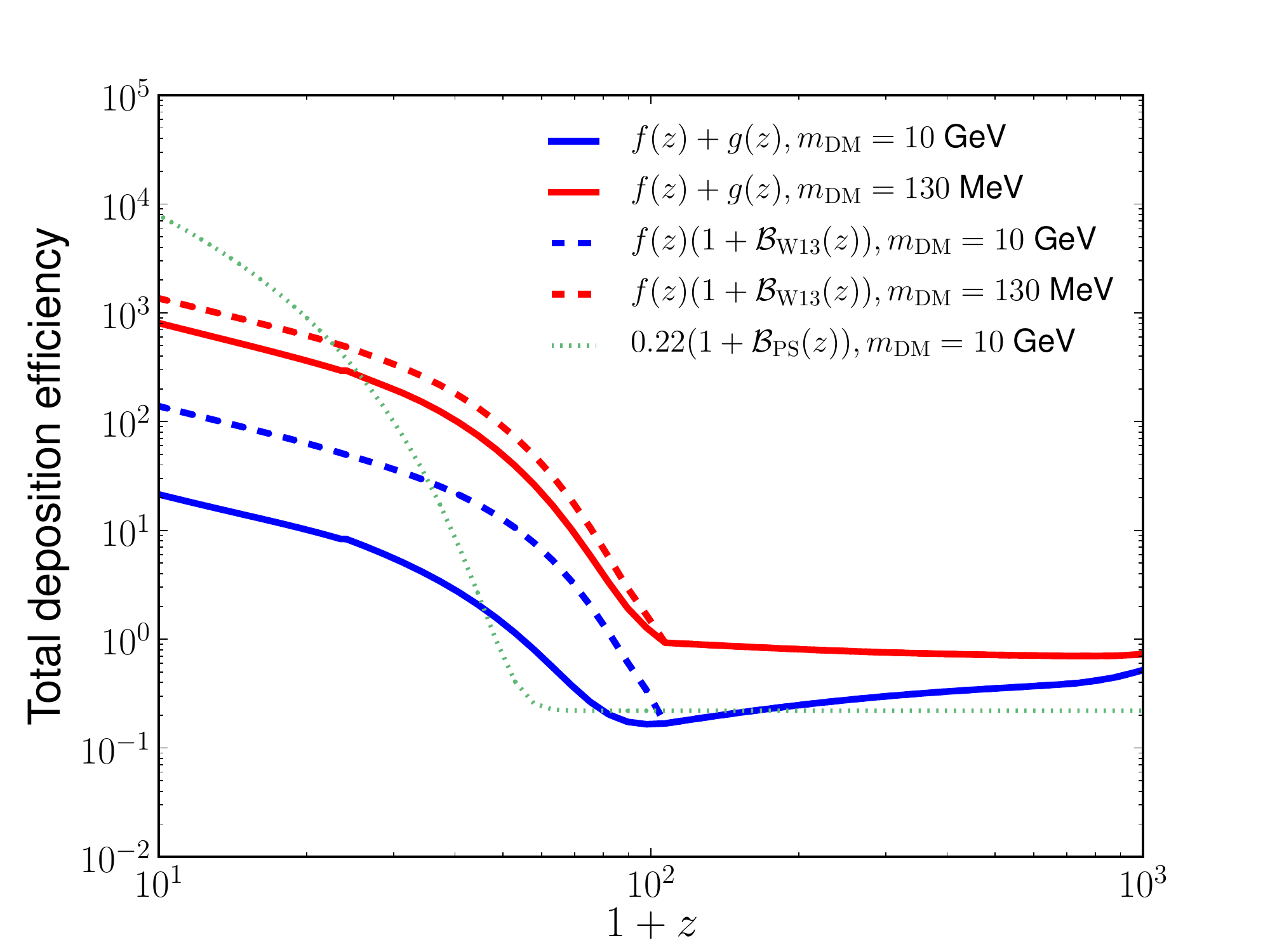}
	\caption{{\it Total efficiencies}, $\sum_c [f_c(z)+g_c(z)]$, corresponding to different benchmark DM models. We depict the efficiencies associated to $m_{\rm DM}=10$~GeV (blue lines) and $m_{\rm DM}=130$~MeV (red lines) for annihilations into $e^+e^-$, following the approach described in the text, i.e., $f(z) + g(z)$ (solid lines) and factoring $[1 + {\cal B}_{\rm W13}(z)]$ out of the integral over the transfer functions, i.e., $f(z) [1+{\cal B}_{\rm W13}(z)]$ (dashed lines). In all these cases we use the P12 concentration-mass relation and the W13 halo mass function with $M_{\rm min} =10^{-12} \, M_\odot$ and adding one level of substructure, i.e., our default DM settings. We also show (green dotted lines) the result for $m_{\rm DM}=10$~GeV, but factoring $[1 + {\cal B}_{\rm PS}(z)]$ out of the integral over the transfer functions and using the PS boost factor with $M_{\rm min} =10^{-9} \, M_\odot$, as defined in Ref.~\cite{Evoli:2014pva}. For the latter case, the $\sum_c f_c$ is assumed to be constant and equal to 0.22 (see Sec.~\ref{sec:comp-with-prev}), corresponding to annihilations into $\mu^+\mu^-$.}
	\label{fig:efficiencies}
\end{figure}

In order to relate the deposited and injected energies when including halo formation, some previous works~\cite{Chuzhoy:2007fg, Cumberbatch:2008rh, Giesen:2012rp, Valdes:2012zv, Evoli:2014pva} have used the parametrization given by Eq.~(\ref{eq:eps}), replacing Eq.~(\ref{eq:injectedEnergy}) by Eq.~(\ref{eq:DMenergy}), and factoring out the halo enhancement ${\cal B}(z)$ and the energy deposition efficiency $f(z)$. However, to self-consistently describe the distinction between energy \textit{injection} (from DM annihilations) and \textit{deposition} (ionization, heating and excitation), the boost ${\cal B}(z)$ should rather be included inside the integral over the transfer functions~\cite{Slatyer:2012yq}:
\begin{equation}
\int dz' \, T^c(z,z',E) \rightarrow \int dz' \, [1 + {\cal B}(z)] \, T^c(z,z',E) ~.
\end{equation}
For practicality, we separate this integral into the smooth part, $f_c(z)$, and the halo part, $g_c(z)$. In this language, the DM energy deposition rate per baryon can be written as~\cite{Lopez-Honorez:2013lcm, Diamanti:2013bia}
\begin{equation}
 \frac{1}{n_{b}} \left(\frac{dE_c (z)}{dtdV}\right)_{\rm deposited} = p_{\textrm{DM}} \, \left[f_c(z) + g_c(z)\right] \, (1+z)^3 ~,
\label{eq:hrcosmorec} 
\end{equation}
where 
\begin{equation}
p_{\rm DM}= 1.5 \times 10^{-24} \, \left(\frac{100 \, {\rm GeV}}{m_{\rm DM}} \right)  \, \left(\frac{\left< \sigma v \right>}{3\times 10^{-26} {\rm cm}^3/{\rm s}}\right) \, \left(\frac{\Omega_{\rm DM}h^2}{0.13}\right)^2 \, {\rm eV/s} ~.
\end{equation}

Throughout the rest of the paper, as our default DM settings to compute DM annihilation signals, we use the P12 concentration-mass relation, the W13 halo mass function with $M_{\rm min} =10^{-12} \, M_\odot$, add one level of substructure, and assume DM annihilations with a branching ratio of 100\% into $e^+e^-$.

In Fig.~\ref{fig:efficiencies}, we show the total DM energy deposition efficiency $\sum_c \left[f_c(z) + g_c(z)\right]$ as a function of redshift for several DM scenarios. We consider annihilations with our default DM settings for two different DM masses: $m_{\rm DM}=10$ GeV (blue lines) and $m_{\rm DM}=130$~MeV (red lines). We illustrate the differences between the results obtained following the approach described above, i.e., $f(z) + g(z)$ (solid lines) and when $[1 + {\cal B}_{\rm W13}(z)]$ is factored out of the integral over the transfer functions (dashed lines), in which $g(z)= f(z) {\cal B}_{\rm W13}(z)$. The latter gives rise to the overestimation of the DM energy deposition rate after halos start to form. However, the lower the DM mass, the closer the result is to the on-the-spot approximation and thus, the less important this overestimation is~\cite{Slatyer:2012yq}. In the mass interval $m_{\rm DM} = [10$~GeV$ - 1$~TeV$]$, the differences increase from one to two orders of magnitude, but for $m_{\rm DM} \lesssim 100$~MeV the differences are below a factor of two. Moreover, it turns out that the fraction of injected energy proceeding into ionization, heating and excitation presents a maximum for injection energies of $\sim 100$~MeV~\cite{Slatyer:2012yq, Slatyer:2015kla} and therefore, the maximal total efficiency occurs for $m_{\rm DM} \sim 100$~MeV. As can be seen in Fig.~\ref{fig:efficiencies}, this results into an efficiency for $m_{\rm DM} = 130$~MeV (red solid line) which is a factor of $\sim 40$ higher than that for $m_{\rm DM} = 10$~GeV (blue solid line) at $z \sim 10-30$. On the other hand, if the energy absorption is not correctly accounted for, i.e., if one simply factors $[1 + {\cal B}_{\rm W13}(z)]$ out of the integral over the transfer functions, the difference between the efficiencies for these masses (dashed lines) is reduced to a factor of $\sim 10$. Regardless of whether $[1 + {\cal B}_{\rm W13}(z)]$ is factored out or not, the efficiency for $m_{\rm DM} \sim  100$~MeV is about an order of magnitude higher than that for $m_{\rm DM} \lesssim 10$~MeV. At these lower masses, most of the energy is converted into low-energy continuum photons with energies too low to heat or excite the medium. These results have implications regarding the range of masses for which the largest signal in the $21$~cm observables is expected, as we discuss in the following sections. 

In Fig.~\ref{fig:efficiencies}, we also illustrate the case of annihilations of DM particles with a mass $m_{\rm DM} = 10$~GeV (green dotted lines), that is meant to reproduce the results presented in Ref.~\cite{Evoli:2014pva}. We have used the PS boost factor with $M_{\rm min} =10^{-9} \, M_\odot$, as defined in Ref.~\cite{Evoli:2014pva}. In order to obtain the DM energy deposition rate corresponding to annihilations into $\mu^+\mu^-$, we use\footnote{Notice that in Ref.~\cite{Evoli:2014pva}, the {\tt MEDEA2} code~\cite{Valdes:2009cq} is used to evaluate the fractional energy release into each channel by a single DM annihilation, i.e., the combination $\chi_{\rm c} f(z)$.} an overall constant total efficiency $\sum_c f_c = 0.22$, which is multiplied by the factor $[1+{\cal B}_{\rm PS}(z)]$, i.e., factoring this term outside the integral over the transfer functions. In this case, the total deposition efficiency is two orders of magnitude higher than the total efficiency obtained following the method described above for $m_{\rm DM} = 10$~GeV (blue solid line)\footnote{Recall that the blue and red lines in Fig.~\ref{fig:efficiencies} correspond to the most optimistic case of DM annihilations into $e^+e^-$.} at $z \sim 10-30$. Notice that, for $m_{\rm DM} = 10$~GeV, the differences arise from both, the fact of using different approaches to calculate the energy deposition rate from DM annihilations in halos and the different halo boosts which have been considered (see further discussion in Sec.~\ref{sec:comp-with-prev}).

We end this section on a technical note. The data provided by Ref.~\cite{Slatyer:2015kla} do not extend below $z = 11$, which would involve complex modeling of heating and UV emission during reionization. To bridge this gap, we interpolate an effective $\chi_c(x_e) \equiv f_c(z)/\sum_cf_c(z)$ function, which we map to the ionization fraction $x_e(z)$ in our reionization model below $z = 24$. While a precise determination of $f_c(z)$ and $g_c(z)$ during reionization would require a very detailed (and model-dependent) simulation of a complicated IGM, the approach used here is, in principle, more accurate and appropriate than using the ad-hoc, commonly considered SSCK approximations.

\subsection{DM and 21 cm observables}
\label{sec:dark-matt-annih}

\begin{table}[t]
	\begin{center} \begin{tabular}{| c | c | c |  c | c | c |} 
			\hline \hline
			$m_{\rm DM}$ [GeV] \, & 0.001 & 0.009 & 0.13 & 1.1  & 10 \\[1ex] 
			$\left<\sigma v\right>$ [cm$^3$/s] \, & \, $10^{-30}$ \, & \, $10^{-29}$ \, & \, $ 10^{-28}$ \, & \,  $10^{-27}$ \, & \, $10^{-26}$ \, \\
			\hline \hline
		\end{tabular}\end{center}
		\caption{Table of the DM masses and the associated values of the thermal average of the annihilation cross section times the relative velocity considered in this work. These cross sections are chosen after rounding the values of the \emph{Planck} CMB limits~\cite{Ade:2015xua}.}
		\label{tab:msv}
\end{table}

We now address the effects on the 21~cm signal of the injection of electromagnetic energy  into the IGM by the products from DM annihilations. For that purpose, we use Eq.~(\ref{eq:hrcosmorec}) to describe DM energy deposition. In this work, we consider, as our default case, DM annihilations with a branching ratio of 100\% into $e^{+}e^{-}$; of all channels, this choice leads to the largest possible effect on the IGM, and thus on the 21~cm signal. We take values of the annihilation cross section that saturate the limit from the \emph{Planck} CMB analysis, i.e., $p_{\rm ann} = f_{\rm eff}\left<\sigma v\right>/m_{\rm DM}= 4.1\times 10^{-28} \, {\rm cm}^3/{\rm s}/{\rm GeV}$~\cite{Ade:2015xua}. In Tab.~\ref{tab:msv}, we indicate the pairs of masses and annihilation cross sections we consider. The values of $\left<\sigma v\right>$ have been chosen by using an efficiency parameter $f_{\rm eff} $ (in the notation of Ref.~\cite{Ade:2015xua}) equal to our $f(z=600)$, but have been rounded. In addition, when considering the imprint of DM annihilations on the 21~cm signal, unless mentioned otherwise, we have made use of {\tt cosmorec}~\cite{Chluba:2009uv} (following the procedure described in Refs.~\cite{Lopez-Honorez:2013lcm, Diamanti:2013bia}, with the updated $f_c(z)$ and $g_c(z)$ considered here) to generate new tables to initialize $x_e$ and $T_K$ in {\tt 21cmFAST}\footnote{For the $m_{\rm DM} = 130$~MeV case, using our default DM settings, we get a factor of $\sim 2.5$ enhancement of $T_K$ and a factor of $\sim 6$  enhancement of $x_e$ at $z=35$, compared to the default {\tt RECFAST} tables~\cite{Seager:1999bc} available in {\tt 21cmFAST}.}.

\begin{figure}
	\begin{tabular}{c }
		\includegraphics[width=0.9\textwidth]{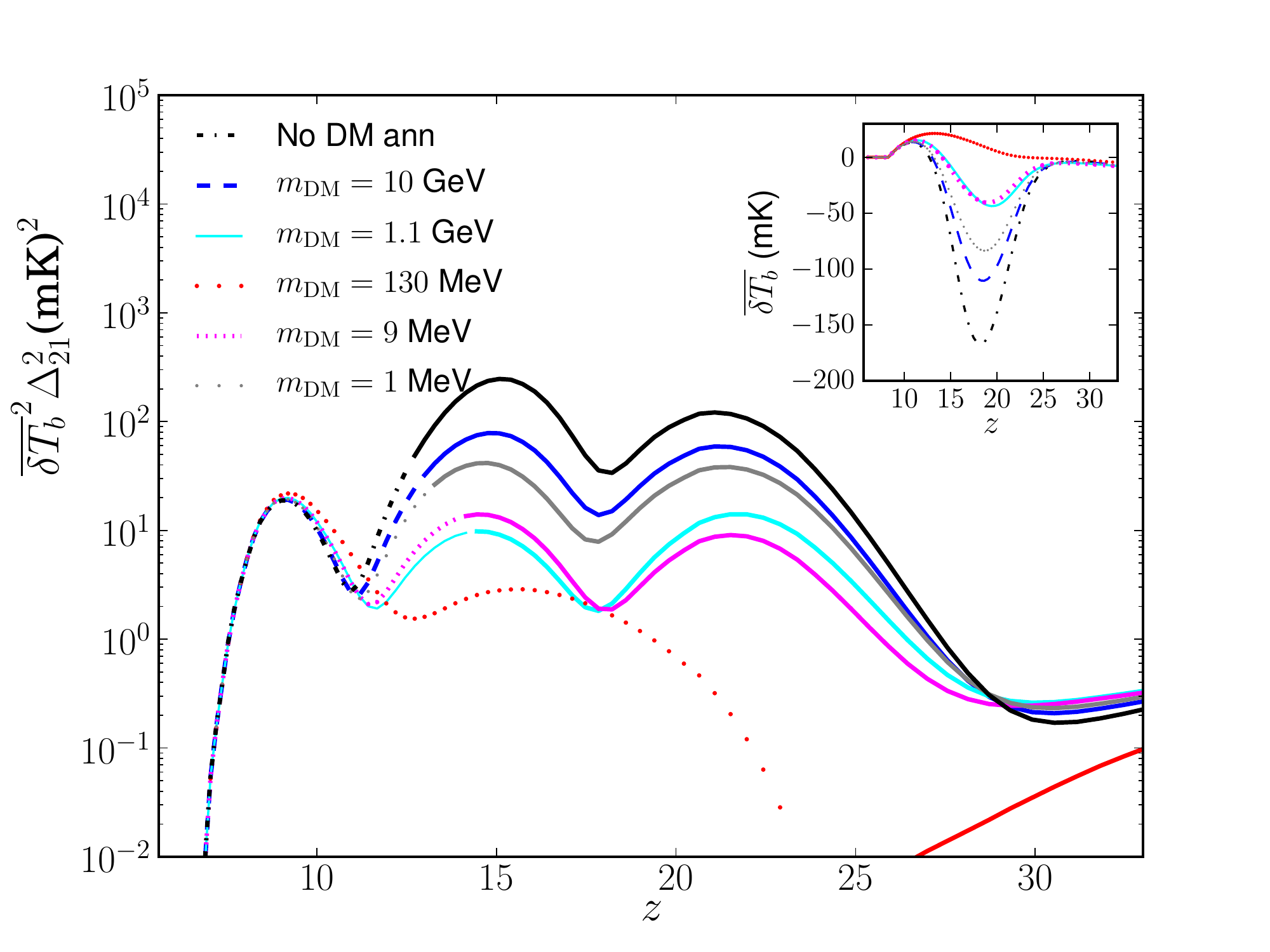} 
	\end{tabular}
	\caption{{\it Effect of varying the DM mass}. Global 21~cm signal (inset) and the associated power spectrum of fluctuations at a scale $k=0.1 \, {\rm Mpc}^{-1}$ (main panel), using our default DM settings, for five different values of the DM mass, with annihilation cross sections saturating the \emph{Planck} CMB limits~\cite{Ade:2015xua} (see Tab.~\ref{tab:msv}). From bottom (top) to top (bottom) in the main panel (inset): $m_{\rm DM} = 130$~MeV (red lines), $1.1$~GeV (cyan lines), $9$~MeV (magenta lines), $1$~MeV (gray lines), $10$~GeV (blue lines) and without DM annihilations (black lines). In the main panel, we distinguish when the signal is in absorption (solid lines) or in emission (non-solid lines).}	
	\label{fig:tbpsdm-M}
\end{figure}

The overall effect of DM annihilations on the 21~cm observables consists of an earlier and more uniform heating of the IGM, as has already been stressed~\cite{Evoli:2014pva}.  In Fig.~\ref{fig:tbpsdm-M}, we illustrate the impact of DM annihilations on the 21~cm global signal (inset) and the associated power spectrum at $k = 0.1 \, {\rm Mpc}^{-1}$ (main panel), using our default DM settings, for different DM masses\footnote{Although not illustrated here, the effects of higher masses, with smaller energy deposition efficiencies, are even weaker than those for $m_{\rm DM} = 10$~GeV.} (from bottom to top in the main panel): $m_{\rm DM} = 130$~MeV (red lines), $1.1$~GeV (cyan lines), $9$~MeV (magenta lines), $1$~MeV (gray lines), $10$~GeV (blue lines), with annihilation cross sections saturating the \emph{Planck} CMB limits~\cite{Ade:2015xua} (see Tab.~\ref{tab:msv}), and without including DM annihilations (black lines). In the main panel, we distinguish when the signal is in absorption (solid lines) or emission (non-solid lines). We see that the annihilations of the $m_{\rm DM} = 130$~MeV DM candidate have the strongest impact. As anticipated above, this is mainly due to the fact that the deposition efficiency is always much larger for $m_{\rm DM} = 130$~MeV in the redshift range relevant to our analysis. On the other hand, the lower the mass, the larger the DM number density, and a priori one would expect a larger effect for lower masses. However, the existing CMB limits follow the same behavior.  As a result, the strongest effect that remains allowed by CMB data is expected to be obtained for $m_{\rm DM} \simeq 130$~MeV and the weakest one for $m_{\rm DM} = 10$~GeV. In the following, we consider these two illustrative cases ($m_{\rm DM} = 130$~MeV and $10$~GeV) as our benchmarks and we compare them to the case without DM annihilations, with our default values for the astrophysical parameters. In summary, DM annihilations suppress the absorption trough in the 21~cm differential brightness temperature before heating from X-ray sources becomes dominant. This effect is seen as a suppression of the X-ray heating peak in the 21~cm power spectrum and can even result in the disappearance of the Ly$\alpha$ pumping peak.

\begin{figure}
	\begin{tabular}{c }
		\includegraphics[width=0.9\textwidth]{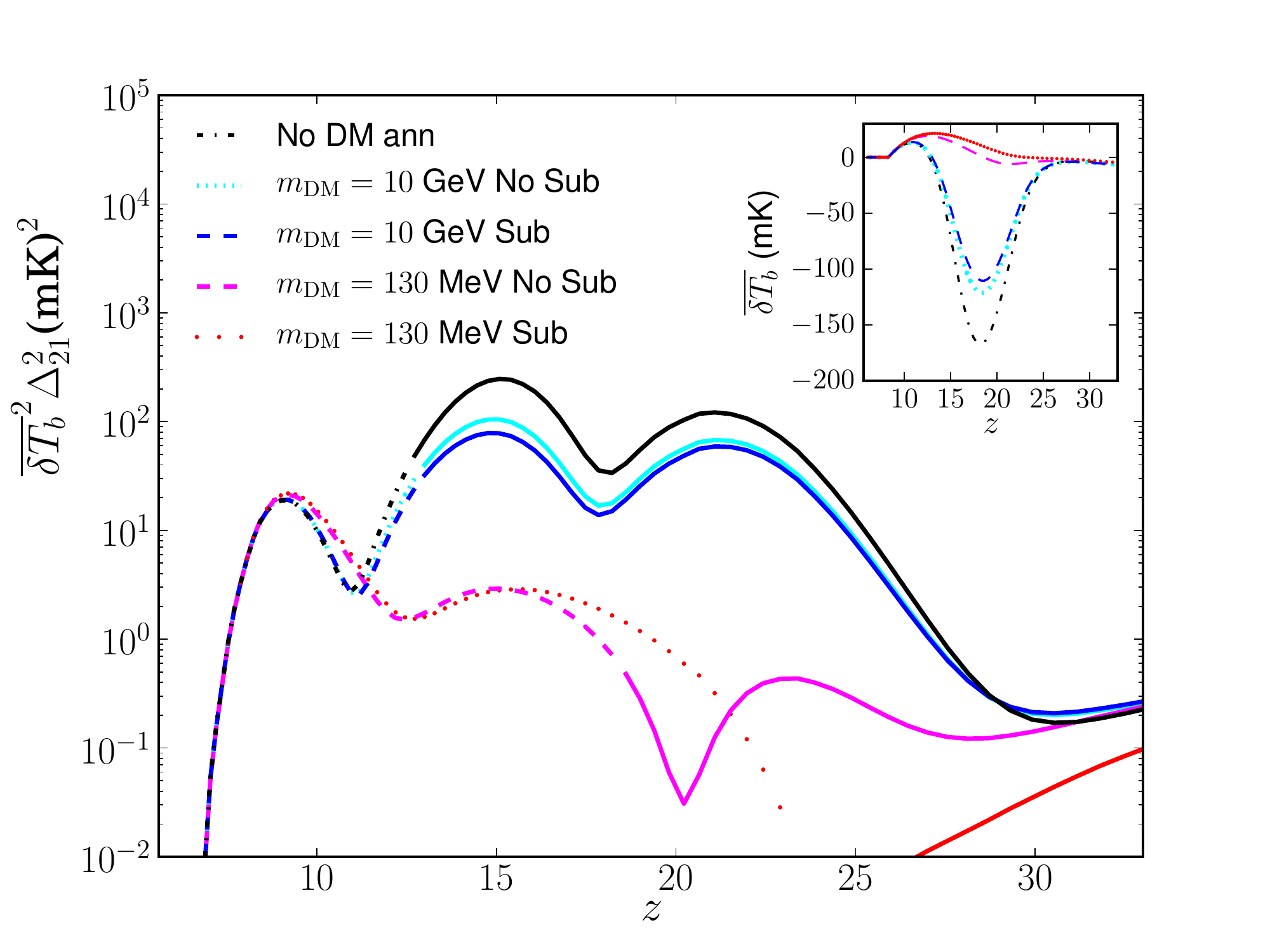} 
	\end{tabular}
	\caption{{\it Effect of the addition of substructure}. Global 21~cm signal (inset) and the associated power spectrum of fluctuations at a scale $k=0.1 \, {\rm Mpc}^{-1}$ (main panel) for $m_{\rm DM} = 10$~GeV, without (cyan lines) and with (blue lines) substructure,  and $m_{\rm DM} = 130$~MeV, without (magenta lines) and with (red lines) substructure, with annihilation cross sections saturating the \emph{Planck} CMB limits~\cite{Ade:2015xua} (see Tab.~\ref{tab:msv}), and using the rest of our default DM settings. The case without DM annihilations is also depicted (black lines). In the main panel, we distinguish when the signal is in absorption (solid lines) or in emission (non-solid lines).}	
	\label{fig:tbpsdm-subs}
\end{figure}

\begin{figure}
	\begin{tabular}{c }
		\includegraphics[width=0.9\textwidth]{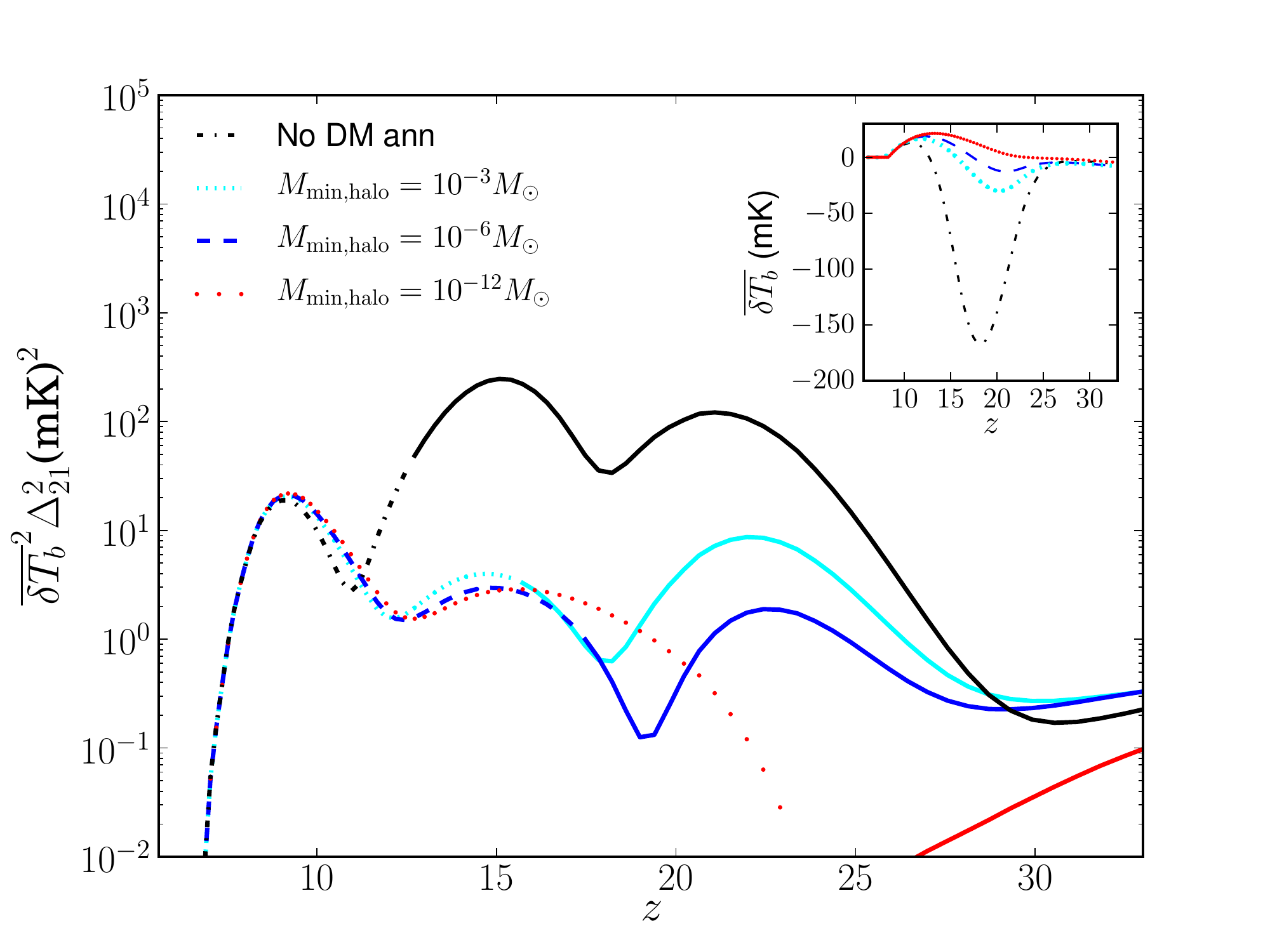} 
	\end{tabular}
	\caption{{\it Effect of varying $M_{\rm min}$}. Global 21~cm signal (inset) and the associated power spectrum of fluctuations at a scale $k=0.1 \, {\rm Mpc}^{-1}$ (main panel) for $m_{\rm DM} = 130$~MeV and $\left<\sigma v\right> = 10^{-28} \, {\rm cm}^3/{\rm s}$. We show the result for three values of the minimum halo mass: $M_{\rm min} = 10^{-12} \, M_\odot$ (red lines), $10^{-6} \, M_\odot$ (blue lines) and $10^{-3} \, M_\odot$ (cyan lines); using the rest of our default DM settings. The case without DM annihilation is also depicted (black lines). In the bottom panel, we distinguish when the signal is in absorption (solid lines) or in emission (non-solid lines).}	
	\label{fig:tbpsdm-h}
\end{figure}

In Fig.~\ref{fig:tbpsdm-subs}, we show the effect of adding substructure to the smooth halo contribution, as described in Sec.~\ref{sec:structure-formation}, on the global 21~cm differential brightness temperature (inset) and the 21~cm power spectrum at $k=0.1 \, {\rm Mpc}^{-1}$ (main panel). Again, in the main panel, we distinguish when the signal is in absorption (solid lines) or emission (non-solid lines). The results are shown for $m_{\rm DM} = 10$~GeV, without (cyan lines) and with (blue lines) substructure, and for $m_{\rm DM} = 130$~MeV, without (magenta lines) and with (red lines) substructure, with annihilation cross sections saturating the \emph{Planck} CMB limits~\cite{Ade:2015xua} (see Tab.~\ref{tab:msv}), and using the rest of our default DM settings. The case without DM annihilations is also displayed (black lines). Whereas the addition of the halo contribution is crucial for $10 \lesssim z \lesssim 30$, the addition of substructure has little impact on our results for $m_{\rm DM} = 10$~GeV, although it is not completely negligible. Nevertheless, the effect is more important for $m_{\rm DM} = 130$~MeV, which is the case with the potentially strongest impact on the signal. In this case, when the contribution from substructures is included, DM annihilations heat the IGM so that $T_S > T_{\rm CMB}$ (i.e., the signal is in emission) for $z \lesssim 20$, resulting in a dramatic drop in the power spectrum between the Ly$\alpha$ pumping and the X-ray heating epochs and in the disappearance of the Ly$\alpha$ pumping peak. When the contribution from substructures is neglected, the suppression of the power spectrum is less pronounced and the Ly$\alpha$ pumping  peak is present. For both DM masses, the X-ray heating peak appears in emission, but the effects of substructures are not very significant for $z \lesssim 15$.

In Fig.~\ref{fig:tbpsdm-h}, we illustrate the dependence of the global 21~cm differential brightness temperature (inset) and the 21~cm power spectrum at $k=0.1 \, {\rm Mpc}^{-1}$ (main panel) on the minimum halo mass, $M_{\rm min}$. We show the results for $m_{\rm DM} = 130$~MeV and $\left<\sigma v\right> = 10^{-28} \, {\rm cm}^3/{\rm s}$, for three values of $M_{\rm min}$: $10^{-12} \, M_\odot$ (red lines), $10^{-6} \, M_\odot$ (blue lines) and $10^{-3} \, M_\odot$ (cyan lines); using the rest of our default DM settings. As usual,  in the main panel, we distinguish when the signal is in absorption (solid lines) or emission (non-solid lines) and show the case in which DM annihilations are not included in black. As expected, the strongest effect is obtained for the smallest $M_{\rm min}$, i.e., for the extreme case of $M_{\rm min} = 10^{-12} \, M_\odot$, although the main differences occur for $z \gtrsim 20$. Indeed, in this case the early onset of DM annihilations overrides the effect of Ly$\alpha$ pumping: the related peak does not develop and the power spectrum starts to be seen in emission at very early times ($z \sim 20-25$).  Finally, by comparing Figs.~\ref{fig:tbpsdm-subs} and~\ref{fig:tbpsdm-h}, we note that the uncertainties in the 21~cm signal caused by our ignorance of $M_{\rm min}$ are more important than the effects due to the presence of substructure.

\begin{figure*}
	\centering
	\hspace{3mm}
	\includegraphics[width=.5\textwidth]{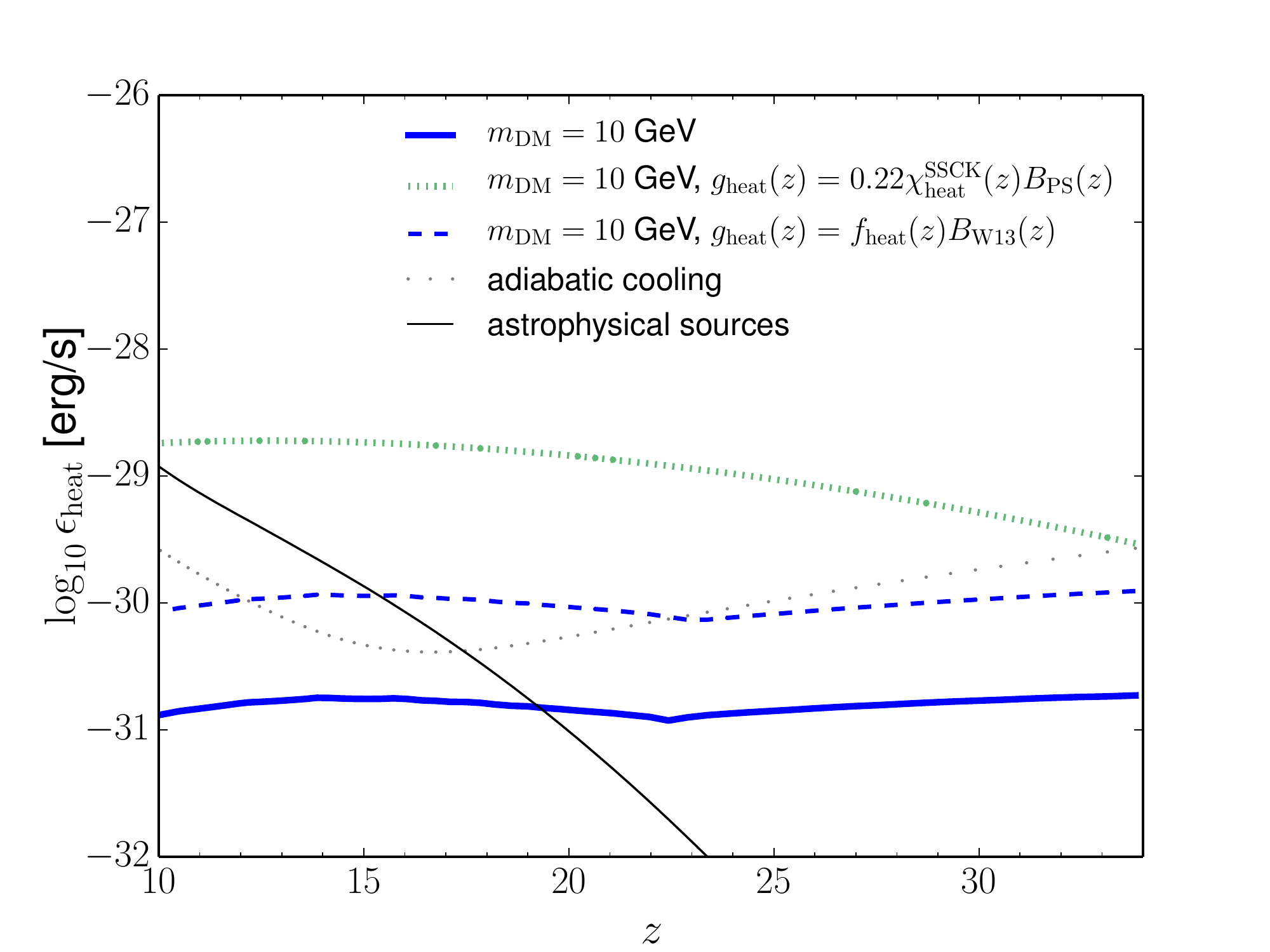} \hspace{-6mm}
	\includegraphics[width=.5\textwidth]{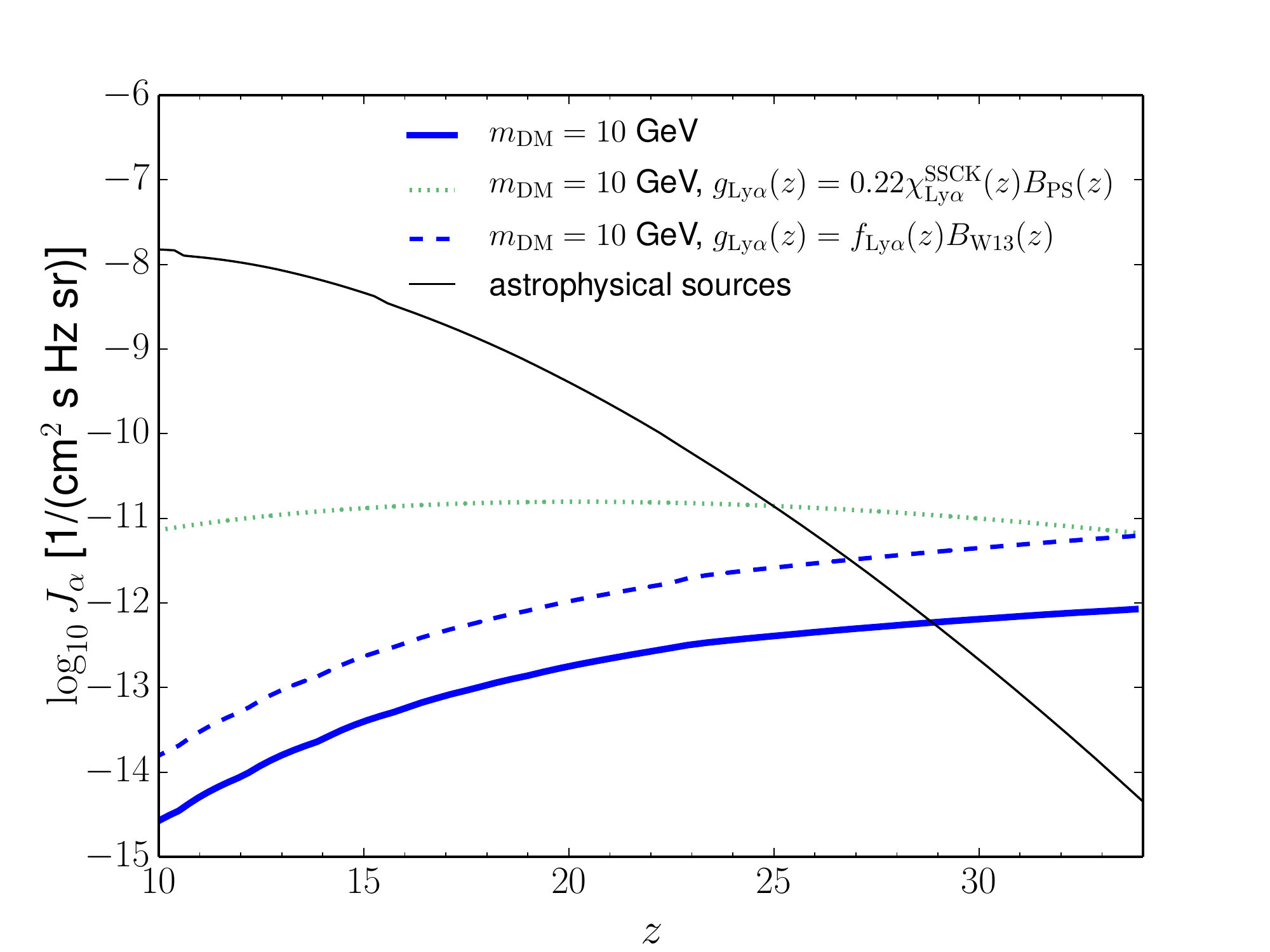} 
	\caption{{\it Energy deposition rate into heat $\epsilon_{\rm heat}$ (left panel) and Ly${\alpha}$ flux (right panel)}, obtained using different approaches. In all DM cases, $m_{\rm DM} = 10$~GeV and $\left<\sigma v\right> = 10^{-26} \, {\rm cm}^3/{\rm s}$ are assumed. We show the results following the approach described in the text (blue solid lines) and factoring $[1 + {\cal B}_{\rm W13}(z)]$ out of the integral over the transfer functions (blue dashed lines). In these cases we use our default DM settings. We also show (green dotted lines) the result of factoring $[1 + {\cal B}_{\rm PS}(z)]$ out of the integral over the transfer functions and using the PS boost factor with $M_{\rm min} =10^{-9} \, M_\odot$, as defined in Ref.~\cite{Evoli:2014pva}, and $\sum_c f_c = 0.22$, corresponding to annihilations into $\mu^+\mu^-$. Our default astrophysical contribution (black solid line) and the adiabatic cooling rate (gray dotted line, only in the left panel) are also shown.}
	\label{fig:epsJ}
\end{figure*}

\subsection{Comparison with previous results}
\label{sec:comp-with-prev}

Early studies about the effects of DM decays and annihilations on the 21~cm signal only considered the contribution from the smooth DM distribution and were mostly concerned with the signal at very low frequencies where the astrophysical background is absent~\cite{Shchekinov:2006eb, Furlanetto:2006wp, Valdes:2007cu} (but where the galactic foreground is huge). Soon after, the importance of halos was pointed out~\cite{Chuzhoy:2007fg, Cumberbatch:2008rh}, although using the parametrization in Eq.~(\ref{eq:eps}) and factoring out the halo enhancement and the deposition efficiency, i.e., effectively using $ g(z)=f(z) {\cal B}(z)$, or equivalently substituting Eq.~(\ref{eq:injectedEnergy}) by Eq.~(\ref{eq:DMenergy}). Following Ref.~\cite{Natarajan:2008pk}, this was corrected using a semi-analytical modeling of photon propagation~\cite{Natarajan:2009bm, Yuan:2009xq}. These works also computed the signal for lower redshifts, and the first estimates of the DM signal over the astrophysical one were presented in Ref.~\cite{Yuan:2009xq}. In Ref.~\cite{Valdes:2012zv}, the effects on the global 21~cm signal produced by DM annihilations were studied and the importance of the gradient $d\overline{\delta T_b}/d\nu$ was stressed as a tool to distinguish a potential DM signal from astrophysical X-ray heating~\cite{Shchekinov:2006eb, Valdes:2007cu}. The latter evolves much faster since the fractional increase of halos which can host galaxies is much larger than the fractional increase of collapsed halos contributing to DM heating. As a result, the inclusion of DM annihilations could significantly reduce this gradient. 

Recently, the impact of DM annihilations on the 21~cm power spectrum~\cite{Furlanetto:2006wp, Natarajan:2009bm} has also been further stressed~\cite{Evoli:2014pva}. It was concluded that, as a result of the early and uniform heating from the DM annihilation products, the power of the X-ray heating peak (middle peak) gets reduced, and could be lower than the peaks associated to the Ly$\alpha$ pumping and the reionization processes. Moreover, it was argued that the X-ray heating peak could even occur in emission due to the earlier DM heating, which would be very challenging to mimic with astrophysics~\cite{Evoli:2014pva}. In light of our results, presented above, we next examine some of these conclusions. 

We first note that the effects of DM annihilations on the 21~cm signal described in those recent studies~\cite{Valdes:2012zv, Evoli:2014pva}, for DM masses in the GeV range and above, are much larger than those obtained in our work. This is partly due to the different treatment in the calculation of the deposited energy into the IGM by the products from DM annihilations in halos. Although these previous studies used an updated calculation of the efficiency of energy deposition (in the smooth background), when treating the energy injection in halos, they adopted the same approximations of some earlier works~\cite{Chuzhoy:2007fg, Cumberbatch:2008rh}. However, a self-consistent treatment needs to properly take into account the fact that the redshift of injection is different from that of deposition and thus, the function $[1 + {\cal B}(z)]$ must be included inside the integral over the transfer functions~\cite{Slatyer:2012yq, Lopez-Honorez:2013lcm, Diamanti:2013bia, Poulin:2015pna} and cannot be factored out. As a consequence, the effects of DM annihilations obtained in our work are milder than in Refs.~\cite{Valdes:2012zv, Evoli:2014pva} and some of them could, in principle, be mimicked by a modest change in some uncertain astrophysical parameters.

To illustrate the origin of the differences between those recent results~\cite{Valdes:2012zv, Evoli:2014pva} and ours, we plot in Fig.~\ref{fig:epsJ} the energy deposition rate into heat per baryon (left panel) and the Ly$\alpha$ flux (right panel), for several cases.  In both panels, we show the contribution from astrophysical sources using our default values for the relevant parameters (black solid lines). Moreover, n the left panel we also show the adiabatic cooling rate (gray dotted line). We consider the case of $m_{\rm DM} = 10$~GeV and $\left< \sigma v \right> = 10^{-26}$~cm$^3$/s.  We first compare the results obtained in this paper using Eq.~(\ref{eq:hrcosmorec}) (blue solid lines), using our default DM settings, as described in Sec.~\ref{sec:structure-formation}, to the ones obtained by factoring $[1+ {\cal B}_{\rm W13}(z)]$ outside the integral over the transfer function, i.e., effectively taking $g_c(z) = f_c(z) \, {\cal B}_{\rm W13}(z)$ (blue dashed lines). As expected from the discussion in Sec.~\ref{sec:structure-formation}, the energy deposition rate into heat per baryon and the Ly$\alpha$ flux are always larger when factoring out $[1+ {\cal B}_{\rm W13}(z)]$, by about an order of magnitude for this DM mass.

We also illustrate (green dotted lines) the case of $m_{\rm DM} = 10$~GeV, annihilations into $\mu^+\mu^-$, $\sum_c f_c = 0.22$, with a cross section $\left< \sigma v \right> = 10^{-26}$~cm$^3$/s, and using the PS boost factor with $M_{\rm min} =10^{-9} \, M_\odot$, as defined in Ref.~\cite{Evoli:2014pva}. As done in Ref.~\cite{Evoli:2014pva}, we also factor $[1+ {\cal B}_{\rm PS}(z)]$ outside the integral over the transfer function, i.e., effectively taking $g_c(z) = 0.22 \, \chi_c^{\rm SSCK}(z) \, {\cal B}_{\rm PS}(z)$, with $c = \{{\rm heat}, {\rm Ly}\alpha\}$. The Ly$\alpha$ fraction is fixed to $\chi_{\mathrm{Ly}\alpha} = \chi_{\rm HI}^{\rm SSCK}+\chi_{\rm HeI}^{\rm SSCK}$ (see, e.g., Ref.~\cite{Galli:2013dna}). The heating rate and the intensity of the Ly$\alpha$ flux obtained with the approach described in this work (blue solid lines) are two orders of magnitude smaller than in the case of Ref.~\cite{Evoli:2014pva} (green dotted lines) at $10 \lesssim z \lesssim 30$. As can be seen by comparing the two blue lines and the green dotted line, the differences arise from both, the distinct approaches followed to calculate the energy deposition from DM annihilations in halos (important for $m_{\rm DM} =10$~GeV) and the different halo boosts which have been considered.

\begin{figure}
	\includegraphics[width=0.9\textwidth]{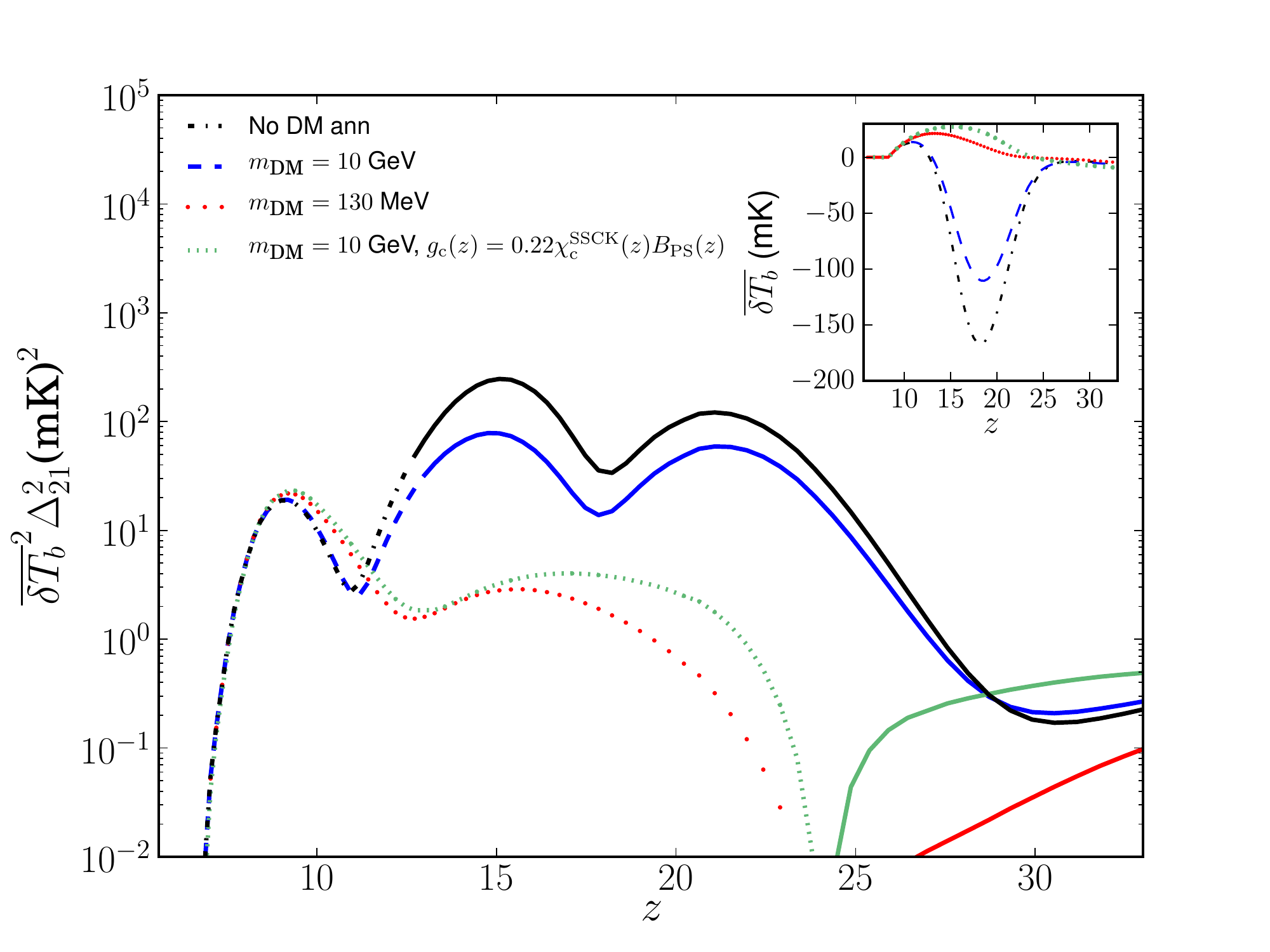} 
	\caption{{\it Dependence on the approach followed to compute the energy deposition rate from DM annihilations inside halos.} Global 21~cm signal (inset) and the associated power spectrum at a scale $k=0.1 \, {\rm Mpc}^{-1}$ (main panel), with annihilation cross sections saturating the \emph{Planck} CMB limits~\cite{Ade:2015xua} (see Tab.~\ref{tab:msv}). We show the results  for $m_{\rm DM} = 10$~GeV (blue solid lines) and for $m_{\rm DM} = 130$~MeV (red lines) following the approach described in the text. In these cases, we use our default DM settings. We also show (green lines) the result of factoring $[1 + {\cal B}(z)]$ out of the integral over the transfer functions, using the PS boost factor with $M_{\rm min} =10^{-9} \, M_\odot$, as defined in Ref.~\cite{Evoli:2014pva}, and $\sum_c f_c = 0.22$, corresponding to annihilations into $\mu^+\mu^-$. The case without DM annihilation is also depicted (black lines). In the main panel, we distinguish when the signal is in absorption (solid lines) or in emission (non-solid lines).}
	\label{fig:J-T-Ps-Evoli}
\end{figure}

Finally, in Fig.~\ref{fig:J-T-Ps-Evoli}, we also compare the results for $m_{\rm DM} = 10$~GeV and $\left<\sigma v\right> = 10^{-26} \, {\rm cm}^3/{\rm s}$ obtained with each of the two approaches, i.e., the one followed here with our default approach (blue lines) and that described in Ref.~\cite{Evoli:2014pva} with $M_{\rm min} = 10^{-9} \, M_\odot$ (green lines), for the global 21~cm signal (inset) and the power spectrum at $k=0.1 \, {\rm Mpc}^{-1}$ (main panel). We also show our results for $m_{\rm DM} = 130$~MeV and $\left<\sigma v\right> = 10^{-28} \, {\rm cm}^3/{\rm s}$ (red lines). As above, in the main panel, we distinguish when the signal is in absorption (solid lines) or in emission (non-solid lines). The differences between the two approaches are very significant. Indeed, it turns out that the results for $m_{\rm DM} = 130$~MeV (red lines) are at a similar level as those obtained following the approach of Ref.~\cite{Evoli:2014pva}, but for $m_{\rm DM} = 10$~GeV (green lines). As can be inferred from this figure and the previous discussion, the main conclusion is that the parametrization of the injection of energy from DM annihilations in halos extensively used in the literature~\cite{Valdes:2012zv, Evoli:2014pva, Chuzhoy:2007fg, Cumberbatch:2008rh}, largely overestimates the impact of DM annihilations on the $21$~cm signal for $m_{\rm DM} = 10$~GeV. Therefore, let us restate the main conclusions of Ref.~\cite{Evoli:2014pva} in view of our results:

\begin{enumerate}
\item {\it With DM annihilations the X-ray heating peak in the 21~cm power could be lower than the other two peaks.} We do not reach this conclusion for the case considered in Ref.~\cite{Evoli:2014pva}, i.e., $m_{\rm DM} = 10$~GeV and annihilations into two leptons\footnote{Recall that we consider DM annihilations into $e^+e^-$, which give rise to a stronger effect on the 21cm signal than DM annihilations into $\mu^+\mu^-$, channel considered in Ref.~\cite{Evoli:2014pva}.} with a cross-section $\left< \sigma v \right> = 10^{-26}$~cm$^3$/s. Similarly, we do not reach this conclusion for $m_{\rm DM} = 1$~MeV, $9$~MeV and $1$~GeV (see Fig.~\ref{fig:tbpsdm-M}). However, we do observe this effect for $m_{\rm DM} = 130$~MeV and $\left< \sigma v \right> = 10^{-28}$~cm$^3$/s, even for $M_{\rm min} = 10^{-3} \, M_\odot$ (see Fig.~\ref{fig:tbpsdm-h}).

\item {\it Models in which DM annihilations heat the IGM to temperatures $T_S > T_{\rm CMB}$ without astrophysical sources result in a dramatic drop in large-scale power between the Ly$\alpha$ pumping and X-ray heating epochs.} This feature is only seen for the most extreme case we consider, i.e., $m_{\rm DM} = 130$~MeV, $\left< \sigma v \right> = 10^{-28}$~cm$^3$/s and $M_{\rm min} = 10^{-12} \, M_\odot$. For all the rest of the cases, including that considered in Ref.~\cite{Evoli:2014pva}, DM annihilations alone do not heat the IGM so that $T_S > T_{\rm CMB}$ between those epochs.

\item {\it The X-ray heating peak could occur when the IGM is already in emission against the CMB.} Similarly to the above points, we do not find this result for DM scenarios as the one studied in Ref.~\cite{Evoli:2014pva}. However, we do reach that conclusion for the most extreme of our cases, $m_{\rm DM} = 130$~MeV, $\left< \sigma v \right> = 10^{-28}$~cm$^3$/s and $M_{\rm min} = 10^{-12} \, M_\odot$, while for less optimistic minimum halo masses, this feature is only \emph{partially} observed (see Figs.~\ref{fig:tbpsdm-M} and~\ref{fig:tbpsdm-h}). For the rest of the vast range of possible DM masses considered in this work, such an effect is absent.
\end{enumerate}

\subsection{Forecast for future experiments}
\label{sec:forecast}

\begin{figure*}
	\includegraphics[width=1\textwidth]{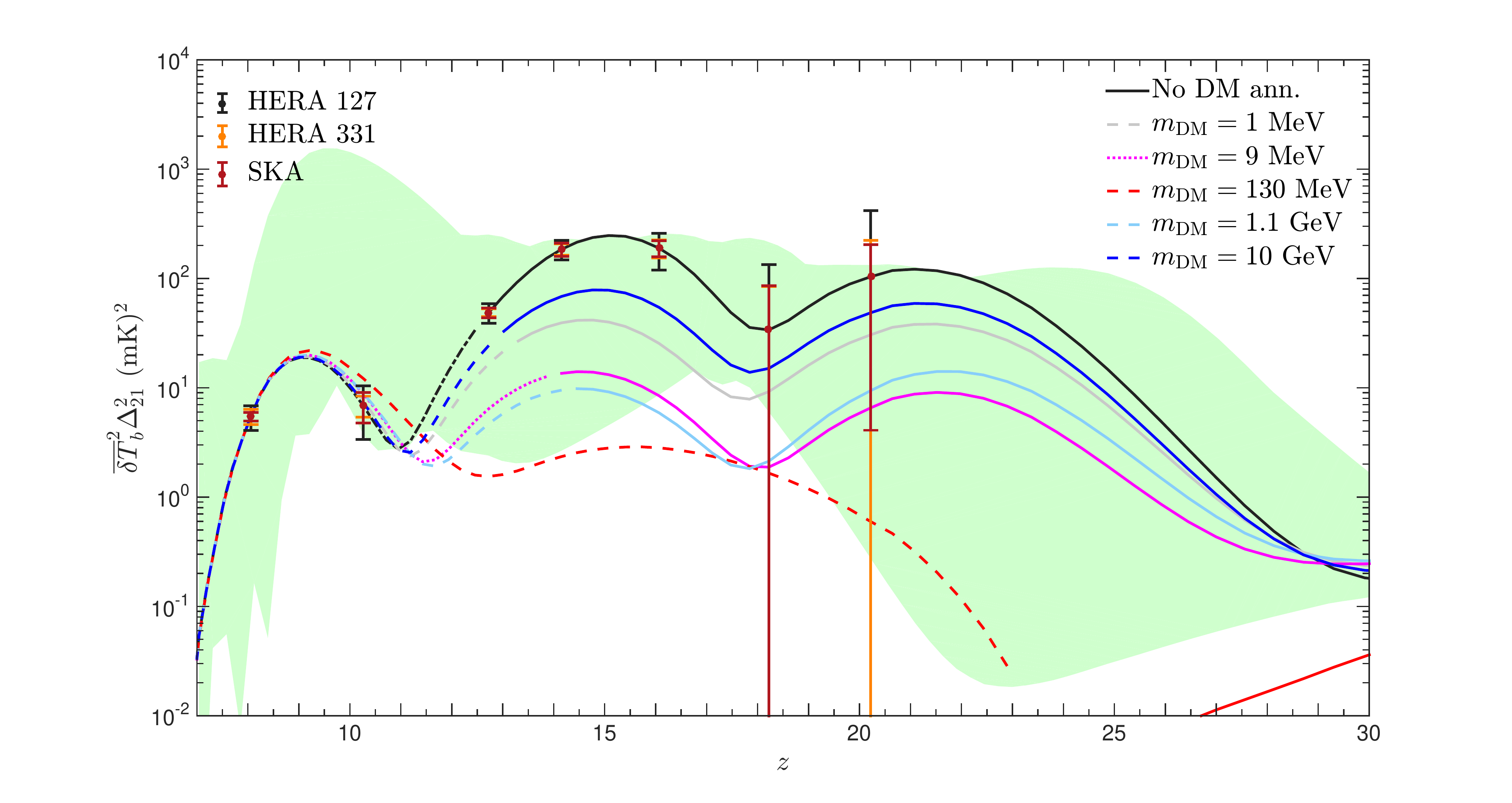} 
	\caption{{\it Future observational prospects: the purely astrophysical case}. Power spectrum at a scale $k=0.1 \, {\rm Mpc}^{-1}$ for the standard scenario as well as for five different values of the DM mass with annihilating cross sections saturating the \emph{Planck} CMB limits~\cite{Ade:2015xua} (see Fig.~\ref{fig:tbpsdm-M}). Black, orange and brown error bars refer to the HERA 127, HERA 331 and SKA-low frequency expected errors, respectively. The shaded green region denotes predictions from all the different astrophysical models shown in the four panels of Fig.~\ref{fig:tbps}.}
	\label{fig:sense}
\end{figure*}

Finally, we conclude the discussion about the effects of DM annihilations on the 21~cm signal by illustrating the potential of future $21$~cm experiments to detect it. In order to do so, we make use of the publicly available code {\tt 21cmSense}\footnote{\url{https://github.com/jpober/21cmSense}}~\cite{Pober:2013jna, Pober:2012zz} (see also Ref.~\cite{Parsons:2011ew}). In these references, a detailed calculation of the sensitivity to the $21$~cm differential brightness temperature power spectrum is developed and described, resulting in a total noise that reads
\begin{equation}
\overline{\Delta^2_{T+S}}(k) =\left(\sum_i \frac{1}{\left(\Delta^2_{N,i}+\overline{\Delta}^2_{21}\right)^{2}}\right)^{-1/2}~,
\end{equation}
with a thermal noise ($N$) contribution plus a sample variance error ($S$), $\overline{\Delta}^2_{21} \equiv \overline{\delta T_b}^2 \Delta^2_{21}$. The thermal noise is computed at each pixel $i$ of a multiple baseline interferometer array and it depends on the solid angle, the integrated observation time, and the temperature of the system (which accounts for both the sky and the receiver temperatures). 

While an almost complete foreground removal may be challenging, even in the context of futuristic 21~cm observations, we make this (very optimistic) assumption with the goal of establishing an upper limit to the expected sensitivity of planned large interferometers. This  allows us to illustrate the difficulty of singling out scenarios with a contribution from DM annihilations from the standard astrophysical signal. We focus here on the future HERA~\cite{Beardsley:2014bea} and SKA-low frequency~\cite{Mellema:2012ht} experiments (see also the recent Ref.~\cite{Ewall-Wice:2015uul}). For the HERA experiment, we  consider both the intermediate and final configurations, with 127 and 331 antennas respectively (denoted by HERA~127 and HERA~331 in what follows), with a diameter of 14~m \cite{reiondotorg}. For SKA-low frequency, we follow the design presented in the SKA System Baseline Design Document~\cite{Mellema:2012ht}, with 911~antennas with a diameter of 35~m. In all cases, we assume a total observing time of 1080~hours and a bandwidth of 8~MHz, both being the default parameters in {\tt 21cmSense}. 

In Fig.~\ref{fig:sense}, we assume our default model for the astrophysical reionization, heating and Ly$\alpha$ parameters ($\zeta_{\rm UV}=31.5$, $\zeta_{\rm X,0}=10^{56}$, stellar emissivity of Pop II normalized to $\sim 4400$ ionizing photons, $f_{\star,0}=0.1$ and $T_{\textrm{vir},0}=10^4$~K) and with the W13 halo mass function. We show the expected errors from the future HERA and SKA experiments for $z=8, \, 10, \, 12, \, 14, \, 16, \, 18$ and $20$ (corresponding to frequencies in the range $[68-158]$~MHz). We also show the expected 21~cm power spectrum in the presence of DM annihilations using our default DM settings, for the same DM masses shown in Fig.~\ref{fig:tbpsdm-M}. The green shaded region illustrates the impact of varying some of the unknown astrophysical parameters (halo mass function, minimum virial temperature, number of X-ray photons per solar mass and number of photons per stellar baryon between Ly$\alpha$ and the Lyman limit) in the ranges shown by all the lines in the four panels of Fig.~\ref{fig:tbps}, in the absence of DM annihilations. Although this uncertainty region is large, the expected errors are quite promising for $z \lesssim 16$. Bearing in mind that the contribution from DM annihilations always reduces the intensity of the power spectrum and the astrophysical uncertainties discussed in Sec.~\ref{sec:impact-21-cm} shift the signal to earlier or later times\footnote{As discussed in Sec.~\ref{sec:evo}, they could also produce some variation in the maximum value of the power spectrum. Indeed, decreasing the Ly$\alpha$ flux does significantly suppress the power spectrum, but for most of the parameters, the change is less pronounced.}, it might be possible to set constraints on a potential contribution from DM annihilations. However, in some extreme astrophysical models~\cite{Valdes:2012zv, Evoli:2014pva} even setting limits would be rather non trivial. In contrast to the conclusions of Ref.~\cite{Evoli:2014pva}, it is apparent from Fig.~\ref{fig:sense} that the identification of DM signatures for $m_{\rm DM} =10$~GeV would be a very challenging task.

\begin{figure*}
	\includegraphics[width=1\textwidth]{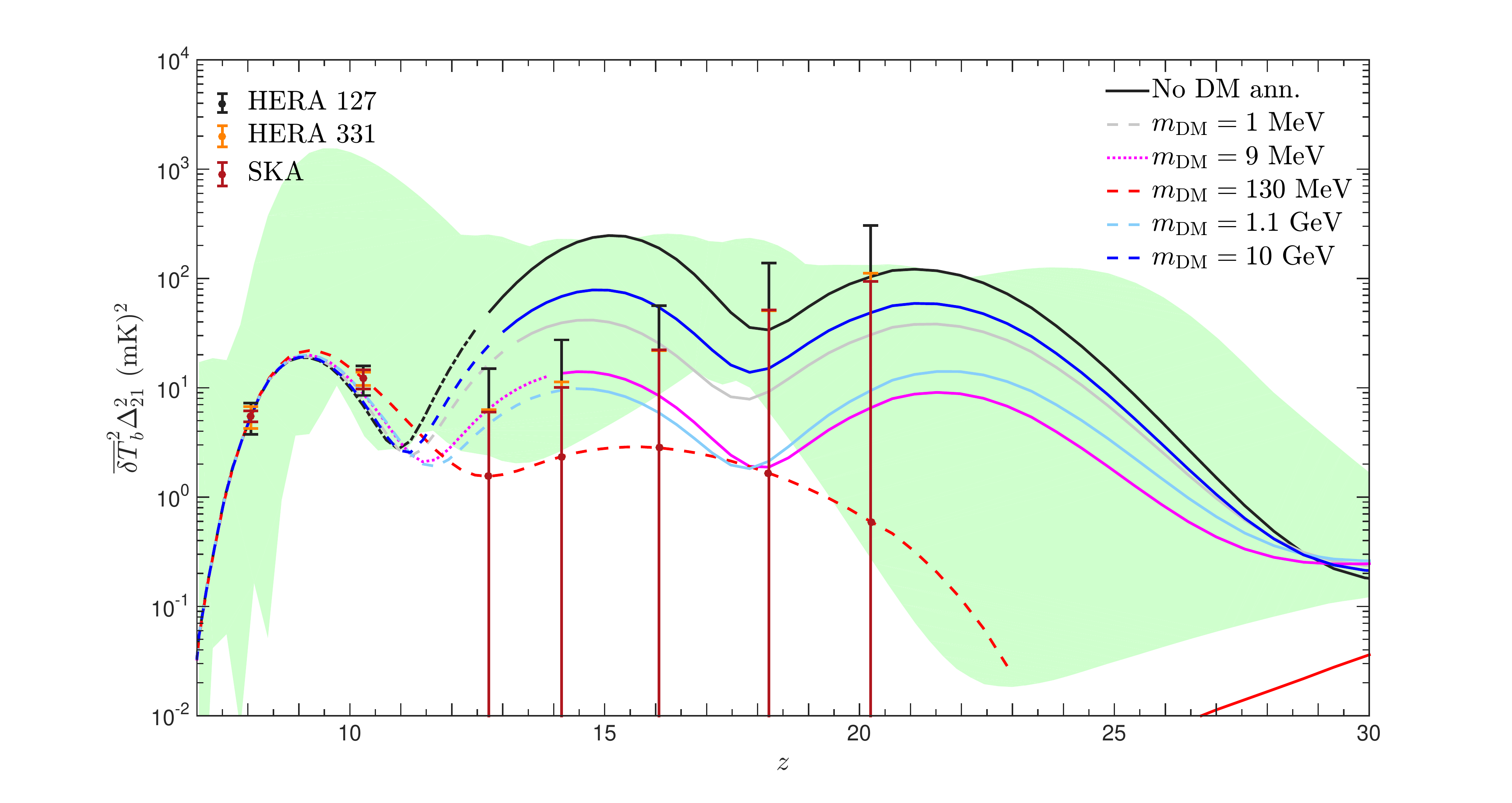} 
	\caption{{\it Future observational prospects: the DM case}. Same as Fig.~\ref{fig:sense} but assuming the signal also include the contribution from DM annihilations for the case with the strongest effects, i.e., for $m_{\rm DM} = 130$~MeV, $\left<\sigma v\right> = 10^{-28} \, {\rm cm}^3/{\rm s}$, with the P12 concentration-mass relation and the W13 halo mass function, with $M_{\rm min} = 10^{-12} \, M_\odot$ and including substructure.}
	\label{fig:senseDM}
\end{figure*}

Finally, in Fig.~\ref{fig:senseDM}, we also show the errors for the same experimental configurations, but assuming the signal also includes the contribution from DM annihilations for the case with the strongest effects, i.e., for $m_{\rm DM} = 130$~MeV, $\left<\sigma v\right> = 10^{-28} \, {\rm cm}^3/{\rm s}$, and using our default DM settings. Although the differences of the 21~cm differential brightness temperature power spectrum with respect to the purely astrophysical case are significant (including uncertainties), the expected errors for the assumed configurations are quite large due to the suppression of the signal.  Nevertheless, such a suppression, along with the fact that the signal is in emission for $z \lesssim 20$, might hint at the presence of DM annihilations.

As can be seen from the green area and the errors in Figs.~\ref{fig:sense} and~\ref{fig:senseDM}, it thus appears that a DM candidate with $m_{\rm DM} \sim 100$~MeV, which annihilates into $e^+e^-$ with $\left< \sigma v \right> = 10^{-28}$~cm$^3$/s, represents the only case which could be unambiguously distinguished from a purely astrophysical signal, even in the presence of uncertainties, as the X-ray heating peak would be suppressed (even for the less optimistic values of the minimum halo mass, see Fig.~\ref{fig:tbpsdm-h}) and it would occur fully in emission. One should keep in mind, however, that these uncertainties are always present: different astrophysical parameters could conspire to reduce the suppression predicted by this scenario. For lower and higher masses, the detection or setting of constraints will be even  more challenging, as the suppression of the second peak with respect to the other two is much smaller and the transition of this peak from absorption to emission no longer occurs.

\section{Conclusions}
\label{sec:conclusions}

Future 21~cm cosmology, based on the detection of the spin flip line of neutral hydrogen, offers a unique window to extract information about some fundamental parameters governing the EoR, to explore the formation of the first stars and to study the dark ages. Different experimental set-ups focus on the detection of the global 21~cm signal (the differential brightness temperature) or its spatial fluctuations (the 21~cm differential brightness temperature power spectrum). 

We have reviewed the well-known signatures of the 21~cm signal as it evolves through three relevant cosmic epochs (Sec.~\ref{sec:21-cm-signal}). First, the so-called Ly$\alpha$ pumping epoch around $z \simeq 20$, when the first astrophysical sources switch on. Then, around $z \simeq 15$, the X-ray heating epoch resulting from galaxy emission, and finally the Universe's reionization at $z \lesssim 10$. Moreover, we have also discussed the effects of the uncertainties on some of the relevant astrophysical parameters (Sec.~\ref{sec:evo}), which modify the specific signatures on the 21~cm signal associated to these different epochs and show large degeneracies among themselves. We have studied the dependence of the 21~cm signal on the abundance of halos suitable to host galaxies. Ionization, heating and excitation of the IGM critically depend on the parametrization for the halo mass function (Sec.~\ref{sec:halo-mass-function}), which dictates the number density of halos per unit mass and therefore, the star formation rate itself. In this work, we have considered three well-known models: the spherical-collapse model of Press-Schechter~\cite{Press:1973iz, Bond:1990iw}, the ellipsoidal-collapse model of Sheth-Tormen~\cite{Sheth:1999mn, Sheth:1999su} and the results from the N-body simulations by Watson'13~\cite{Watson:2012mt}. Some of the previous 21~cm studies rely on a fixed hybrid analytical approach, while here we have used a more recent appraisal to the halo mass function, exploring the effects of the uncertainties on its precise shape. Other crucial parameters we have studied are the minimum halo temperature for hosting star-forming galaxies (Sec.~\ref{sec:halo-viri-temp}), and the efficiency for heating (Sec.~\ref{sec:number-x-ray}) and exciting (Sec.~\ref{sec:Jalpha}) the IGM. Dedicated simulations varying these astrophysical parameters have been performed (Sec.~\ref{sec:impact-21-cm}) and we have discussed their correlations. Overall, these parameters are largely unconstrained and therefore, the precise shape of the signal is quite uncertain.

Following this, we have described the impact of DM annihilations on the 21~cm signal, which is one of the main goals of this work. We discussed how the products from DM annihilations could change the thermal history of the IGM and whether they could lead to distinctive features on the global 21~cm signal and on the 21~cm power spectrum at redshifts $10 \lesssim z \lesssim 30$ (Sec.~\ref{sec:dark-matter-impact}). Our calculations include up-to-date computations of the energy deposition rates by the DM annihilation products (Sec.~\ref{sec:DMdeposition}). More importantly, we have followed an approach that treats the contribution from DM annihilations in halos correctly (Sec.~\ref{sec:structure-formation}). The proper computation of the energy deposition from DM annihilations is crucial for $m_{\rm DM} \gtrsim 1$~GeV, as approximate parametrizations lead to an overestimation of the DM effects and to the misunderstanding of the true underlying DM imprint. In order to illustrate some of the possible signatures from DM annihilations, we have considered different models, by studying in detail the dependence of the signal on the DM mass, on the presence of substructure and on the minimum halo mass (Sec.~\ref{sec:dark-matt-annih}). Given these results, we have also shown that isolating the effects of DM annihilations from the highly uncertain astrophysical background is a difficult task, due to unknowns in the galaxy formation history (heating of the IGM, star formation rate, number of halos as a function of the redshift, among others).  However, even in the presence of such a highly complicated scenario, in which plenty of degeneracies between the DM and astrophysical parameters appear, a recent work~\cite{Evoli:2014pva} pointed out several robust signatures of DM heating on the 21~cm power spectrum. In particular, the authors of that work concluded that an unambiguous signature of DM annihilations could be the observation of the X-ray heating peak in the 21~cm power spectrum to occur in emission against the CMB for DM particles with a mass of $m_{\rm DM} = 10$~GeV. In this work, we have critically examined these claims (in particular, in Sec.~\ref{sec:comp-with-prev}), reaching much less optimistic conclusions. Finally, we have discussed the prospects of detection of DM annihilation signatures by future 21~cm observations, the HERA and SKA-low frequency radio interferometers (Sec.~\ref{sec:forecast}). We have shown that disentangling DM signatures from astrophysical processes would be a challenging task, given the current uncertainties on describing the astrophysical processes, although a more quantitative analysis is left for future work. 

Throughout this paper, the values of the annihilation cross sections have been chosen accordingly to the most recent limits set by the \emph{Planck} CMB data~\cite{Ade:2015xua}. We have analyzed a vast range of DM masses, from $m_{\rm DM} =1$~MeV to 10~GeV. The impact of higher masses, although not illustrated here, has also been explored, resulting in very weak signatures, even weaker than those for the $m_{\rm DM} = 10$~GeV case. We find that only DM candidates with $m_{\rm DM} \sim 100$~MeV could give rise to unambiguous features on the 21~cm signal, such that the effects from DM annihilations could be disentangled from the astrophysical background. For $m_{\rm DM} \sim 100$~MeV, the DM heating efficiency and the Ly$\alpha$ flux turn out to be maximal and the power spectrum is largely suppressed, with the X-ray heating peak occurring fully in emission. For other DM masses, however, the expectations are less optimistic and the identification of a signal from energy injection from DM annihilations would be a challenging task for future measurements of the 21~cm power spectrum at $10 \lesssim z \lesssim 30$. As discussed in detail in this work, this is in contrast to some results found in the recent literature~\cite{Evoli:2014pva}.

On the other hand, DM annihilations could also lead to changes in the thermal and ionization history of the Universe at higher redshifts ($z \gtrsim 50$)~\cite{Shchekinov:2006eb, Furlanetto:2006wp, Valdes:2007cu}, which would also lead to modifications of the 21~cm differential brightness temperature and its power spectrum. While measuring these low frequencies with terrestrial radio interferometers is very challenging, future space or Moon-based interferometers may provide measurements in this range of redshifts and could be sensitive to DM annihilations, free from astrophysical uncertainties. A dedicated analysis of this very interesting option will be carried out elsewhere.

\section*{Acknowledgments}

We would like to thank J.~Pober for clarifications on the use of {\tt 21cmSense} and for providing us with the antenna configuration for the HERA and SKA arrays. We are grateful to A.~Mesinger for clarifications about {\tt 21cmFast} and C.~Evoli for discussions about the results of Ref.~\cite{Evoli:2014pva}. LLH was supported in part by FWO-Vlaanderen with the post doctoral fellowship project 1271513 and the project G020714N, by the Belgian Federal Science Policy Office through the Interuniversity Attraction Pole P7/37 and by the Vrije Universiteit Brussel through the Strategic Research Program \textit{High-Energy Physics}. OM is supported by PROMETEO II/2014/050 and by the Spanish Grant FPA2014--57816-P of the MINECO. AM has been supported by the Funda\c{c}\~ao para a Ci\^encia e a Tecnologia (FCT) of Portugal and thanks IFIC for hospitality. SPR is supported by a Ram\'on y Cajal contract, by the Spanish MINECO under Grant FPA2014-54459-P and by the Generalitat Valenciana under Grant PROMETEOII/2014/049. OM and SPR are also partially supported by the MINECO Grant SEV-2014-0398 and by PITN-GA-2011-289442-INVISIBLES. AM and SPR are also partially supported by the Portuguese FCT through the CFTP-FCT Unit 777 (PEst-OE/FIS/UI0777/2013).

\bibliographystyle{apsrev4-1}
\bibliography{bib21}

\begin{thebibliography}{193}%
\makeatletter
\providecommand \@ifxundefined [1]{%
 \@ifx{#1\undefined}
}%
\providecommand \@ifnum [1]{%
 \ifnum #1\expandafter \@firstoftwo
 \else \expandafter \@secondoftwo
 \fi
}%
\providecommand \@ifx [1]{%
 \ifx #1\expandafter \@firstoftwo
 \else \expandafter \@secondoftwo
 \fi
}%
\providecommand \natexlab [1]{#1}%
\providecommand \enquote  [1]{``#1''}%
\providecommand \bibnamefont  [1]{#1}%
\providecommand \bibfnamefont [1]{#1}%
\providecommand \citenamefont [1]{#1}%
\providecommand \href@noop [0]{\@secondoftwo}%
\providecommand \href [0]{\begingroup \@sanitize@url \@href}%
\providecommand \@href[1]{\@@startlink{#1}\@@href}%
\providecommand \@@href[1]{\endgroup#1\@@endlink}%
\providecommand \@sanitize@url [0]{\catcode `\\12\catcode `\$12\catcode
  `\&12\catcode `\#12\catcode `\^12\catcode `\_12\catcode `\%12\relax}%
\providecommand \@@startlink[1]{}%
\providecommand \@@endlink[0]{}%
\providecommand \url  [0]{\begingroup\@sanitize@url \@url }%
\providecommand \@url [1]{\endgroup\@href {#1}{\urlprefix }}%
\providecommand \urlprefix  [0]{URL }%
\providecommand \Eprint [0]{\href }%
\providecommand \doibase [0]{http://dx.doi.org/}%
\providecommand \selectlanguage [0]{\@gobble}%
\providecommand \bibinfo  [0]{\@secondoftwo}%
\providecommand \bibfield  [0]{\@secondoftwo}%
\providecommand \translation [1]{[#1]}%
\providecommand \BibitemOpen [0]{}%
\providecommand \bibitemStop [0]{}%
\providecommand \bibitemNoStop [0]{.\EOS\space}%
\providecommand \EOS [0]{\spacefactor3000\relax}%
\providecommand \BibitemShut  [1]{\csname bibitem#1\endcsname}%
\let\auto@bib@innerbib\@empty
\bibitem [{\citenamefont {Ananthakrishnan}(1995)}]{Ananthakrishnan:1995}%
  \BibitemOpen
  \bibfield  {author} {\bibinfo {author} {\bibfnamefont {S.}~\bibnamefont
  {Ananthakrishnan}},\ }\href@noop {} {\bibfield  {journal} {\bibinfo
  {journal} {J, Astrophys. Astr. Suppl.}\ }\textbf {\bibinfo {volume} {16}},\
  \bibinfo {pages} {427} (\bibinfo {year} {1995})}\BibitemShut {NoStop}%
\bibitem [{\citenamefont {Paciga}\ \emph {et~al.}(2011)\citenamefont {Paciga}
  \emph {et~al.}}]{Paciga:2011}%
  \BibitemOpen
  \bibfield  {author} {\bibinfo {author} {\bibfnamefont {G.}~\bibnamefont
  {Paciga}} \emph {et~al.},\ }\href {\doibase 10.1111/j.1365-2966.2011.18208.x}
  {\bibfield  {journal} {\bibinfo  {journal} {Mon. Not. Roy. Astron. Soc.}\
  }\textbf {\bibinfo {volume} {413}},\ \bibinfo {pages} {1174} (\bibinfo {year}
  {2011})},\ \Eprint {http://arxiv.org/abs/1006.1351} {arXiv:1006.1351}
  \BibitemShut {NoStop}%
\bibitem [{\citenamefont {van Haarlem}\ \emph {et~al.}(2013)\citenamefont {van
  Haarlem} \emph {et~al.}}]{vanHaarlem:2013dsa}%
  \BibitemOpen
  \bibfield  {author} {\bibinfo {author} {\bibfnamefont {M.~P.}\ \bibnamefont
  {van Haarlem}} \emph {et~al.},\ }\href {\doibase 10.1051/0004-6361/201220873}
  {\bibfield  {journal} {\bibinfo  {journal} {Astron. Astrophys.}\ }\textbf
  {\bibinfo {volume} {556}},\ \bibinfo {pages} {A2} (\bibinfo {year} {2013})},\
  \Eprint {http://arxiv.org/abs/1305.3550} {arXiv:1305.3550 [astro-ph.IM]}
  \BibitemShut {NoStop}%
\bibitem [{\citenamefont {Tingay}\ \emph {et~al.}(2013)\citenamefont {Tingay}
  \emph {et~al.}}]{Tingay:2012ps}%
  \BibitemOpen
  \bibfield  {author} {\bibinfo {author} {\bibfnamefont {S.~J.}\ \bibnamefont
  {Tingay}} \emph {et~al.},\ }\href {\doibase 10.1017/pasa.2012.007} {\bibfield
   {journal} {\bibinfo  {journal} {Publ. Astron. Soc. Austral.}\ }\textbf
  {\bibinfo {volume} {30}},\ \bibinfo {pages} {7} (\bibinfo {year} {2013})},\
  \Eprint {http://arxiv.org/abs/1206.6945} {arXiv:1206.6945 [astro-ph.IM]}
  \BibitemShut {NoStop}%
\bibitem [{\citenamefont {Parsons}\ \emph {et~al.}(2010)\citenamefont {Parsons}
  \emph {et~al.}}]{Parsons:2009in}%
  \BibitemOpen
  \bibfield  {author} {\bibinfo {author} {\bibfnamefont {A.~R.}\ \bibnamefont
  {Parsons}} \emph {et~al.},\ }\href {\doibase 10.1088/0004-6256/139/4/1468}
  {\bibfield  {journal} {\bibinfo  {journal} {Astron. J.}\ }\textbf {\bibinfo
  {volume} {139}},\ \bibinfo {pages} {1468} (\bibinfo {year} {2010})},\ \Eprint
  {http://arxiv.org/abs/0904.2334} {arXiv:0904.2334 [astro-ph.CO]} \BibitemShut
  {NoStop}%
\bibitem [{\citenamefont {Paciga}\ \emph {et~al.}(2013)\citenamefont {Paciga}
  \emph {et~al.}}]{Paciga:2013fj}%
  \BibitemOpen
  \bibfield  {author} {\bibinfo {author} {\bibfnamefont {G.}~\bibnamefont
  {Paciga}} \emph {et~al.},\ }\href {\doibase 10.1093/mnras/stt753} {\bibfield
  {journal} {\bibinfo  {journal} {Mon. Not. Roy. Astron. Soc.}\ }\textbf
  {\bibinfo {volume} {433}},\ \bibinfo {pages} {639} (\bibinfo {year}
  {2013})},\ \Eprint {http://arxiv.org/abs/1301.5906} {arXiv:1301.5906
  [astro-ph.CO]} \BibitemShut {NoStop}%
\bibitem [{\citenamefont {Dillon}\ \emph {et~al.}(2014)\citenamefont {Dillon}
  \emph {et~al.}}]{Dillon:2013rfa}%
  \BibitemOpen
  \bibfield  {author} {\bibinfo {author} {\bibfnamefont {J.~S.}\ \bibnamefont
  {Dillon}} \emph {et~al.},\ }\href {\doibase 10.1103/PhysRevD.89.023002}
  {\bibfield  {journal} {\bibinfo  {journal} {Phys. Rev.}\ }\textbf {\bibinfo
  {volume} {D89}},\ \bibinfo {pages} {023002} (\bibinfo {year} {2014})},\
  \Eprint {http://arxiv.org/abs/1304.4229} {arXiv:1304.4229 [astro-ph.CO]}
  \BibitemShut {NoStop}%
\bibitem [{\citenamefont {Jacobs}\ \emph {et~al.}(2015)\citenamefont {Jacobs}
  \emph {et~al.}}]{Pober:2014aca}%
  \BibitemOpen
  \bibfield  {author} {\bibinfo {author} {\bibfnamefont {D.~C.}\ \bibnamefont
  {Jacobs}} \emph {et~al.},\ }\href {\doibase 10.1088/0004-637X/801/1/51}
  {\bibfield  {journal} {\bibinfo  {journal} {Astrophys. J.}\ }\textbf
  {\bibinfo {volume} {801}},\ \bibinfo {pages} {51} (\bibinfo {year} {2015})},\
  \Eprint {http://arxiv.org/abs/1408.3389} {arXiv:1408.3389 [astro-ph.CO]}
  \BibitemShut {NoStop}%
\bibitem [{\citenamefont {Ali}\ \emph {et~al.}(2015)\citenamefont {Ali} \emph
  {et~al.}}]{Ali:2015uua}%
  \BibitemOpen
  \bibfield  {author} {\bibinfo {author} {\bibfnamefont {Z.~S.}\ \bibnamefont
  {Ali}} \emph {et~al.},\ }\href {\doibase 10.1088/0004-637X/809/1/61}
  {\bibfield  {journal} {\bibinfo  {journal} {Astrophys. J.}\ }\textbf
  {\bibinfo {volume} {809}},\ \bibinfo {pages} {61} (\bibinfo {year} {2015})},\
  \Eprint {http://arxiv.org/abs/1502.06016} {arXiv:1502.06016 [astro-ph.CO]}
  \BibitemShut {NoStop}%
\bibitem [{rei()}]{reiondotorg}%
  \BibitemOpen
  \href@noop {} {\enquote {\bibinfo {title} {Hera - the hydrogen epoch of
  reionization array},}\ }\bibinfo {howpublished}
  {\url{http://reionization.org}},\ \bibinfo {note} {accessed:
  2016-03-14}\BibitemShut {NoStop}%
\bibitem [{\citenamefont {Mellema}\ \emph {et~al.}(2013)\citenamefont {Mellema}
  \emph {et~al.}}]{Mellema:2012ht}%
  \BibitemOpen
  \bibfield  {author} {\bibinfo {author} {\bibfnamefont {G.}~\bibnamefont
  {Mellema}} \emph {et~al.},\ }\href {\doibase 10.1007/s10686-013-9334-5}
  {\bibfield  {journal} {\bibinfo  {journal} {Exper. Astron.}\ }\textbf
  {\bibinfo {volume} {36}},\ \bibinfo {pages} {235} (\bibinfo {year} {2013})},\
  \Eprint {http://arxiv.org/abs/1210.0197} {arXiv:1210.0197 [astro-ph.CO]}
  \BibitemShut {NoStop}%
\bibitem [{\citenamefont {Bowman}\ and\ \citenamefont
  {Rogers}(2010)}]{Bowman:2012hf}%
  \BibitemOpen
  \bibfield  {author} {\bibinfo {author} {\bibfnamefont {J.~D.}\ \bibnamefont
  {Bowman}}\ and\ \bibinfo {author} {\bibfnamefont {A.~E.~E.}\ \bibnamefont
  {Rogers}},\ }\href {\doibase 10.1038/nature09601} {\bibfield  {journal}
  {\bibinfo  {journal} {Nature}\ }\textbf {\bibinfo {volume} {468}},\ \bibinfo
  {pages} {796} (\bibinfo {year} {2010})},\ \Eprint
  {http://arxiv.org/abs/1209.1117} {arXiv:1209.1117 [astro-ph.CO]} \BibitemShut
  {NoStop}%
\bibitem [{\citenamefont {Greenhill}\ and\ \citenamefont
  {Bernardi}(2012)}]{Greenhill:2012mn}%
  \BibitemOpen
  \bibfield  {author} {\bibinfo {author} {\bibfnamefont {L.~J.}\ \bibnamefont
  {Greenhill}}\ and\ \bibinfo {author} {\bibfnamefont {G.}~\bibnamefont
  {Bernardi}},\ }\href@noop {} {\  (\bibinfo {year} {2012})},\ \Eprint
  {http://arxiv.org/abs/1201.1700} {arXiv:1201.1700 [astro-ph.CO]} \BibitemShut
  {NoStop}%
\bibitem [{\citenamefont {Burns}\ \emph {et~al.}(2012)\citenamefont {Burns}
  \emph {et~al.}}]{Burns:2011wf}%
  \BibitemOpen
  \bibfield  {author} {\bibinfo {author} {\bibfnamefont {J.~O.}\ \bibnamefont
  {Burns}} \emph {et~al.},\ }\href {\doibase 10.1016/j.asr.2011.10.014}
  {\bibfield  {journal} {\bibinfo  {journal} {Adv. Space Res.}\ }\textbf
  {\bibinfo {volume} {49}},\ \bibinfo {pages} {433} (\bibinfo {year} {2012})},\
  \Eprint {http://arxiv.org/abs/1106.5194} {arXiv:1106.5194 [astro-ph.CO]}
  \BibitemShut {NoStop}%
\bibitem [{\citenamefont {Villaescusa-Navarro}\ \emph
  {et~al.}(2014)\citenamefont {Villaescusa-Navarro}, \citenamefont {Viel},
  \citenamefont {Datta},\ and\ \citenamefont
  {Choudhury}}]{Villaescusa-Navarro:2014cma}%
  \BibitemOpen
  \bibfield  {author} {\bibinfo {author} {\bibfnamefont {F.}~\bibnamefont
  {Villaescusa-Navarro}}, \bibinfo {author} {\bibfnamefont {M.}~\bibnamefont
  {Viel}}, \bibinfo {author} {\bibfnamefont {K.~K.}\ \bibnamefont {Datta}}, \
  and\ \bibinfo {author} {\bibfnamefont {T.~R.}\ \bibnamefont {Choudhury}},\
  }\href {\doibase 10.1088/1475-7516/2014/09/050} {\bibfield  {journal}
  {\bibinfo  {journal} {JCAP}\ }\textbf {\bibinfo {volume} {1409}},\ \bibinfo
  {pages} {050} (\bibinfo {year} {2014})},\ \Eprint
  {http://arxiv.org/abs/1405.6713} {arXiv:1405.6713 [astro-ph.CO]} \BibitemShut
  {NoStop}%
\bibitem [{\citenamefont {Wyithe}\ \emph {et~al.}(2008)\citenamefont {Wyithe},
  \citenamefont {Loeb},\ and\ \citenamefont {Geil}}]{Wyithe:2007rq}%
  \BibitemOpen
  \bibfield  {author} {\bibinfo {author} {\bibfnamefont {S.}~\bibnamefont
  {Wyithe}}, \bibinfo {author} {\bibfnamefont {A.}~\bibnamefont {Loeb}}, \ and\
  \bibinfo {author} {\bibfnamefont {P.}~\bibnamefont {Geil}},\ }\href {\doibase
  10.1111/j.1365-2966.2007.12631.x} {\bibfield  {journal} {\bibinfo  {journal}
  {Mon. Not. Roy. Astron. Soc.}\ }\textbf {\bibinfo {volume} {383}},\ \bibinfo
  {pages} {1195} (\bibinfo {year} {2008})},\ \Eprint
  {http://arxiv.org/abs/0709.2955} {arXiv:0709.2955 [astro-ph]} \BibitemShut
  {NoStop}%
\bibitem [{\citenamefont {Chang}\ \emph {et~al.}(2008)\citenamefont {Chang},
  \citenamefont {Pen}, \citenamefont {Peterson},\ and\ \citenamefont
  {McDonald}}]{Chang:2007xk}%
  \BibitemOpen
  \bibfield  {author} {\bibinfo {author} {\bibfnamefont {T.-C.}\ \bibnamefont
  {Chang}}, \bibinfo {author} {\bibfnamefont {U.-L.}\ \bibnamefont {Pen}},
  \bibinfo {author} {\bibfnamefont {J.~B.}\ \bibnamefont {Peterson}}, \ and\
  \bibinfo {author} {\bibfnamefont {P.}~\bibnamefont {McDonald}},\ }\href
  {\doibase 10.1103/PhysRevLett.100.091303} {\bibfield  {journal} {\bibinfo
  {journal} {Phys. Rev. Lett.}\ }\textbf {\bibinfo {volume} {100}},\ \bibinfo
  {pages} {091303} (\bibinfo {year} {2008})},\ \Eprint
  {http://arxiv.org/abs/0709.3672} {arXiv:0709.3672 [astro-ph]} \BibitemShut
  {NoStop}%
\bibitem [{\citenamefont {Loeb}\ and\ \citenamefont
  {Wyithe}(2008)}]{Loeb:2008hg}%
  \BibitemOpen
  \bibfield  {author} {\bibinfo {author} {\bibfnamefont {A.}~\bibnamefont
  {Loeb}}\ and\ \bibinfo {author} {\bibfnamefont {S.}~\bibnamefont {Wyithe}},\
  }\href {\doibase 10.1103/PhysRevLett.100.161301} {\bibfield  {journal}
  {\bibinfo  {journal} {Phys. Rev. Lett.}\ }\textbf {\bibinfo {volume} {100}},\
  \bibinfo {pages} {161301} (\bibinfo {year} {2008})},\ \Eprint
  {http://arxiv.org/abs/0801.1677} {arXiv:0801.1677 [astro-ph]} \BibitemShut
  {NoStop}%
\bibitem [{\citenamefont {Chang}\ \emph {et~al.}(2016)\citenamefont {Chang}
  \emph {et~al.}}]{Chang:2016}%
  \BibitemOpen
  \bibfield  {author} {\bibinfo {author} {\bibfnamefont {T.-C.}\ \bibnamefont
  {Chang}} \emph {et~al.} (\bibinfo {collaboration} {GBT-HIM Team}),\
  }\href@noop {} {\enquote {\bibinfo {title} {{21-cm Intensity Mapping}},}\ }
  (\bibinfo {year} {2016})\BibitemShut {NoStop}%
\bibitem [{CHI(2012)}]{CHIME}%
  \BibitemOpen
  \href@noop {} {\enquote {\bibinfo {title} {{CHIME Overview}},}\ }\bibinfo
  {howpublished} {\url{http://chime.phas.ubc.ca/CHIME_overview.pdf}} (\bibinfo
  {year} {2012})\BibitemShut {NoStop}%
\bibitem [{\citenamefont {Chen}(2015{\natexlab{a}})}]{Chen:2015}%
  \BibitemOpen
  \bibfield  {author} {\bibinfo {author} {\bibfnamefont {X.}~\bibnamefont
  {Chen}},\ }\href@noop {} {\bibfield  {journal} {\bibinfo  {journal} {IAU
  General Assembly}\ }\textbf {\bibinfo {volume} {22}},\ \bibinfo {eid}
  {2252187} (\bibinfo {year} {2015}{\natexlab{a}})}\BibitemShut {NoStop}%
\bibitem [{\citenamefont {Dewdney}\ \emph {et~al.}(2015)\citenamefont {Dewdney}
  \emph {et~al.}}]{SKA}%
  \BibitemOpen
  \bibfield  {author} {\bibinfo {author} {\bibfnamefont {P.}~\bibnamefont
  {Dewdney}} \emph {et~al.} (\bibinfo {collaboration} {SKA Collaboration}),\
  }\href@noop {} {\enquote {\bibinfo {title} {{SKA baseline description}},}\
  }\bibinfo {howpublished}
  {\url{https://www.skatelescope.org/wp-content/uploads/2014/03/SKA-TEL-SKO-0000308_SKA1_System_Baseline_v2_DescriptionRev01-part-1-signed.pdf}}
  (\bibinfo {year} {2015})\BibitemShut {NoStop}%
\bibitem [{\citenamefont {Bull}\ \emph {et~al.}(2015)\citenamefont {Bull},
  \citenamefont {Ferreira}, \citenamefont {Patel},\ and\ \citenamefont
  {Santos}}]{Bull:2014rha}%
  \BibitemOpen
  \bibfield  {author} {\bibinfo {author} {\bibfnamefont {P.}~\bibnamefont
  {Bull}}, \bibinfo {author} {\bibfnamefont {P.~G.}\ \bibnamefont {Ferreira}},
  \bibinfo {author} {\bibfnamefont {P.}~\bibnamefont {Patel}}, \ and\ \bibinfo
  {author} {\bibfnamefont {M.~G.}\ \bibnamefont {Santos}},\ }\href {\doibase
  10.1088/0004-637X/803/1/21} {\bibfield  {journal} {\bibinfo  {journal}
  {Astrophys. J.}\ }\textbf {\bibinfo {volume} {803}},\ \bibinfo {pages} {21}
  (\bibinfo {year} {2015})},\ \Eprint {http://arxiv.org/abs/1405.1452}
  {arXiv:1405.1452 [astro-ph.CO]} \BibitemShut {NoStop}%
\bibitem [{\citenamefont {Scott}\ and\ \citenamefont
  {Rees}(1990)}]{Scott:1990}%
  \BibitemOpen
  \bibfield  {author} {\bibinfo {author} {\bibfnamefont {D.}~\bibnamefont
  {Scott}}\ and\ \bibinfo {author} {\bibfnamefont {M.~J.}\ \bibnamefont
  {Rees}},\ }\href@noop {} {\bibfield  {journal} {\bibinfo  {journal} {Mon.
  Not. Roy. Astron. Soc.}\ }\textbf {\bibinfo {volume} {247}},\ \bibinfo
  {pages} {510} (\bibinfo {year} {1990})}\BibitemShut {NoStop}%
\bibitem [{\citenamefont {Tozzi}\ \emph {et~al.}(2000)\citenamefont {Tozzi},
  \citenamefont {Madau}, \citenamefont {Meiksin},\ and\ \citenamefont
  {Rees}}]{Tozzi:1999zh}%
  \BibitemOpen
  \bibfield  {author} {\bibinfo {author} {\bibfnamefont {P.}~\bibnamefont
  {Tozzi}}, \bibinfo {author} {\bibfnamefont {P.}~\bibnamefont {Madau}},
  \bibinfo {author} {\bibfnamefont {A.}~\bibnamefont {Meiksin}}, \ and\
  \bibinfo {author} {\bibfnamefont {M.~J.}\ \bibnamefont {Rees}},\ }\href
  {\doibase 10.1086/308196} {\bibfield  {journal} {\bibinfo  {journal}
  {Astrophys. J.}\ }\textbf {\bibinfo {volume} {528}},\ \bibinfo {pages} {597}
  (\bibinfo {year} {2000})},\ \Eprint {http://arxiv.org/abs/astro-ph/9903139}
  {arXiv:astro-ph/9903139 [astro-ph]} \BibitemShut {NoStop}%
\bibitem [{\citenamefont {Iliev}\ \emph {et~al.}(2002)\citenamefont {Iliev},
  \citenamefont {Shapiro}, \citenamefont {Ferrara},\ and\ \citenamefont
  {Martel}}]{Iliev:2002gj}%
  \BibitemOpen
  \bibfield  {author} {\bibinfo {author} {\bibfnamefont {I.~T.}\ \bibnamefont
  {Iliev}}, \bibinfo {author} {\bibfnamefont {P.~R.}\ \bibnamefont {Shapiro}},
  \bibinfo {author} {\bibfnamefont {A.}~\bibnamefont {Ferrara}}, \ and\
  \bibinfo {author} {\bibfnamefont {H.}~\bibnamefont {Martel}},\ }\href
  {\doibase 10.1086/341869} {\bibfield  {journal} {\bibinfo  {journal}
  {Astrophys. J.}\ }\textbf {\bibinfo {volume} {572}},\ \bibinfo {pages} {123}
  (\bibinfo {year} {2002})},\ \Eprint {http://arxiv.org/abs/astro-ph/0202410}
  {arXiv:astro-ph/0202410 [astro-ph]} \BibitemShut {NoStop}%
\bibitem [{\citenamefont {Barkana}\ and\ \citenamefont
  {Loeb}(2005{\natexlab{a}})}]{Barkana:2005xu}%
  \BibitemOpen
  \bibfield  {author} {\bibinfo {author} {\bibfnamefont {R.}~\bibnamefont
  {Barkana}}\ and\ \bibinfo {author} {\bibfnamefont {A.}~\bibnamefont {Loeb}},\
  }\href {\doibase 10.1111/j.1745-3933.2005.00079.x} {\bibfield  {journal}
  {\bibinfo  {journal} {Mon. Not. Roy. Astron. Soc.}\ }\textbf {\bibinfo
  {volume} {363}},\ \bibinfo {pages} {L36} (\bibinfo {year}
  {2005}{\natexlab{a}})},\ \Eprint {http://arxiv.org/abs/astro-ph/0502083}
  {arXiv:astro-ph/0502083 [astro-ph]} \BibitemShut {NoStop}%
\bibitem [{\citenamefont {Barkana}\ and\ \citenamefont
  {Loeb}(2005{\natexlab{b}})}]{Barkana:2004zy}%
  \BibitemOpen
  \bibfield  {author} {\bibinfo {author} {\bibfnamefont {R.}~\bibnamefont
  {Barkana}}\ and\ \bibinfo {author} {\bibfnamefont {A.}~\bibnamefont {Loeb}},\
  }\href {\doibase 10.1086/430599} {\bibfield  {journal} {\bibinfo  {journal}
  {Astrophys. J.}\ }\textbf {\bibinfo {volume} {624}},\ \bibinfo {pages} {L65}
  (\bibinfo {year} {2005}{\natexlab{b}})},\ \Eprint
  {http://arxiv.org/abs/astro-ph/0409572} {arXiv:astro-ph/0409572 [astro-ph]}
  \BibitemShut {NoStop}%
\bibitem [{\citenamefont {Bowman}\ \emph {et~al.}(2007)\citenamefont {Bowman},
  \citenamefont {Morales},\ and\ \citenamefont {Hewitt}}]{Bowman:2005hj}%
  \BibitemOpen
  \bibfield  {author} {\bibinfo {author} {\bibfnamefont {J.~D.}\ \bibnamefont
  {Bowman}}, \bibinfo {author} {\bibfnamefont {M.~F.}\ \bibnamefont {Morales}},
  \ and\ \bibinfo {author} {\bibfnamefont {J.~N.}\ \bibnamefont {Hewitt}},\
  }\href {\doibase 10.1086/516560} {\bibfield  {journal} {\bibinfo  {journal}
  {Astrophys. J.}\ }\textbf {\bibinfo {volume} {661}},\ \bibinfo {pages} {1}
  (\bibinfo {year} {2007})},\ \Eprint {http://arxiv.org/abs/astro-ph/0512262}
  {arXiv:astro-ph/0512262 [astro-ph]} \BibitemShut {NoStop}%
\bibitem [{\citenamefont {McQuinn}\ \emph {et~al.}(2006)\citenamefont
  {McQuinn}, \citenamefont {Zahn}, \citenamefont {Zaldarriaga}, \citenamefont
  {Hernquist},\ and\ \citenamefont {Furlanetto}}]{McQuinn:2005hk}%
  \BibitemOpen
  \bibfield  {author} {\bibinfo {author} {\bibfnamefont {M.}~\bibnamefont
  {McQuinn}}, \bibinfo {author} {\bibfnamefont {O.}~\bibnamefont {Zahn}},
  \bibinfo {author} {\bibfnamefont {M.}~\bibnamefont {Zaldarriaga}}, \bibinfo
  {author} {\bibfnamefont {L.}~\bibnamefont {Hernquist}}, \ and\ \bibinfo
  {author} {\bibfnamefont {S.~R.}\ \bibnamefont {Furlanetto}},\ }\href
  {\doibase 10.1086/505167} {\bibfield  {journal} {\bibinfo  {journal}
  {Astrophys. J.}\ }\textbf {\bibinfo {volume} {653}},\ \bibinfo {pages} {815}
  (\bibinfo {year} {2006})},\ \Eprint {http://arxiv.org/abs/astro-ph/0512263}
  {arXiv:astro-ph/0512263 [astro-ph]} \BibitemShut {NoStop}%
\bibitem [{\citenamefont {Santos}\ and\ \citenamefont
  {Cooray}(2006)}]{Santos:2006fp}%
  \BibitemOpen
  \bibfield  {author} {\bibinfo {author} {\bibfnamefont {M.~G.}\ \bibnamefont
  {Santos}}\ and\ \bibinfo {author} {\bibfnamefont {A.}~\bibnamefont
  {Cooray}},\ }\href {\doibase 10.1103/PhysRevD.74.083517} {\bibfield
  {journal} {\bibinfo  {journal} {Phys. Rev.}\ }\textbf {\bibinfo {volume}
  {D74}},\ \bibinfo {pages} {083517} (\bibinfo {year} {2006})},\ \Eprint
  {http://arxiv.org/abs/astro-ph/0605677} {arXiv:astro-ph/0605677 [astro-ph]}
  \BibitemShut {NoStop}%
\bibitem [{\citenamefont {Mao}\ \emph {et~al.}(2008)\citenamefont {Mao},
  \citenamefont {Tegmark}, \citenamefont {McQuinn}, \citenamefont
  {Zaldarriaga},\ and\ \citenamefont {Zahn}}]{Mao:2008ug}%
  \BibitemOpen
  \bibfield  {author} {\bibinfo {author} {\bibfnamefont {Y.}~\bibnamefont
  {Mao}}, \bibinfo {author} {\bibfnamefont {M.}~\bibnamefont {Tegmark}},
  \bibinfo {author} {\bibfnamefont {M.}~\bibnamefont {McQuinn}}, \bibinfo
  {author} {\bibfnamefont {M.}~\bibnamefont {Zaldarriaga}}, \ and\ \bibinfo
  {author} {\bibfnamefont {O.}~\bibnamefont {Zahn}},\ }\href {\doibase
  10.1103/PhysRevD.78.023529} {\bibfield  {journal} {\bibinfo  {journal} {Phys.
  Rev.}\ }\textbf {\bibinfo {volume} {D78}},\ \bibinfo {pages} {023529}
  (\bibinfo {year} {2008})},\ \Eprint {http://arxiv.org/abs/0802.1710}
  {arXiv:0802.1710 [astro-ph]} \BibitemShut {NoStop}%
\bibitem [{\citenamefont {Visbal}\ \emph {et~al.}(2009)\citenamefont {Visbal},
  \citenamefont {Loeb},\ and\ \citenamefont {Wyithe}}]{Visbal:2008rg}%
  \BibitemOpen
  \bibfield  {author} {\bibinfo {author} {\bibfnamefont {E.}~\bibnamefont
  {Visbal}}, \bibinfo {author} {\bibfnamefont {A.}~\bibnamefont {Loeb}}, \ and\
  \bibinfo {author} {\bibfnamefont {J.~S.~B.}\ \bibnamefont {Wyithe}},\ }\href
  {\doibase 10.1088/1475-7516/2009/10/030} {\bibfield  {journal} {\bibinfo
  {journal} {JCAP}\ }\textbf {\bibinfo {volume} {0910}},\ \bibinfo {pages}
  {030} (\bibinfo {year} {2009})},\ \Eprint {http://arxiv.org/abs/0812.0419}
  {arXiv:0812.0419 [astro-ph]} \BibitemShut {NoStop}%
\bibitem [{\citenamefont {Clesse}\ \emph {et~al.}(2012)\citenamefont {Clesse},
  \citenamefont {Lopez-Honorez}, \citenamefont {Ringeval}, \citenamefont
  {Tashiro},\ and\ \citenamefont {Tytgat}}]{Clesse:2012th}%
  \BibitemOpen
  \bibfield  {author} {\bibinfo {author} {\bibfnamefont {S.}~\bibnamefont
  {Clesse}}, \bibinfo {author} {\bibfnamefont {L.}~\bibnamefont
  {Lopez-Honorez}}, \bibinfo {author} {\bibfnamefont {C.}~\bibnamefont
  {Ringeval}}, \bibinfo {author} {\bibfnamefont {H.}~\bibnamefont {Tashiro}}, \
  and\ \bibinfo {author} {\bibfnamefont {M.~H.~G.}\ \bibnamefont {Tytgat}},\
  }\href {\doibase 10.1103/PhysRevD.86.123506} {\bibfield  {journal} {\bibinfo
  {journal} {Phys. Rev.}\ }\textbf {\bibinfo {volume} {D86}},\ \bibinfo {pages}
  {123506} (\bibinfo {year} {2012})},\ \Eprint {http://arxiv.org/abs/1208.4277}
  {arXiv:1208.4277 [astro-ph.CO]} \BibitemShut {NoStop}%
\bibitem [{\citenamefont {Liu}\ \emph {et~al.}(2016)\citenamefont {Liu} \emph
  {et~al.}}]{Liu:2015txa}%
  \BibitemOpen
  \bibfield  {author} {\bibinfo {author} {\bibfnamefont {A.}~\bibnamefont
  {Liu}} \emph {et~al.},\ }\href {\doibase 10.1103/PhysRevD.93.043013}
  {\bibfield  {journal} {\bibinfo  {journal} {Phys. Rev.}\ }\textbf {\bibinfo
  {volume} {D93}},\ \bibinfo {pages} {043013} (\bibinfo {year} {2016})},\
  \Eprint {http://arxiv.org/abs/1509.08463} {arXiv:1509.08463 [astro-ph.CO]}
  \BibitemShut {NoStop}%
\bibitem [{\citenamefont {Liu}\ and\ \citenamefont
  {Parsons}(2016)}]{Liu:2015gaa}%
  \BibitemOpen
  \bibfield  {author} {\bibinfo {author} {\bibfnamefont {A.}~\bibnamefont
  {Liu}}\ and\ \bibinfo {author} {\bibfnamefont {A.~R.}\ \bibnamefont
  {Parsons}},\ }\href {\doibase 10.1093/mnras/stw071} {\bibfield  {journal}
  {\bibinfo  {journal} {Mon. Not. Roy. Astron. Soc.}\ }\textbf {\bibinfo
  {volume} {457}},\ \bibinfo {pages} {1864} (\bibinfo {year} {2016})},\ \Eprint
  {http://arxiv.org/abs/1510.08815} {arXiv:1510.08815 [astro-ph.CO]}
  \BibitemShut {NoStop}%
\bibitem [{\citenamefont {Loeb}\ and\ \citenamefont
  {Zaldarriaga}(2004)}]{Loeb:2003ya}%
  \BibitemOpen
  \bibfield  {author} {\bibinfo {author} {\bibfnamefont {A.}~\bibnamefont
  {Loeb}}\ and\ \bibinfo {author} {\bibfnamefont {M.}~\bibnamefont
  {Zaldarriaga}},\ }\href {\doibase 10.1103/PhysRevLett.92.211301} {\bibfield
  {journal} {\bibinfo  {journal} {Phys. Rev. Lett.}\ }\textbf {\bibinfo
  {volume} {92}},\ \bibinfo {pages} {211301} (\bibinfo {year} {2004})},\
  \Eprint {http://arxiv.org/abs/astro-ph/0312134} {arXiv:astro-ph/0312134
  [astro-ph]} \BibitemShut {NoStop}%
\bibitem [{\citenamefont {Pritchard}\ and\ \citenamefont
  {Pierpaoli}(2008)}]{Pritchard:2008wy}%
  \BibitemOpen
  \bibfield  {author} {\bibinfo {author} {\bibfnamefont {J.~R.}\ \bibnamefont
  {Pritchard}}\ and\ \bibinfo {author} {\bibfnamefont {E.}~\bibnamefont
  {Pierpaoli}},\ }\href {\doibase 10.1103/PhysRevD.78.065009} {\bibfield
  {journal} {\bibinfo  {journal} {Phys. Rev.}\ }\textbf {\bibinfo {volume}
  {D78}},\ \bibinfo {pages} {065009} (\bibinfo {year} {2008})},\ \Eprint
  {http://arxiv.org/abs/0805.1920} {arXiv:0805.1920 [astro-ph]} \BibitemShut
  {NoStop}%
\bibitem [{\citenamefont {Shimabukuro}\ \emph {et~al.}(2014)\citenamefont
  {Shimabukuro}, \citenamefont {Ichiki}, \citenamefont {Inoue},\ and\
  \citenamefont {Yokoyama}}]{Shimabukuro:2014ava}%
  \BibitemOpen
  \bibfield  {author} {\bibinfo {author} {\bibfnamefont {H.}~\bibnamefont
  {Shimabukuro}}, \bibinfo {author} {\bibfnamefont {K.}~\bibnamefont {Ichiki}},
  \bibinfo {author} {\bibfnamefont {S.}~\bibnamefont {Inoue}}, \ and\ \bibinfo
  {author} {\bibfnamefont {S.}~\bibnamefont {Yokoyama}},\ }\href {\doibase
  10.1103/PhysRevD.90.083003} {\bibfield  {journal} {\bibinfo  {journal} {Phys.
  Rev.}\ }\textbf {\bibinfo {volume} {D90}},\ \bibinfo {pages} {083003}
  (\bibinfo {year} {2014})},\ \Eprint {http://arxiv.org/abs/1403.1605}
  {arXiv:1403.1605 [astro-ph.CO]} \BibitemShut {NoStop}%
\bibitem [{\citenamefont {Oyama}\ \emph {et~al.}(2013)\citenamefont {Oyama},
  \citenamefont {Shimizu},\ and\ \citenamefont {Kohri}}]{Oyama:2012tq}%
  \BibitemOpen
  \bibfield  {author} {\bibinfo {author} {\bibfnamefont {Y.}~\bibnamefont
  {Oyama}}, \bibinfo {author} {\bibfnamefont {A.}~\bibnamefont {Shimizu}}, \
  and\ \bibinfo {author} {\bibfnamefont {K.}~\bibnamefont {Kohri}},\ }\href
  {\doibase 10.1016/j.physletb.2012.12.053} {\bibfield  {journal} {\bibinfo
  {journal} {Phys. Lett.}\ }\textbf {\bibinfo {volume} {B718}},\ \bibinfo
  {pages} {1186} (\bibinfo {year} {2013})},\ \Eprint
  {http://arxiv.org/abs/1205.5223} {arXiv:1205.5223 [astro-ph.CO]} \BibitemShut
  {NoStop}%
\bibitem [{\citenamefont {Villaescusa-Navarro}\ \emph
  {et~al.}(2015)\citenamefont {Villaescusa-Navarro}, \citenamefont {Bull},\
  and\ \citenamefont {Viel}}]{Villaescusa-Navarro:2015cca}%
  \BibitemOpen
  \bibfield  {author} {\bibinfo {author} {\bibfnamefont {F.}~\bibnamefont
  {Villaescusa-Navarro}}, \bibinfo {author} {\bibfnamefont {P.}~\bibnamefont
  {Bull}}, \ and\ \bibinfo {author} {\bibfnamefont {M.}~\bibnamefont {Viel}},\
  }\href@noop {} {\  (\bibinfo {year} {2015})},\ \Eprint
  {http://arxiv.org/abs/1507.05102} {arXiv:1507.05102 [astro-ph.CO]}
  \BibitemShut {NoStop}%
\bibitem [{\citenamefont {Oyama}\ \emph {et~al.}(2016)\citenamefont {Oyama},
  \citenamefont {Kohri},\ and\ \citenamefont {Hazumi}}]{Oyama:2015gma}%
  \BibitemOpen
  \bibfield  {author} {\bibinfo {author} {\bibfnamefont {Y.}~\bibnamefont
  {Oyama}}, \bibinfo {author} {\bibfnamefont {K.}~\bibnamefont {Kohri}}, \ and\
  \bibinfo {author} {\bibfnamefont {M.}~\bibnamefont {Hazumi}},\ }\href
  {\doibase 10.1088/1475-7516/2016/02/008} {\bibfield  {journal} {\bibinfo
  {journal} {JCAP}\ }\textbf {\bibinfo {volume} {1602}},\ \bibinfo {pages}
  {008} (\bibinfo {year} {2016})},\ \Eprint {http://arxiv.org/abs/1510.03806}
  {arXiv:1510.03806 [astro-ph.CO]} \BibitemShut {NoStop}%
\bibitem [{\citenamefont {Zurek}\ and\ \citenamefont
  {Hogan}(2007)}]{Zurek:2007gn}%
  \BibitemOpen
  \bibfield  {author} {\bibinfo {author} {\bibfnamefont {K.~M.}\ \bibnamefont
  {Zurek}}\ and\ \bibinfo {author} {\bibfnamefont {C.~J.}\ \bibnamefont
  {Hogan}},\ }\href {\doibase 10.1103/PhysRevD.76.063002} {\bibfield  {journal}
  {\bibinfo  {journal} {Phys. Rev.}\ }\textbf {\bibinfo {volume} {D76}},\
  \bibinfo {pages} {063002} (\bibinfo {year} {2007})},\ \Eprint
  {http://arxiv.org/abs/astro-ph/0703624} {arXiv:astro-ph/0703624 [ASTRO-PH]}
  \BibitemShut {NoStop}%
\bibitem [{\citenamefont {Mesinger}\ \emph {et~al.}(2014)\citenamefont
  {Mesinger}, \citenamefont {Ewall-Wice},\ and\ \citenamefont
  {Hewitt}}]{Mesinger:2013nua}%
  \BibitemOpen
  \bibfield  {author} {\bibinfo {author} {\bibfnamefont {A.}~\bibnamefont
  {Mesinger}}, \bibinfo {author} {\bibfnamefont {A.}~\bibnamefont
  {Ewall-Wice}}, \ and\ \bibinfo {author} {\bibfnamefont {J.}~\bibnamefont
  {Hewitt}},\ }\href {\doibase 10.1093/mnras/stu125} {\bibfield  {journal}
  {\bibinfo  {journal} {Mon. Not. Roy. Astron. Soc.}\ }\textbf {\bibinfo
  {volume} {439}},\ \bibinfo {pages} {3262} (\bibinfo {year} {2014})},\ \Eprint
  {http://arxiv.org/abs/1310.0465} {arXiv:1310.0465 [astro-ph.CO]}\BibitemShut
  {NoStop}%
\bibitem [{\citenamefont {Sitwell}\ \emph {et~al.}(2014)\citenamefont
  {Sitwell}, \citenamefont {Mesinger}, \citenamefont {Ma},\ and\ \citenamefont
  {Sigurdson}}]{Sitwell:2013fpa}%
  \BibitemOpen
  \bibfield  {author} {\bibinfo {author} {\bibfnamefont {M.}~\bibnamefont
  {Sitwell}}, \bibinfo {author} {\bibfnamefont {A.}~\bibnamefont {Mesinger}},
  \bibinfo {author} {\bibfnamefont {Y.-Z.}\ \bibnamefont {Ma}}, \ and\ \bibinfo
  {author} {\bibfnamefont {K.}~\bibnamefont {Sigurdson}},\ }\href {\doibase
  10.1093/mnras/stt2392} {\bibfield  {journal} {\bibinfo  {journal} {Mon. Not.
  Roy. Astron. Soc.}\ }\textbf {\bibinfo {volume} {438}},\ \bibinfo {pages}
  {2664} (\bibinfo {year} {2014})},\ \Eprint {http://arxiv.org/abs/1310.0029}
  {arXiv:1310.0029 [astro-ph.CO]} \BibitemShut {NoStop}%
\bibitem [{\citenamefont {Sekiguchi}\ and\ \citenamefont
  {Tashiro}(2014)}]{Sekiguchi:2014wfa}%
  \BibitemOpen
  \bibfield  {author} {\bibinfo {author} {\bibfnamefont {T.}~\bibnamefont
  {Sekiguchi}}\ and\ \bibinfo {author} {\bibfnamefont {H.}~\bibnamefont
  {Tashiro}},\ }\href {\doibase 10.1088/1475-7516/2014/08/007} {\bibfield
  {journal} {\bibinfo  {journal} {JCAP}\ }\textbf {\bibinfo {volume} {1408}},\
  \bibinfo {pages} {007} (\bibinfo {year} {2014})},\ \Eprint
  {http://arxiv.org/abs/1401.5563} {arXiv:1401.5563 [astro-ph.CO]} \BibitemShut
  {NoStop}%
\bibitem [{\citenamefont {Carucci}\ \emph {et~al.}(2015)\citenamefont
  {Carucci}, \citenamefont {Villaescusa-Navarro}, \citenamefont {Viel},\ and\
  \citenamefont {Lapi}}]{Carucci:2015bra}%
  \BibitemOpen
  \bibfield  {author} {\bibinfo {author} {\bibfnamefont {I.~P.}\ \bibnamefont
  {Carucci}}, \bibinfo {author} {\bibfnamefont {F.}~\bibnamefont
  {Villaescusa-Navarro}}, \bibinfo {author} {\bibfnamefont {M.}~\bibnamefont
  {Viel}}, \ and\ \bibinfo {author} {\bibfnamefont {A.}~\bibnamefont {Lapi}},\
  }\href {\doibase 10.1088/1475-7516/2015/07/047} {\bibfield  {journal}
  {\bibinfo  {journal} {JCAP}\ }\textbf {\bibinfo {volume} {1507}},\ \bibinfo
  {pages} {047} (\bibinfo {year} {2015})},\ \Eprint
  {http://arxiv.org/abs/1502.06961} {arXiv:1502.06961 [astro-ph.CO]}
  \BibitemShut {NoStop}%
\bibitem [{\citenamefont {Morales}\ and\ \citenamefont
  {Wyithe}(2010)}]{Morales:2009gs}%
  \BibitemOpen
  \bibfield  {author} {\bibinfo {author} {\bibfnamefont {M.~F.}\ \bibnamefont
  {Morales}}\ and\ \bibinfo {author} {\bibfnamefont {J.~S.~B.}\ \bibnamefont
  {Wyithe}},\ }\href {\doibase 10.1146/annurev-astro-081309-130936} {\bibfield
  {journal} {\bibinfo  {journal} {Ann. Rev. Astron. Astrophys.}\ }\textbf
  {\bibinfo {volume} {48}},\ \bibinfo {pages} {127} (\bibinfo {year} {2010})},\
  \Eprint {http://arxiv.org/abs/0910.3010} {arXiv:0910.3010 [astro-ph.CO]}
  \BibitemShut {NoStop}%
\bibitem [{\citenamefont {Archidiacono}\ \emph {et~al.}(2014)\citenamefont
  {Archidiacono}, \citenamefont {Lopez-Honorez},\ and\ \citenamefont
  {Mena}}]{Archidiacono:2014msa}%
  \BibitemOpen
  \bibfield  {author} {\bibinfo {author} {\bibfnamefont {M.}~\bibnamefont
  {Archidiacono}}, \bibinfo {author} {\bibfnamefont {L.}~\bibnamefont
  {Lopez-Honorez}}, \ and\ \bibinfo {author} {\bibfnamefont {O.}~\bibnamefont
  {Mena}},\ }\href {\doibase 10.1103/PhysRevD.90.123016} {\bibfield  {journal}
  {\bibinfo  {journal} {Phys. Rev.}\ }\textbf {\bibinfo {volume} {D90}},\
  \bibinfo {pages} {123016} (\bibinfo {year} {2014})},\ \Eprint
  {http://arxiv.org/abs/1409.1802} {arXiv:1409.1802 [astro-ph.CO]} \BibitemShut
  {NoStop}%
\bibitem [{\citenamefont {Chen}(2015{\natexlab{b}})}]{Chen:2015vdi}%
  \BibitemOpen
  \bibfield  {author} {\bibinfo {author} {\bibfnamefont {X.}~\bibnamefont
  {Chen}},\ }\href {\doibase 10.1142/S0217751X15450116} {\bibfield  {journal}
  {\bibinfo  {journal} {Int. J. Mod. Phys.}\ }\textbf {\bibinfo {volume}
  {A30}},\ \bibinfo {pages} {1545011} (\bibinfo {year}
  {2015}{\natexlab{b}})}\BibitemShut {NoStop}%
\bibitem [{\citenamefont {Cooray}(2006)}]{Cooray:2006km}%
  \BibitemOpen
  \bibfield  {author} {\bibinfo {author} {\bibfnamefont {A.}~\bibnamefont
  {Cooray}},\ }\href {\doibase 10.1103/PhysRevLett.97.261301} {\bibfield
  {journal} {\bibinfo  {journal} {Phys. Rev. Lett.}\ }\textbf {\bibinfo
  {volume} {97}},\ \bibinfo {pages} {261301} (\bibinfo {year} {2006})},\
  \Eprint {http://arxiv.org/abs/astro-ph/0610257} {arXiv:astro-ph/0610257
  [astro-ph]} \BibitemShut {NoStop}%
\bibitem [{\citenamefont {Pillepich}\ \emph {et~al.}(2007)\citenamefont
  {Pillepich}, \citenamefont {Porciani},\ and\ \citenamefont
  {Matarrese}}]{Pillepich:2006fj}%
  \BibitemOpen
  \bibfield  {author} {\bibinfo {author} {\bibfnamefont {A.}~\bibnamefont
  {Pillepich}}, \bibinfo {author} {\bibfnamefont {C.}~\bibnamefont {Porciani}},
  \ and\ \bibinfo {author} {\bibfnamefont {S.}~\bibnamefont {Matarrese}},\
  }\href {\doibase 10.1086/517963} {\bibfield  {journal} {\bibinfo  {journal}
  {Astrophys. J.}\ }\textbf {\bibinfo {volume} {662}},\ \bibinfo {pages} {1}
  (\bibinfo {year} {2007})},\ \Eprint {http://arxiv.org/abs/astro-ph/0611126}
  {arXiv:astro-ph/0611126 [astro-ph]} \BibitemShut {NoStop}%
\bibitem [{\citenamefont {Joudaki}\ \emph {et~al.}(2011)\citenamefont
  {Joudaki}, \citenamefont {Dore}, \citenamefont {Ferramacho}, \citenamefont
  {Kaplinghat},\ and\ \citenamefont {Santos}}]{Joudaki:2011sv}%
  \BibitemOpen
  \bibfield  {author} {\bibinfo {author} {\bibfnamefont {S.}~\bibnamefont
  {Joudaki}}, \bibinfo {author} {\bibfnamefont {O.}~\bibnamefont {Dore}},
  \bibinfo {author} {\bibfnamefont {L.}~\bibnamefont {Ferramacho}}, \bibinfo
  {author} {\bibfnamefont {M.}~\bibnamefont {Kaplinghat}}, \ and\ \bibinfo
  {author} {\bibfnamefont {M.~G.}\ \bibnamefont {Santos}},\ }\href {\doibase
  10.1103/PhysRevLett.107.131304} {\bibfield  {journal} {\bibinfo  {journal}
  {Phys. Rev. Lett.}\ }\textbf {\bibinfo {volume} {107}},\ \bibinfo {pages}
  {131304} (\bibinfo {year} {2011})},\ \Eprint {http://arxiv.org/abs/1105.1773}
  {arXiv:1105.1773 [astro-ph.CO]} \BibitemShut {NoStop}%
\bibitem [{\citenamefont {Tashiro}\ and\ \citenamefont
  {Ho}(2013)}]{Tashiro:2012wr}%
  \BibitemOpen
  \bibfield  {author} {\bibinfo {author} {\bibfnamefont {H.}~\bibnamefont
  {Tashiro}}\ and\ \bibinfo {author} {\bibfnamefont {S.}~\bibnamefont {Ho}},\
  }\href {\doibase 10.1093/mnras/stt191} {\bibfield  {journal} {\bibinfo
  {journal} {Mon. Not. Roy. Astron. Soc.}\ }\textbf {\bibinfo {volume} {431}},\
  \bibinfo {pages} {2017} (\bibinfo {year} {2013})},\ \Eprint
  {http://arxiv.org/abs/1205.0563} {arXiv:1205.0563 [astro-ph.CO]} \BibitemShut
  {NoStop}%
\bibitem [{\citenamefont {Chongchitnan}\ and\ \citenamefont
  {Silk}(2012)}]{Chongchitnan:2012we}%
  \BibitemOpen
  \bibfield  {author} {\bibinfo {author} {\bibfnamefont {S.}~\bibnamefont
  {Chongchitnan}}\ and\ \bibinfo {author} {\bibfnamefont {J.}~\bibnamefont
  {Silk}},\ }\href {\doibase 10.1111/j.1745-3933.2012.01315.x} {\bibfield
  {journal} {\bibinfo  {journal} {Mon. Not. Roy. Astron. Soc.}\ }\textbf
  {\bibinfo {volume} {426}},\ \bibinfo {pages} {L21} (\bibinfo {year}
  {2012})},\ \Eprint {http://arxiv.org/abs/1205.6799} {arXiv:1205.6799
  [astro-ph.CO]} \BibitemShut {NoStop}%
\bibitem [{\citenamefont {Chongchitnan}(2013)}]{Chongchitnan:2013oxa}%
  \BibitemOpen
  \bibfield  {author} {\bibinfo {author} {\bibfnamefont {S.}~\bibnamefont
  {Chongchitnan}},\ }\href {\doibase 10.1088/1475-7516/2013/03/037} {\bibfield
  {journal} {\bibinfo  {journal} {JCAP}\ }\textbf {\bibinfo {volume} {1303}},\
  \bibinfo {pages} {037} (\bibinfo {year} {2013})},\ \Eprint
  {http://arxiv.org/abs/1303.4387} {arXiv:1303.4387 [astro-ph.CO]} \BibitemShut
  {NoStop}%
\bibitem [{\citenamefont {Takeuchi}\ and\ \citenamefont
  {Chongchitnan}(2014)}]{Takeuchi:2013hza}%
  \BibitemOpen
  \bibfield  {author} {\bibinfo {author} {\bibfnamefont {Y.}~\bibnamefont
  {Takeuchi}}\ and\ \bibinfo {author} {\bibfnamefont {S.}~\bibnamefont
  {Chongchitnan}},\ }\href {\doibase 10.1093/mnras/stu059} {\bibfield
  {journal} {\bibinfo  {journal} {Mon. Not. Roy. Astron. Soc.}\ }\textbf
  {\bibinfo {volume} {439}},\ \bibinfo {pages} {1125} (\bibinfo {year}
  {2014})},\ \Eprint {http://arxiv.org/abs/1311.2585} {arXiv:1311.2585
  [astro-ph.CO]} \BibitemShut {NoStop}%
\bibitem [{\citenamefont {Sekiguchi}\ \emph {et~al.}(2014)\citenamefont
  {Sekiguchi}, \citenamefont {Tashiro}, \citenamefont {Silk},\ and\
  \citenamefont {Sugiyama}}]{Sekiguchi:2013lma}%
  \BibitemOpen
  \bibfield  {author} {\bibinfo {author} {\bibfnamefont {T.}~\bibnamefont
  {Sekiguchi}}, \bibinfo {author} {\bibfnamefont {H.}~\bibnamefont {Tashiro}},
  \bibinfo {author} {\bibfnamefont {J.}~\bibnamefont {Silk}}, \ and\ \bibinfo
  {author} {\bibfnamefont {N.}~\bibnamefont {Sugiyama}},\ }\href {\doibase
  10.1088/1475-7516/2014/03/001} {\bibfield  {journal} {\bibinfo  {journal}
  {JCAP}\ }\textbf {\bibinfo {volume} {1403}},\ \bibinfo {pages} {001}
  (\bibinfo {year} {2014})},\ \Eprint {http://arxiv.org/abs/1311.3294}
  {arXiv:1311.3294 [astro-ph.CO]} \BibitemShut {NoStop}%
\bibitem [{\citenamefont {Mack}\ and\ \citenamefont
  {Wesley}(2008)}]{Mack:2008nv}%
  \BibitemOpen
  \bibfield  {author} {\bibinfo {author} {\bibfnamefont {K.~J.}\ \bibnamefont
  {Mack}}\ and\ \bibinfo {author} {\bibfnamefont {D.~H.}\ \bibnamefont
  {Wesley}},\ }\href@noop {} {\  (\bibinfo {year} {2008})},\ \Eprint
  {http://arxiv.org/abs/0805.1531} {arXiv:0805.1531 [astro-ph]} \BibitemShut
  {NoStop}%
\bibitem [{\citenamefont {Tashiro}\ and\ \citenamefont
  {Sugiyama}(2013)}]{Tashiro:2012qe}%
  \BibitemOpen
  \bibfield  {author} {\bibinfo {author} {\bibfnamefont {H.}~\bibnamefont
  {Tashiro}}\ and\ \bibinfo {author} {\bibfnamefont {N.}~\bibnamefont
  {Sugiyama}},\ }\href {\doibase 10.1093/mnras/stt1493} {\bibfield  {journal}
  {\bibinfo  {journal} {Mon. Not. Roy. Astron. Soc.}\ }\textbf {\bibinfo
  {volume} {435}},\ \bibinfo {pages} {3001} (\bibinfo {year} {2013})},\ \Eprint
  {http://arxiv.org/abs/1207.6405} {arXiv:1207.6405 [astro-ph.CO]} \BibitemShut
  {NoStop}%
\bibitem [{\citenamefont {Hall}\ \emph {et~al.}(2013)\citenamefont {Hall},
  \citenamefont {Bonvin},\ and\ \citenamefont {Challinor}}]{Hall:2012wd}%
  \BibitemOpen
  \bibfield  {author} {\bibinfo {author} {\bibfnamefont {A.}~\bibnamefont
  {Hall}}, \bibinfo {author} {\bibfnamefont {C.}~\bibnamefont {Bonvin}}, \ and\
  \bibinfo {author} {\bibfnamefont {A.}~\bibnamefont {Challinor}},\ }\href
  {\doibase 10.1103/PhysRevD.87.064026} {\bibfield  {journal} {\bibinfo
  {journal} {Phys. Rev.}\ }\textbf {\bibinfo {volume} {D87}},\ \bibinfo {pages}
  {064026} (\bibinfo {year} {2013})},\ \Eprint {http://arxiv.org/abs/1212.0728}
  {arXiv:1212.0728 [astro-ph.CO]} \BibitemShut {NoStop}%
\bibitem [{\citenamefont {Brax}\ \emph {et~al.}(2013)\citenamefont {Brax},
  \citenamefont {Clesse},\ and\ \citenamefont {Davis}}]{Brax:2012cr}%
  \BibitemOpen
  \bibfield  {author} {\bibinfo {author} {\bibfnamefont {P.}~\bibnamefont
  {Brax}}, \bibinfo {author} {\bibfnamefont {S.}~\bibnamefont {Clesse}}, \ and\
  \bibinfo {author} {\bibfnamefont {A.-C.}\ \bibnamefont {Davis}},\ }\href
  {\doibase 10.1088/1475-7516/2013/01/003} {\bibfield  {journal} {\bibinfo
  {journal} {JCAP}\ }\textbf {\bibinfo {volume} {1301}},\ \bibinfo {pages}
  {003} (\bibinfo {year} {2013})},\ \Eprint {http://arxiv.org/abs/1207.1273}
  {arXiv:1207.1273 [astro-ph.CO]} \BibitemShut {NoStop}%
\bibitem [{\citenamefont {Chen}\ and\ \citenamefont
  {Kamionkowski}(2004)}]{Chen:2003gz}%
  \BibitemOpen
  \bibfield  {author} {\bibinfo {author} {\bibfnamefont {X.-L.}\ \bibnamefont
  {Chen}}\ and\ \bibinfo {author} {\bibfnamefont {M.}~\bibnamefont
  {Kamionkowski}},\ }\href {\doibase 10.1103/PhysRevD.70.043502} {\bibfield
  {journal} {\bibinfo  {journal} {Phys. Rev.}\ }\textbf {\bibinfo {volume}
  {D70}},\ \bibinfo {pages} {043502} (\bibinfo {year} {2004})},\ \Eprint
  {http://arxiv.org/abs/astro-ph/0310473} {arXiv:astro-ph/0310473 [astro-ph]}
  \BibitemShut {NoStop}%
\bibitem [{\citenamefont {Hansen}\ and\ \citenamefont
  {Haiman}(2004)}]{Hansen:2003yj}%
  \BibitemOpen
  \bibfield  {author} {\bibinfo {author} {\bibfnamefont {S.~H.}\ \bibnamefont
  {Hansen}}\ and\ \bibinfo {author} {\bibfnamefont {Z.}~\bibnamefont
  {Haiman}},\ }\href {\doibase 10.1086/379636} {\bibfield  {journal} {\bibinfo
  {journal} {Astrophys.J.}\ }\textbf {\bibinfo {volume} {600}},\ \bibinfo
  {pages} {26} (\bibinfo {year} {2004})},\ \Eprint
  {http://arxiv.org/abs/astro-ph/0305126} {arXiv:astro-ph/0305126 [astro-ph]}
  \BibitemShut {NoStop}%
\bibitem [{\citenamefont {Pierpaoli}(2004)}]{Pierpaoli:2003rz}%
  \BibitemOpen
  \bibfield  {author} {\bibinfo {author} {\bibfnamefont {E.}~\bibnamefont
  {Pierpaoli}},\ }\href {\doibase 10.1103/PhysRevLett.92.031301} {\bibfield
  {journal} {\bibinfo  {journal} {Phys.Rev.Lett.}\ }\textbf {\bibinfo {volume}
  {92}},\ \bibinfo {pages} {031301} (\bibinfo {year} {2004})},\ \Eprint
  {http://arxiv.org/abs/astro-ph/0310375} {arXiv:astro-ph/0310375 [astro-ph]}
  \BibitemShut {NoStop}%
\bibitem [{\citenamefont {Padmanabhan}\ and\ \citenamefont
  {Finkbeiner}(2005)}]{Padmanabhan:2005es}%
  \BibitemOpen
  \bibfield  {author} {\bibinfo {author} {\bibfnamefont {N.}~\bibnamefont
  {Padmanabhan}}\ and\ \bibinfo {author} {\bibfnamefont {D.~P.}\ \bibnamefont
  {Finkbeiner}},\ }\href {\doibase 10.1103/PhysRevD.72.023508} {\bibfield
  {journal} {\bibinfo  {journal} {Phys. Rev.}\ }\textbf {\bibinfo {volume}
  {D72}},\ \bibinfo {pages} {023508} (\bibinfo {year} {2005})},\ \Eprint
  {http://arxiv.org/abs/astro-ph/0503486} {arXiv:astro-ph/0503486 [astro-ph]}
  \BibitemShut {NoStop}%
\bibitem [{\citenamefont {Shchekinov}\ and\ \citenamefont
  {Vasiliev}(2007)}]{Shchekinov:2006eb}%
  \BibitemOpen
  \bibfield  {author} {\bibinfo {author} {\bibfnamefont {Y.~A.}\ \bibnamefont
  {Shchekinov}}\ and\ \bibinfo {author} {\bibfnamefont {E.~O.}\ \bibnamefont
  {Vasiliev}},\ }\href {\doibase 10.1111/j.1365-2966.2007.11715.x} {\bibfield
  {journal} {\bibinfo  {journal} {Mon. Not. Roy. Astron. Soc.}\ }\textbf
  {\bibinfo {volume} {379}},\ \bibinfo {pages} {1003} (\bibinfo {year}
  {2007})},\ \Eprint {http://arxiv.org/abs/astro-ph/0604231}
  {arXiv:astro-ph/0604231 [astro-ph]} \BibitemShut {NoStop}%
\bibitem [{\citenamefont {Furlanetto}\ \emph
  {et~al.}(2006{\natexlab{a}})\citenamefont {Furlanetto}, \citenamefont {Oh},\
  and\ \citenamefont {Pierpaoli}}]{Furlanetto:2006wp}%
  \BibitemOpen
  \bibfield  {author} {\bibinfo {author} {\bibfnamefont {S.~R.}\ \bibnamefont
  {Furlanetto}}, \bibinfo {author} {\bibfnamefont {S.~P.}\ \bibnamefont {Oh}},
  \ and\ \bibinfo {author} {\bibfnamefont {E.}~\bibnamefont {Pierpaoli}},\
  }\href {\doibase 10.1103/PhysRevD.74.103502} {\bibfield  {journal} {\bibinfo
  {journal} {Phys. Rev.}\ }\textbf {\bibinfo {volume} {D74}},\ \bibinfo {pages}
  {103502} (\bibinfo {year} {2006}{\natexlab{a}})},\ \Eprint
  {http://arxiv.org/abs/astro-ph/0608385} {arXiv:astro-ph/0608385 [astro-ph]}
  \BibitemShut {NoStop}%
\bibitem [{\citenamefont {Zhang}\ \emph {et~al.}(2006)\citenamefont {Zhang},
  \citenamefont {Chen}, \citenamefont {Lei},\ and\ \citenamefont
  {Si}}]{Zhang:2006fr}%
  \BibitemOpen
  \bibfield  {author} {\bibinfo {author} {\bibfnamefont {L.}~\bibnamefont
  {Zhang}}, \bibinfo {author} {\bibfnamefont {X.-L.}\ \bibnamefont {Chen}},
  \bibinfo {author} {\bibfnamefont {Y.-A.}\ \bibnamefont {Lei}}, \ and\
  \bibinfo {author} {\bibfnamefont {Z.-G.}\ \bibnamefont {Si}},\ }\href
  {\doibase 10.1103/PhysRevD.74.103519} {\bibfield  {journal} {\bibinfo
  {journal} {Phys. Rev.}\ }\textbf {\bibinfo {volume} {D74}},\ \bibinfo {pages}
  {103519} (\bibinfo {year} {2006})},\ \Eprint
  {http://arxiv.org/abs/astro-ph/0603425} {arXiv:astro-ph/0603425 [astro-ph]}
  \BibitemShut {NoStop}%
\bibitem [{\citenamefont {Mapelli}\ \emph {et~al.}(2006)\citenamefont
  {Mapelli}, \citenamefont {Ferrara},\ and\ \citenamefont
  {Pierpaoli}}]{Mapelli:2006ej}%
  \BibitemOpen
  \bibfield  {author} {\bibinfo {author} {\bibfnamefont {M.}~\bibnamefont
  {Mapelli}}, \bibinfo {author} {\bibfnamefont {A.}~\bibnamefont {Ferrara}}, \
  and\ \bibinfo {author} {\bibfnamefont {E.}~\bibnamefont {Pierpaoli}},\ }\href
  {\doibase 10.1111/j.1365-2966.2006.10408.x} {\bibfield  {journal} {\bibinfo
  {journal} {Mon. Not. Roy. Astron. Soc.}\ }\textbf {\bibinfo {volume} {369}},\
  \bibinfo {pages} {1719} (\bibinfo {year} {2006})},\ \Eprint
  {http://arxiv.org/abs/astro-ph/0603237} {arXiv:astro-ph/0603237 [astro-ph]}
  \BibitemShut {NoStop}%
\bibitem [{\citenamefont {Zhang}\ \emph {et~al.}(2007)\citenamefont {Zhang},
  \citenamefont {Chen}, \citenamefont {Kamionkowski}, \citenamefont {Si},\ and\
  \citenamefont {Zheng}}]{Zhang:2007zzh}%
  \BibitemOpen
  \bibfield  {author} {\bibinfo {author} {\bibfnamefont {L.}~\bibnamefont
  {Zhang}}, \bibinfo {author} {\bibfnamefont {X.}~\bibnamefont {Chen}},
  \bibinfo {author} {\bibfnamefont {M.}~\bibnamefont {Kamionkowski}}, \bibinfo
  {author} {\bibfnamefont {Z.-g.}\ \bibnamefont {Si}}, \ and\ \bibinfo {author}
  {\bibfnamefont {Z.}~\bibnamefont {Zheng}},\ }\href {\doibase
  10.1103/PhysRevD.76.061301} {\bibfield  {journal} {\bibinfo  {journal} {Phys.
  Rev.}\ }\textbf {\bibinfo {volume} {D76}},\ \bibinfo {pages} {061301}
  (\bibinfo {year} {2007})},\ \Eprint {http://arxiv.org/abs/0704.2444}
  {arXiv:0704.2444 [astro-ph]}\BibitemShut {NoStop}%
\bibitem [{\citenamefont {Natarajan}\ and\ \citenamefont
  {Schwarz}(2008)}]{Natarajan:2008pk}%
  \BibitemOpen
  \bibfield  {author} {\bibinfo {author} {\bibfnamefont {A.}~\bibnamefont
  {Natarajan}}\ and\ \bibinfo {author} {\bibfnamefont {D.~J.}\ \bibnamefont
  {Schwarz}},\ }\href {\doibase 10.1103/PhysRevD.78.103524,
  10.1103/PhysRevD.81.089905} {\bibfield  {journal} {\bibinfo  {journal} {Phys.
  Rev.}\ }\textbf {\bibinfo {volume} {D78}},\ \bibinfo {pages} {103524}
  (\bibinfo {year} {2008})},\ \bibinfo {note} {[Erratum: Phys.
  Rev.D81,089905(2010)]},\ \Eprint {http://arxiv.org/abs/0805.3945}
  {arXiv:0805.3945 [astro-ph]} \BibitemShut {NoStop}%
\bibitem [{\citenamefont {Natarajan}\ and\ \citenamefont
  {Schwarz}(2009)}]{Natarajan:2009bm}%
  \BibitemOpen
  \bibfield  {author} {\bibinfo {author} {\bibfnamefont {A.}~\bibnamefont
  {Natarajan}}\ and\ \bibinfo {author} {\bibfnamefont {D.~J.}\ \bibnamefont
  {Schwarz}},\ }\href {\doibase 10.1103/PhysRevD.80.043529} {\bibfield
  {journal} {\bibinfo  {journal} {Phys. Rev.}\ }\textbf {\bibinfo {volume}
  {D80}},\ \bibinfo {pages} {043529} (\bibinfo {year} {2009})},\ \Eprint
  {http://arxiv.org/abs/0903.4485} {arXiv:0903.4485 [astro-ph.CO]} \BibitemShut
  {NoStop}%
\bibitem [{\citenamefont {Belikov}\ and\ \citenamefont
  {Hooper}(2009)}]{Belikov:2009qx}%
  \BibitemOpen
  \bibfield  {author} {\bibinfo {author} {\bibfnamefont {A.~V.}\ \bibnamefont
  {Belikov}}\ and\ \bibinfo {author} {\bibfnamefont {D.}~\bibnamefont
  {Hooper}},\ }\href {\doibase 10.1103/PhysRevD.80.035007} {\bibfield
  {journal} {\bibinfo  {journal} {Phys. Rev.}\ }\textbf {\bibinfo {volume}
  {D80}},\ \bibinfo {pages} {035007} (\bibinfo {year} {2009})},\ \Eprint
  {http://arxiv.org/abs/0904.1210} {arXiv:0904.1210 [hep-ph]} \BibitemShut
  {NoStop}%
\bibitem [{\citenamefont {Galli}\ \emph {et~al.}(2009)\citenamefont {Galli},
  \citenamefont {Iocco}, \citenamefont {Bertone},\ and\ \citenamefont
  {Melchiorri}}]{Galli:2009zc}%
  \BibitemOpen
  \bibfield  {author} {\bibinfo {author} {\bibfnamefont {S.}~\bibnamefont
  {Galli}}, \bibinfo {author} {\bibfnamefont {F.}~\bibnamefont {Iocco}},
  \bibinfo {author} {\bibfnamefont {G.}~\bibnamefont {Bertone}}, \ and\
  \bibinfo {author} {\bibfnamefont {A.}~\bibnamefont {Melchiorri}},\ }\href
  {\doibase 10.1103/PhysRevD.80.023505} {\bibfield  {journal} {\bibinfo
  {journal} {Phys. Rev.}\ }\textbf {\bibinfo {volume} {D80}},\ \bibinfo {pages}
  {023505} (\bibinfo {year} {2009})},\ \Eprint {http://arxiv.org/abs/0905.0003}
  {arXiv:0905.0003 [astro-ph.CO]} \BibitemShut {NoStop}%
\bibitem [{\citenamefont {Slatyer}\ \emph {et~al.}(2009)\citenamefont
  {Slatyer}, \citenamefont {Padmanabhan},\ and\ \citenamefont
  {Finkbeiner}}]{Slatyer:2009yq}%
  \BibitemOpen
  \bibfield  {author} {\bibinfo {author} {\bibfnamefont {T.~R.}\ \bibnamefont
  {Slatyer}}, \bibinfo {author} {\bibfnamefont {N.}~\bibnamefont
  {Padmanabhan}}, \ and\ \bibinfo {author} {\bibfnamefont {D.~P.}\ \bibnamefont
  {Finkbeiner}},\ }\href {\doibase 10.1103/PhysRevD.80.043526} {\bibfield
  {journal} {\bibinfo  {journal} {Phys. Rev.}\ }\textbf {\bibinfo {volume}
  {D80}},\ \bibinfo {pages} {043526} (\bibinfo {year} {2009})},\ \Eprint
  {http://arxiv.org/abs/0906.1197} {arXiv:0906.1197 [astro-ph.CO]} \BibitemShut
  {NoStop}%
\bibitem [{\citenamefont {Huetsi}\ \emph {et~al.}(2009)\citenamefont {Huetsi},
  \citenamefont {Hektor},\ and\ \citenamefont {Raidal}}]{Huetsi:2009ex}%
  \BibitemOpen
  \bibfield  {author} {\bibinfo {author} {\bibfnamefont {G.}~\bibnamefont
  {Huetsi}}, \bibinfo {author} {\bibfnamefont {A.}~\bibnamefont {Hektor}}, \
  and\ \bibinfo {author} {\bibfnamefont {M.}~\bibnamefont {Raidal}},\ }\href
  {\doibase 10.1051/0004-6361/200912760} {\bibfield  {journal} {\bibinfo
  {journal} {Astron. Astrophys.}\ }\textbf {\bibinfo {volume} {505}},\ \bibinfo
  {pages} {999} (\bibinfo {year} {2009})},\ \Eprint
  {http://arxiv.org/abs/0906.4550} {arXiv:0906.4550 [astro-ph.CO]} \BibitemShut
  {NoStop}%
\bibitem [{\citenamefont {Cirelli}\ \emph {et~al.}(2009)\citenamefont
  {Cirelli}, \citenamefont {Iocco},\ and\ \citenamefont
  {Panci}}]{Cirelli:2009bb}%
  \BibitemOpen
  \bibfield  {author} {\bibinfo {author} {\bibfnamefont {M.}~\bibnamefont
  {Cirelli}}, \bibinfo {author} {\bibfnamefont {F.}~\bibnamefont {Iocco}}, \
  and\ \bibinfo {author} {\bibfnamefont {P.}~\bibnamefont {Panci}},\ }\href
  {\doibase 10.1088/1475-7516/2009/10/009} {\bibfield  {journal} {\bibinfo
  {journal} {JCAP}\ }\textbf {\bibinfo {volume} {0910}},\ \bibinfo {pages}
  {009} (\bibinfo {year} {2009})},\ \Eprint {http://arxiv.org/abs/0907.0719}
  {arXiv:0907.0719 [astro-ph.CO]} \BibitemShut {NoStop}%
\bibitem [{\citenamefont {Kanzaki}\ \emph {et~al.}(2010)\citenamefont
  {Kanzaki}, \citenamefont {Kawasaki},\ and\ \citenamefont
  {Nakayama}}]{Kanzaki:2009hf}%
  \BibitemOpen
  \bibfield  {author} {\bibinfo {author} {\bibfnamefont {T.}~\bibnamefont
  {Kanzaki}}, \bibinfo {author} {\bibfnamefont {M.}~\bibnamefont {Kawasaki}}, \
  and\ \bibinfo {author} {\bibfnamefont {K.}~\bibnamefont {Nakayama}},\ }\href
  {\doibase 10.1143/PTP.123.853} {\bibfield  {journal} {\bibinfo  {journal}
  {Prog. Theor. Phys.}\ }\textbf {\bibinfo {volume} {123}},\ \bibinfo {pages}
  {853} (\bibinfo {year} {2010})},\ \Eprint {http://arxiv.org/abs/0907.3985}
  {arXiv:0907.3985 [astro-ph.CO]} \BibitemShut {NoStop}%
\bibitem [{\citenamefont {Hisano}\ \emph {et~al.}(2011)\citenamefont {Hisano}
  \emph {et~al.}}]{Hisano:2011dc}%
  \BibitemOpen
  \bibfield  {author} {\bibinfo {author} {\bibfnamefont {J.}~\bibnamefont
  {Hisano}} \emph {et~al.},\ }\href {\doibase 10.1103/PhysRevD.83.123511}
  {\bibfield  {journal} {\bibinfo  {journal} {Phys. Rev.}\ }\textbf {\bibinfo
  {volume} {D83}},\ \bibinfo {pages} {123511} (\bibinfo {year} {2011})},\
  \Eprint {http://arxiv.org/abs/1102.4658} {arXiv:1102.4658 [hep-ph]}
  \BibitemShut {NoStop}%
\bibitem [{\citenamefont {Galli}\ \emph {et~al.}(2011)\citenamefont {Galli},
  \citenamefont {Iocco}, \citenamefont {Bertone},\ and\ \citenamefont
  {Melchiorri}}]{Galli:2011rz}%
  \BibitemOpen
  \bibfield  {author} {\bibinfo {author} {\bibfnamefont {S.}~\bibnamefont
  {Galli}}, \bibinfo {author} {\bibfnamefont {F.}~\bibnamefont {Iocco}},
  \bibinfo {author} {\bibfnamefont {G.}~\bibnamefont {Bertone}}, \ and\
  \bibinfo {author} {\bibfnamefont {A.}~\bibnamefont {Melchiorri}},\ }\href
  {\doibase 10.1103/PhysRevD.84.027302} {\bibfield  {journal} {\bibinfo
  {journal} {Phys. Rev.}\ }\textbf {\bibinfo {volume} {D84}},\ \bibinfo {pages}
  {027302} (\bibinfo {year} {2011})},\ \Eprint {http://arxiv.org/abs/1106.1528}
  {arXiv:1106.1528 [astro-ph.CO]} \BibitemShut {NoStop}%
\bibitem [{\citenamefont {Finkbeiner}\ \emph {et~al.}(2012)\citenamefont
  {Finkbeiner}, \citenamefont {Galli}, \citenamefont {Lin},\ and\ \citenamefont
  {Slatyer}}]{Finkbeiner:2011dx}%
  \BibitemOpen
  \bibfield  {author} {\bibinfo {author} {\bibfnamefont {D.~P.}\ \bibnamefont
  {Finkbeiner}}, \bibinfo {author} {\bibfnamefont {S.}~\bibnamefont {Galli}},
  \bibinfo {author} {\bibfnamefont {T.}~\bibnamefont {Lin}}, \ and\ \bibinfo
  {author} {\bibfnamefont {T.~R.}\ \bibnamefont {Slatyer}},\ }\href {\doibase
  10.1103/PhysRevD.85.043522} {\bibfield  {journal} {\bibinfo  {journal} {Phys.
  Rev.}\ }\textbf {\bibinfo {volume} {D85}},\ \bibinfo {pages} {043522}
  (\bibinfo {year} {2012})},\ \Eprint {http://arxiv.org/abs/1109.6322}
  {arXiv:1109.6322 [astro-ph.CO]} \BibitemShut {NoStop}%
\bibitem [{\citenamefont {Giesen}\ \emph {et~al.}(2012)\citenamefont {Giesen},
  \citenamefont {Lesgourgues}, \citenamefont {Audren},\ and\ \citenamefont
  {Ali-Haimoud}}]{Giesen:2012rp}%
  \BibitemOpen
  \bibfield  {author} {\bibinfo {author} {\bibfnamefont {G.}~\bibnamefont
  {Giesen}}, \bibinfo {author} {\bibfnamefont {J.}~\bibnamefont {Lesgourgues}},
  \bibinfo {author} {\bibfnamefont {B.}~\bibnamefont {Audren}}, \ and\ \bibinfo
  {author} {\bibfnamefont {Y.}~\bibnamefont {Ali-Haimoud}},\ }\href {\doibase
  10.1088/1475-7516/2012/12/008} {\bibfield  {journal} {\bibinfo  {journal}
  {JCAP}\ }\textbf {\bibinfo {volume} {1212}},\ \bibinfo {pages} {008}
  (\bibinfo {year} {2012})},\ \Eprint {http://arxiv.org/abs/1209.0247}
  {arXiv:1209.0247 [astro-ph.CO]} \BibitemShut {NoStop}%
\bibitem [{\citenamefont {Slatyer}(2013)}]{Slatyer:2012yq}%
  \BibitemOpen
  \bibfield  {author} {\bibinfo {author} {\bibfnamefont {T.~R.}\ \bibnamefont
  {Slatyer}},\ }\href {\doibase 10.1103/PhysRevD.87.123513} {\bibfield
  {journal} {\bibinfo  {journal} {Phys. Rev.}\ }\textbf {\bibinfo {volume}
  {D87}},\ \bibinfo {pages} {123513} (\bibinfo {year} {2013})},\ \Eprint
  {http://arxiv.org/abs/1211.0283} {arXiv:1211.0283 [astro-ph.CO]} \BibitemShut
  {NoStop}%
\bibitem [{\citenamefont {Lopez-Honorez}\ \emph {et~al.}(2013)\citenamefont
  {Lopez-Honorez}, \citenamefont {Mena}, \citenamefont {Palomares-Ruiz},\ and\
  \citenamefont {Vincent}}]{Lopez-Honorez:2013lcm}%
  \BibitemOpen
  \bibfield  {author} {\bibinfo {author} {\bibfnamefont {L.}~\bibnamefont
  {Lopez-Honorez}}, \bibinfo {author} {\bibfnamefont {O.}~\bibnamefont {Mena}},
  \bibinfo {author} {\bibfnamefont {S.}~\bibnamefont {Palomares-Ruiz}}, \ and\
  \bibinfo {author} {\bibfnamefont {A.~C.}\ \bibnamefont {Vincent}},\ }\href
  {\doibase 10.1088/1475-7516/2013/07/046} {\bibfield  {journal} {\bibinfo
  {journal} {JCAP}\ }\textbf {\bibinfo {volume} {1307}},\ \bibinfo {pages}
  {046} (\bibinfo {year} {2013})},\ \Eprint {http://arxiv.org/abs/1303.5094}
  {arXiv:1303.5094 [astro-ph.CO]} \BibitemShut {NoStop}%
\bibitem [{\citenamefont {Galli}\ \emph {et~al.}(2013)\citenamefont {Galli},
  \citenamefont {Slatyer}, \citenamefont {Valdes},\ and\ \citenamefont
  {Iocco}}]{Galli:2013dna}%
  \BibitemOpen
  \bibfield  {author} {\bibinfo {author} {\bibfnamefont {S.}~\bibnamefont
  {Galli}}, \bibinfo {author} {\bibfnamefont {T.~R.}\ \bibnamefont {Slatyer}},
  \bibinfo {author} {\bibfnamefont {M.}~\bibnamefont {Valdes}}, \ and\ \bibinfo
  {author} {\bibfnamefont {F.}~\bibnamefont {Iocco}},\ }\href {\doibase
  10.1103/PhysRevD.88.063502} {\bibfield  {journal} {\bibinfo  {journal} {Phys.
  Rev.}\ }\textbf {\bibinfo {volume} {D88}},\ \bibinfo {pages} {063502}
  (\bibinfo {year} {2013})},\ \Eprint {http://arxiv.org/abs/1306.0563}
  {arXiv:1306.0563 [astro-ph.CO]} \BibitemShut {NoStop}%
\bibitem [{\citenamefont {Diamanti}\ \emph {et~al.}(2014)\citenamefont
  {Diamanti}, \citenamefont {Lopez-Honorez}, \citenamefont {Mena},
  \citenamefont {Palomares-Ruiz},\ and\ \citenamefont
  {Vincent}}]{Diamanti:2013bia}%
  \BibitemOpen
  \bibfield  {author} {\bibinfo {author} {\bibfnamefont {R.}~\bibnamefont
  {Diamanti}}, \bibinfo {author} {\bibfnamefont {L.}~\bibnamefont
  {Lopez-Honorez}}, \bibinfo {author} {\bibfnamefont {O.}~\bibnamefont {Mena}},
  \bibinfo {author} {\bibfnamefont {S.}~\bibnamefont {Palomares-Ruiz}}, \ and\
  \bibinfo {author} {\bibfnamefont {A.~C.}\ \bibnamefont {Vincent}},\ }\href
  {\doibase 10.1088/1475-7516/2014/02/017} {\bibfield  {journal} {\bibinfo
  {journal} {JCAP}\ }\textbf {\bibinfo {volume} {1402}},\ \bibinfo {pages}
  {017} (\bibinfo {year} {2014})},\ \Eprint {http://arxiv.org/abs/1308.2578}
  {arXiv:1308.2578 [astro-ph.CO]} \BibitemShut {NoStop}%
\bibitem [{\citenamefont {Madhavacheril}\ \emph {et~al.}(2014)\citenamefont
  {Madhavacheril}, \citenamefont {Sehgal},\ and\ \citenamefont
  {Slatyer}}]{Madhavacheril:2013cna}%
  \BibitemOpen
  \bibfield  {author} {\bibinfo {author} {\bibfnamefont {M.~S.}\ \bibnamefont
  {Madhavacheril}}, \bibinfo {author} {\bibfnamefont {N.}~\bibnamefont
  {Sehgal}}, \ and\ \bibinfo {author} {\bibfnamefont {T.~R.}\ \bibnamefont
  {Slatyer}},\ }\href {\doibase 10.1103/PhysRevD.89.103508} {\bibfield
  {journal} {\bibinfo  {journal} {Phys. Rev.}\ }\textbf {\bibinfo {volume}
  {D89}},\ \bibinfo {pages} {103508} (\bibinfo {year} {2014})},\ \Eprint
  {http://arxiv.org/abs/1310.3815} {arXiv:1310.3815 [astro-ph.CO]} \BibitemShut
  {NoStop}%
\bibitem [{\citenamefont {Ade}\ \emph {et~al.}(2015)\citenamefont {Ade} \emph
  {et~al.}}]{Ade:2015xua}%
  \BibitemOpen
  \bibfield  {author} {\bibinfo {author} {\bibfnamefont {P.~A.~R.}\
  \bibnamefont {Ade}} \emph {et~al.} (\bibinfo {collaboration} {Planck
  Collaboration}),\ }\href@noop {} {\  (\bibinfo {year} {2015})},\ \Eprint
  {http://arxiv.org/abs/1502.01589} {arXiv:1502.01589 [astro-ph.CO]}
  \BibitemShut {NoStop}%
\bibitem [{\citenamefont {Slatyer}(2016)}]{Slatyer:2015jla}%
  \BibitemOpen
  \bibfield  {author} {\bibinfo {author} {\bibfnamefont {T.~R.}\ \bibnamefont
  {Slatyer}},\ }\href {\doibase 10.1103/PhysRevD.93.023527} {\bibfield
  {journal} {\bibinfo  {journal} {Phys. Rev.}\ }\textbf {\bibinfo {volume}
  {D93}},\ \bibinfo {pages} {023527} (\bibinfo {year} {2016})},\ \Eprint
  {http://arxiv.org/abs/1506.03811} {arXiv:1506.03811 [hep-ph]} \BibitemShut
  {NoStop}%
\bibitem [{\citenamefont {Kawasaki}\ \emph {et~al.}(2015)\citenamefont
  {Kawasaki}, \citenamefont {Nakayama},\ and\ \citenamefont
  {Sekiguchi}}]{Kawasaki:2015peu}%
  \BibitemOpen
  \bibfield  {author} {\bibinfo {author} {\bibfnamefont {M.}~\bibnamefont
  {Kawasaki}}, \bibinfo {author} {\bibfnamefont {K.}~\bibnamefont {Nakayama}},
  \ and\ \bibinfo {author} {\bibfnamefont {T.}~\bibnamefont {Sekiguchi}},\
  }\href@noop {} {\  (\bibinfo {year} {2015})},\ \Eprint
  {http://arxiv.org/abs/1512.08015} {arXiv:1512.08015 [astro-ph.CO]}
  \BibitemShut {NoStop}%
\bibitem [{\citenamefont {Valdes}\ \emph {et~al.}(2007)\citenamefont {Valdes},
  \citenamefont {Ferrara}, \citenamefont {Mapelli},\ and\ \citenamefont
  {Ripamonti}}]{Valdes:2007cu}%
  \BibitemOpen
  \bibfield  {author} {\bibinfo {author} {\bibfnamefont {M.}~\bibnamefont
  {Valdes}}, \bibinfo {author} {\bibfnamefont {A.}~\bibnamefont {Ferrara}},
  \bibinfo {author} {\bibfnamefont {M.}~\bibnamefont {Mapelli}}, \ and\
  \bibinfo {author} {\bibfnamefont {E.}~\bibnamefont {Ripamonti}},\ }\href
  {\doibase 10.1111/j.1365-2966.2007.11594.x} {\bibfield  {journal} {\bibinfo
  {journal} {Mon. Not. Roy. Astron. Soc.}\ }\textbf {\bibinfo {volume} {377}},\
  \bibinfo {pages} {245} (\bibinfo {year} {2007})},\ \Eprint
  {http://arxiv.org/abs/astro-ph/0701301} {arXiv:astro-ph/0701301 [astro-ph]}
  \BibitemShut {NoStop}%
\bibitem [{\citenamefont {Chuzhoy}(2008)}]{Chuzhoy:2007fg}%
  \BibitemOpen
  \bibfield  {author} {\bibinfo {author} {\bibfnamefont {L.}~\bibnamefont
  {Chuzhoy}},\ }\href {\doibase 10.1086/589504} {\bibfield  {journal} {\bibinfo
   {journal} {Astrophys. J.}\ }\textbf {\bibinfo {volume} {679}},\ \bibinfo
  {pages} {L65} (\bibinfo {year} {2008})},\ \Eprint
  {http://arxiv.org/abs/0710.1856} {arXiv:0710.1856 [astro-ph]} \BibitemShut
  {NoStop}%
\bibitem [{\citenamefont {Cumberbatch}\ \emph {et~al.}(2010)\citenamefont
  {Cumberbatch}, \citenamefont {Lattanzi},\ and\ \citenamefont
  {Silk}}]{Cumberbatch:2008rh}%
  \BibitemOpen
  \bibfield  {author} {\bibinfo {author} {\bibfnamefont {D.~T.}\ \bibnamefont
  {Cumberbatch}}, \bibinfo {author} {\bibfnamefont {M.}~\bibnamefont
  {Lattanzi}}, \ and\ \bibinfo {author} {\bibfnamefont {J.}~\bibnamefont
  {Silk}},\ }\href {\doibase 10.1103/PhysRevD.82.103508} {\bibfield  {journal}
  {\bibinfo  {journal} {Phys. Rev.}\ }\textbf {\bibinfo {volume} {D82}},\
  \bibinfo {pages} {103508} (\bibinfo {year} {2010})},\ \Eprint
  {http://arxiv.org/abs/0808.0881} {arXiv:0808.0881 [astro-ph]} \BibitemShut
  {NoStop}%
\bibitem [{\citenamefont {Yuan}\ \emph {et~al.}(2010)\citenamefont {Yuan},
  \citenamefont {Yue}, \citenamefont {Bi}, \citenamefont {Chen},\ and\
  \citenamefont {Zhang}}]{Yuan:2009xq}%
  \BibitemOpen
  \bibfield  {author} {\bibinfo {author} {\bibfnamefont {Q.}~\bibnamefont
  {Yuan}}, \bibinfo {author} {\bibfnamefont {B.}~\bibnamefont {Yue}}, \bibinfo
  {author} {\bibfnamefont {X.-J.}\ \bibnamefont {Bi}}, \bibinfo {author}
  {\bibfnamefont {X.}~\bibnamefont {Chen}}, \ and\ \bibinfo {author}
  {\bibfnamefont {X.}~\bibnamefont {Zhang}},\ }\href {\doibase
  10.1088/1475-7516/2010/10/023} {\bibfield  {journal} {\bibinfo  {journal}
  {JCAP}\ }\textbf {\bibinfo {volume} {1010}},\ \bibinfo {pages} {023}
  (\bibinfo {year} {2010})},\ \Eprint {http://arxiv.org/abs/0912.2504}
  {arXiv:0912.2504 [astro-ph.CO]} \BibitemShut {NoStop}%
\bibitem [{\citenamefont {Valdes}\ \emph {et~al.}(2013)\citenamefont {Valdes},
  \citenamefont {Evoli}, \citenamefont {Mesinger}, \citenamefont {Ferrara},\
  and\ \citenamefont {Yoshida}}]{Valdes:2012zv}%
  \BibitemOpen
  \bibfield  {author} {\bibinfo {author} {\bibfnamefont {M.}~\bibnamefont
  {Valdes}}, \bibinfo {author} {\bibfnamefont {C.}~\bibnamefont {Evoli}},
  \bibinfo {author} {\bibfnamefont {A.}~\bibnamefont {Mesinger}}, \bibinfo
  {author} {\bibfnamefont {A.}~\bibnamefont {Ferrara}}, \ and\ \bibinfo
  {author} {\bibfnamefont {N.}~\bibnamefont {Yoshida}},\ }\href {\doibase
  10.1093/mnras/sts458} {\bibfield  {journal} {\bibinfo  {journal} {Mon. Not.
  Roy. Astron. Soc.}\ }\textbf {\bibinfo {volume} {429}},\ \bibinfo {pages}
  {1705} (\bibinfo {year} {2013})},\ \Eprint {http://arxiv.org/abs/1209.2120}
  {arXiv:1209.2120 [astro-ph.CO]} \BibitemShut {NoStop}%
\bibitem [{\citenamefont {Evoli}\ \emph {et~al.}(2014)\citenamefont {Evoli},
  \citenamefont {Mesinger},\ and\ \citenamefont {Ferrara}}]{Evoli:2014pva}%
  \BibitemOpen
  \bibfield  {author} {\bibinfo {author} {\bibfnamefont {C.}~\bibnamefont
  {Evoli}}, \bibinfo {author} {\bibfnamefont {A.}~\bibnamefont {Mesinger}}, \
  and\ \bibinfo {author} {\bibfnamefont {A.}~\bibnamefont {Ferrara}},\ }\href
  {\doibase 10.1088/1475-7516/2014/11/024} {\bibfield  {journal} {\bibinfo
  {journal} {JCAP}\ }\textbf {\bibinfo {volume} {1411}},\ \bibinfo {pages}
  {024} (\bibinfo {year} {2014})},\ \Eprint {http://arxiv.org/abs/1408.1109}
  {arXiv:1408.1109 [astro-ph.HE]} \BibitemShut {NoStop}%
\bibitem [{\citenamefont {{Wouthuysen}}(1952)}]{Wouthuysen:1952}%
  \BibitemOpen
  \bibfield  {author} {\bibinfo {author} {\bibfnamefont {S.~A.}\ \bibnamefont
  {{Wouthuysen}}},\ }\href {\doibase 10.1086/106661} {\bibfield  {journal}
  {\bibinfo  {journal} {Astrophys. J.}\ }\textbf {\bibinfo {volume} {57}},\
  \bibinfo {pages} {31} (\bibinfo {year} {1952})}\BibitemShut {NoStop}%
\bibitem [{\citenamefont {{Field}}(1958)}]{Field:1958}%
  \BibitemOpen
  \bibfield  {author} {\bibinfo {author} {\bibfnamefont {G.~B.}\ \bibnamefont
  {{Field}}},\ }\href {\doibase 10.1109/JRPROC.1958.286741} {\bibfield
  {journal} {\bibinfo  {journal} {Proceedings of the IRE}\ }\textbf {\bibinfo
  {volume} {46}},\ \bibinfo {pages} {240} (\bibinfo {year} {1958})}\BibitemShut
  {NoStop}%
\bibitem [{\citenamefont {Hirata}(2006)}]{Hirata:2005mz}%
  \BibitemOpen
  \bibfield  {author} {\bibinfo {author} {\bibfnamefont {C.~M.}\ \bibnamefont
  {Hirata}},\ }\href {\doibase 10.1111/j.1365-2966.2005.09949.x} {\bibfield
  {journal} {\bibinfo  {journal} {Mon. Not. Roy. Astron. Soc.}\ }\textbf
  {\bibinfo {volume} {367}},\ \bibinfo {pages} {259} (\bibinfo {year}
  {2006})},\ \Eprint {http://arxiv.org/abs/astro-ph/0507102}
  {arXiv:astro-ph/0507102 [astro-ph]} \BibitemShut {NoStop}%
\bibitem [{\citenamefont {Madau}\ \emph {et~al.}(1997)\citenamefont {Madau},
  \citenamefont {Meiksin},\ and\ \citenamefont {Rees}}]{Madau:1996cs}%
  \BibitemOpen
  \bibfield  {author} {\bibinfo {author} {\bibfnamefont {P.}~\bibnamefont
  {Madau}}, \bibinfo {author} {\bibfnamefont {A.}~\bibnamefont {Meiksin}}, \
  and\ \bibinfo {author} {\bibfnamefont {M.~J.}\ \bibnamefont {Rees}},\ }\href
  {\doibase 10.1086/303549} {\bibfield  {journal} {\bibinfo  {journal}
  {Astrophys. J.}\ }\textbf {\bibinfo {volume} {475}},\ \bibinfo {pages} {429}
  (\bibinfo {year} {1997})},\ \Eprint {http://arxiv.org/abs/astro-ph/9608010}
  {arXiv:astro-ph/9608010 [astro-ph]} \BibitemShut {NoStop}%
\bibitem [{\citenamefont {Furlanetto}\ \emph
  {et~al.}(2006{\natexlab{b}})\citenamefont {Furlanetto}, \citenamefont {Oh},\
  and\ \citenamefont {Briggs}}]{Furlanetto:2006jb}%
  \BibitemOpen
  \bibfield  {author} {\bibinfo {author} {\bibfnamefont {S.}~\bibnamefont
  {Furlanetto}}, \bibinfo {author} {\bibfnamefont {S.~P.}\ \bibnamefont {Oh}},
  \ and\ \bibinfo {author} {\bibfnamefont {F.}~\bibnamefont {Briggs}},\ }\href
  {\doibase 10.1016/j.physrep.2006.08.002} {\bibfield  {journal} {\bibinfo
  {journal} {Phys. Rept.}\ }\textbf {\bibinfo {volume} {433}},\ \bibinfo
  {pages} {181} (\bibinfo {year} {2006}{\natexlab{b}})},\ \Eprint
  {http://arxiv.org/abs/astro-ph/0608032} {arXiv:astro-ph/0608032 [astro-ph]}
  \BibitemShut {NoStop}%
\bibitem [{\citenamefont {Pritchard}\ and\ \citenamefont
  {Loeb}(2012)}]{Pritchard:2011xb}%
  \BibitemOpen
  \bibfield  {author} {\bibinfo {author} {\bibfnamefont {J.~R.}\ \bibnamefont
  {Pritchard}}\ and\ \bibinfo {author} {\bibfnamefont {A.}~\bibnamefont
  {Loeb}},\ }\href {\doibase 10.1088/0034-4885/75/8/086901} {\bibfield
  {journal} {\bibinfo  {journal} {Rept.Prog.Phys.}\ }\textbf {\bibinfo {volume}
  {75}},\ \bibinfo {pages} {086901} (\bibinfo {year} {2012})},\ \Eprint
  {http://arxiv.org/abs/1109.6012} {arXiv:1109.6012 [astro-ph.CO]} \BibitemShut
  {NoStop}%
\bibitem [{\citenamefont {Furlanetto}(2015)}]{Furlanetto:2015apc}%
  \BibitemOpen
  \bibfield  {author} {\bibinfo {author} {\bibfnamefont {S.~R.}\ \bibnamefont
  {Furlanetto}},\ }\href@noop {} {\  (\bibinfo {year} {2015})},\ \Eprint
  {http://arxiv.org/abs/1511.01131} {arXiv:1511.01131 [astro-ph.CO]}
  \BibitemShut {NoStop}%
\bibitem [{\citenamefont {Press}\ and\ \citenamefont
  {Schechter}(1974)}]{Press:1973iz}%
  \BibitemOpen
  \bibfield  {author} {\bibinfo {author} {\bibfnamefont {W.~H.}\ \bibnamefont
  {Press}}\ and\ \bibinfo {author} {\bibfnamefont {P.}~\bibnamefont
  {Schechter}},\ }\href {\doibase 10.1086/152650} {\bibfield  {journal}
  {\bibinfo  {journal} {Astrophys. J.}\ }\textbf {\bibinfo {volume} {187}},\
  \bibinfo {pages} {425} (\bibinfo {year} {1974})}\BibitemShut {NoStop}%
\bibitem [{\citenamefont {Bond}\ \emph {et~al.}(1991)\citenamefont {Bond},
  \citenamefont {Cole}, \citenamefont {Efstathiou},\ and\ \citenamefont
  {Kaiser}}]{Bond:1990iw}%
  \BibitemOpen
  \bibfield  {author} {\bibinfo {author} {\bibfnamefont {J.~R.}\ \bibnamefont
  {Bond}}, \bibinfo {author} {\bibfnamefont {S.}~\bibnamefont {Cole}}, \bibinfo
  {author} {\bibfnamefont {G.}~\bibnamefont {Efstathiou}}, \ and\ \bibinfo
  {author} {\bibfnamefont {N.}~\bibnamefont {Kaiser}},\ }\href {\doibase
  10.1086/170520} {\bibfield  {journal} {\bibinfo  {journal} {Astrophys. J.}\
  }\textbf {\bibinfo {volume} {379}},\ \bibinfo {pages} {440} (\bibinfo {year}
  {1991})}\BibitemShut {NoStop}%
\bibitem [{\citenamefont {Watson}\ \emph {et~al.}(2013)\citenamefont {Watson}
  \emph {et~al.}}]{Watson:2012mt}%
  \BibitemOpen
  \bibfield  {author} {\bibinfo {author} {\bibfnamefont {W.~A.}\ \bibnamefont
  {Watson}} \emph {et~al.},\ }\href {\doibase 10.1093/mnras/stt791} {\bibfield
  {journal} {\bibinfo  {journal} {Mon. Not. Roy. Astron. Soc.}\ }\textbf
  {\bibinfo {volume} {433}},\ \bibinfo {pages} {1230} (\bibinfo {year}
  {2013})},\ \Eprint {http://arxiv.org/abs/1212.0095} {arXiv:1212.0095
  [astro-ph.CO]} \BibitemShut {NoStop}%
\bibitem [{\citenamefont {Sheth}\ and\ \citenamefont
  {Tormen}(1999)}]{Sheth:1999mn}%
  \BibitemOpen
  \bibfield  {author} {\bibinfo {author} {\bibfnamefont {R.~K.}\ \bibnamefont
  {Sheth}}\ and\ \bibinfo {author} {\bibfnamefont {G.}~\bibnamefont {Tormen}},\
  }\href {\doibase 10.1046/j.1365-8711.1999.02692.x} {\bibfield  {journal}
  {\bibinfo  {journal} {Mon. Not. Roy. Astron. Soc.}\ }\textbf {\bibinfo
  {volume} {308}},\ \bibinfo {pages} {119} (\bibinfo {year} {1999})},\ \Eprint
  {http://arxiv.org/abs/astro-ph/9901122} {arXiv:astro-ph/9901122 [astro-ph]}
  \BibitemShut {NoStop}%
\bibitem [{\citenamefont {Sheth}\ \emph {et~al.}(2001)\citenamefont {Sheth},
  \citenamefont {Mo},\ and\ \citenamefont {Tormen}}]{Sheth:1999su}%
  \BibitemOpen
  \bibfield  {author} {\bibinfo {author} {\bibfnamefont {R.~K.}\ \bibnamefont
  {Sheth}}, \bibinfo {author} {\bibfnamefont {H.~J.}\ \bibnamefont {Mo}}, \
  and\ \bibinfo {author} {\bibfnamefont {G.}~\bibnamefont {Tormen}},\ }\href
  {\doibase 10.1046/j.1365-8711.2001.04006.x} {\bibfield  {journal} {\bibinfo
  {journal} {Mon. Not. Roy. Astron. Soc.}\ }\textbf {\bibinfo {volume} {323}},\
  \bibinfo {pages} {1} (\bibinfo {year} {2001})},\ \Eprint
  {http://arxiv.org/abs/astro-ph/9907024} {arXiv:astro-ph/9907024 [astro-ph]}
  \BibitemShut {NoStop}%
\bibitem [{\citenamefont {Barkana}\ and\ \citenamefont
  {Loeb}(2005{\natexlab{c}})}]{Barkana:2004vb}%
  \BibitemOpen
  \bibfield  {author} {\bibinfo {author} {\bibfnamefont {R.}~\bibnamefont
  {Barkana}}\ and\ \bibinfo {author} {\bibfnamefont {A.}~\bibnamefont {Loeb}},\
  }\href {\doibase 10.1086/429954} {\bibfield  {journal} {\bibinfo  {journal}
  {Astrophys. J.}\ }\textbf {\bibinfo {volume} {626}},\ \bibinfo {pages} {1}
  (\bibinfo {year} {2005}{\natexlab{c}})},\ \Eprint
  {http://arxiv.org/abs/astro-ph/0410129} {arXiv:astro-ph/0410129 [astro-ph]}
  \BibitemShut {NoStop}%
\bibitem [{\citenamefont {Mesinger}\ \emph {et~al.}(2011)\citenamefont
  {Mesinger}, \citenamefont {Furlanetto},\ and\ \citenamefont
  {Cen}}]{Mesinger:2010ne}%
  \BibitemOpen
  \bibfield  {author} {\bibinfo {author} {\bibfnamefont {A.}~\bibnamefont
  {Mesinger}}, \bibinfo {author} {\bibfnamefont {S.}~\bibnamefont
  {Furlanetto}}, \ and\ \bibinfo {author} {\bibfnamefont {R.}~\bibnamefont
  {Cen}},\ }\href {\doibase 10.1111/j.1365-2966.2010.17731.x} {\bibfield
  {journal} {\bibinfo  {journal} {Mon. Not. Roy. Astron. Soc.}\ }\textbf
  {\bibinfo {volume} {411}},\ \bibinfo {pages} {955} (\bibinfo {year}
  {2011})},\ \Eprint {http://arxiv.org/abs/1003.3878} {arXiv:1003.3878
  [astro-ph.CO]} \BibitemShut {NoStop}%
\bibitem [{\citenamefont {Pober}\ \emph
  {et~al.}(2013{\natexlab{a}})\citenamefont {Pober} \emph
  {et~al.}}]{Pober:2013ig}%
  \BibitemOpen
  \bibfield  {author} {\bibinfo {author} {\bibfnamefont {J.~C.}\ \bibnamefont
  {Pober}} \emph {et~al.},\ }\href {\doibase 10.1088/2041-8205/768/2/L36}
  {\bibfield  {journal} {\bibinfo  {journal} {Astrophys. J.}\ }\textbf
  {\bibinfo {volume} {768}},\ \bibinfo {pages} {L36} (\bibinfo {year}
  {2013}{\natexlab{a}})},\ \Eprint {http://arxiv.org/abs/1301.7099}
  {arXiv:1301.7099 [astro-ph.CO]} \BibitemShut {NoStop}%
\bibitem [{\citenamefont {Pritchard}\ and\ \citenamefont
  {Furlanetto}(2007)}]{Pritchard:2006sq}%
  \BibitemOpen
  \bibfield  {author} {\bibinfo {author} {\bibfnamefont {J.~R.}\ \bibnamefont
  {Pritchard}}\ and\ \bibinfo {author} {\bibfnamefont {S.~R.}\ \bibnamefont
  {Furlanetto}},\ }\href {\doibase 10.1111/j.1365-2966.2007.11519.x} {\bibfield
   {journal} {\bibinfo  {journal} {Mon. Not. Roy. Astron. Soc.}\ }\textbf
  {\bibinfo {volume} {376}},\ \bibinfo {pages} {1680} (\bibinfo {year}
  {2007})},\ \Eprint {http://arxiv.org/abs/astro-ph/0607234}
  {arXiv:astro-ph/0607234 [astro-ph]} \BibitemShut {NoStop}%
\bibitem [{\citenamefont {Baek}\ \emph {et~al.}(2010)\citenamefont {Baek},
  \citenamefont {Semelin}, \citenamefont {Di~Matteo}, \citenamefont {Revaz},\
  and\ \citenamefont {Combes}}]{Baek:2010cm}%
  \BibitemOpen
  \bibfield  {author} {\bibinfo {author} {\bibfnamefont {S.}~\bibnamefont
  {Baek}}, \bibinfo {author} {\bibfnamefont {B.}~\bibnamefont {Semelin}},
  \bibinfo {author} {\bibfnamefont {P.}~\bibnamefont {Di~Matteo}}, \bibinfo
  {author} {\bibfnamefont {Y.}~\bibnamefont {Revaz}}, \ and\ \bibinfo {author}
  {\bibfnamefont {F.}~\bibnamefont {Combes}},\ }\href {\doibase
  10.1051/0004-6361/201014347} {\bibfield  {journal} {\bibinfo  {journal}
  {Astron. Astrophys.}\ }\textbf {\bibinfo {volume} {523}},\ \bibinfo {pages}
  {A4} (\bibinfo {year} {2010})},\ \Eprint {http://arxiv.org/abs/1003.0834}
  {arXiv:1003.0834 [astro-ph.CO]} \BibitemShut {NoStop}%
\bibitem [{\citenamefont {Venkatesan}\ \emph {et~al.}(2001)\citenamefont
  {Venkatesan}, \citenamefont {Giroux},\ and\ \citenamefont
  {Shull}}]{Venkatesan:2001cd}%
  \BibitemOpen
  \bibfield  {author} {\bibinfo {author} {\bibfnamefont {A.}~\bibnamefont
  {Venkatesan}}, \bibinfo {author} {\bibfnamefont {M.~L.}\ \bibnamefont
  {Giroux}}, \ and\ \bibinfo {author} {\bibfnamefont {J.~M.}\ \bibnamefont
  {Shull}},\ }\href {\doibase 10.1086/323691} {\bibfield  {journal} {\bibinfo
  {journal} {Astrophys. J.}\ }\textbf {\bibinfo {volume} {563}},\ \bibinfo
  {pages} {1} (\bibinfo {year} {2001})},\ \Eprint
  {http://arxiv.org/abs/astro-ph/0108168} {arXiv:astro-ph/0108168 [astro-ph]}
  \BibitemShut {NoStop}%
\bibitem [{\citenamefont {Chen}\ and\ \citenamefont
  {Miralda-Escude}(2004)}]{Chen:2003gc}%
  \BibitemOpen
  \bibfield  {author} {\bibinfo {author} {\bibfnamefont {X.-L.}\ \bibnamefont
  {Chen}}\ and\ \bibinfo {author} {\bibfnamefont {J.}~\bibnamefont
  {Miralda-Escude}},\ }\href {\doibase 10.1086/380829} {\bibfield  {journal}
  {\bibinfo  {journal} {Astrophys. J.}\ }\textbf {\bibinfo {volume} {602}},\
  \bibinfo {pages} {1} (\bibinfo {year} {2004})},\ \Eprint
  {http://arxiv.org/abs/astro-ph/0303395} {arXiv:astro-ph/0303395 [astro-ph]}
  \BibitemShut {NoStop}%
\bibitem [{\citenamefont {Zaroubi}\ \emph {et~al.}(2007)\citenamefont
  {Zaroubi}, \citenamefont {Thomas}, \citenamefont {Sugiyama},\ and\
  \citenamefont {Silk}}]{Zaroubi:2006fx}%
  \BibitemOpen
  \bibfield  {author} {\bibinfo {author} {\bibfnamefont {S.}~\bibnamefont
  {Zaroubi}}, \bibinfo {author} {\bibfnamefont {R.~M.}\ \bibnamefont {Thomas}},
  \bibinfo {author} {\bibfnamefont {N.}~\bibnamefont {Sugiyama}}, \ and\
  \bibinfo {author} {\bibfnamefont {J.}~\bibnamefont {Silk}},\ }\href {\doibase
  10.1111/j.1365-2966.2006.11361.x} {\bibfield  {journal} {\bibinfo  {journal}
  {Mon. Not. Roy. Astron. Soc.}\ }\textbf {\bibinfo {volume} {375}},\ \bibinfo
  {pages} {1269} (\bibinfo {year} {2007})},\ \Eprint
  {http://arxiv.org/abs/astro-ph/0609151} {arXiv:astro-ph/0609151 [astro-ph]}
  \BibitemShut {NoStop}%
\bibitem [{\citenamefont {Chen}\ and\ \citenamefont
  {Miralda-Escude}(2008)}]{Chen:2006zr}%
  \BibitemOpen
  \bibfield  {author} {\bibinfo {author} {\bibfnamefont {X.-L.}\ \bibnamefont
  {Chen}}\ and\ \bibinfo {author} {\bibfnamefont {J.}~\bibnamefont
  {Miralda-Escude}},\ }\href {\doibase 10.1086/528941} {\bibfield  {journal}
  {\bibinfo  {journal} {Astrophys. J.}\ }\textbf {\bibinfo {volume} {684}},\
  \bibinfo {pages} {18} (\bibinfo {year} {2008})},\ \Eprint
  {http://arxiv.org/abs/astro-ph/0605439} {arXiv:astro-ph/0605439 [astro-ph]}
  \BibitemShut {NoStop}%
\bibitem [{\citenamefont {Mesinger}\ \emph {et~al.}(2013)\citenamefont
  {Mesinger}, \citenamefont {Ferrara},\ and\ \citenamefont
  {Spiegel}}]{Mesinger:2012ys}%
  \BibitemOpen
  \bibfield  {author} {\bibinfo {author} {\bibfnamefont {A.}~\bibnamefont
  {Mesinger}}, \bibinfo {author} {\bibfnamefont {A.}~\bibnamefont {Ferrara}}, \
  and\ \bibinfo {author} {\bibfnamefont {D.~S.}\ \bibnamefont {Spiegel}},\
  }\href {\doibase 10.1093/mnras/stt198} {\bibfield  {journal} {\bibinfo
  {journal} {Mon. Not. Roy. Astron. Soc.}\ }\textbf {\bibinfo {volume} {431}},\
  \bibinfo {pages} {621} (\bibinfo {year} {2013})},\ \Eprint
  {http://arxiv.org/abs/1210.7319} {arXiv:1210.7319 [astro-ph.CO]}\BibitemShut
  {NoStop}%
\bibitem [{\citenamefont {Christian}\ and\ \citenamefont
  {Loeb}(2013)}]{Christian:2013gma}%
  \BibitemOpen
  \bibfield  {author} {\bibinfo {author} {\bibfnamefont {P.}~\bibnamefont
  {Christian}}\ and\ \bibinfo {author} {\bibfnamefont {A.}~\bibnamefont
  {Loeb}},\ }\href {\doibase 10.1088/1475-7516/2013/09/014} {\bibfield
  {journal} {\bibinfo  {journal} {JCAP}\ }\textbf {\bibinfo {volume} {1309}},\
  \bibinfo {pages} {014} (\bibinfo {year} {2013})},\ \Eprint
  {http://arxiv.org/abs/1305.5541} {arXiv:1305.5541 [astro-ph.CO]} \BibitemShut
  {NoStop}%
\bibitem [{\citenamefont {Mirocha}\ \emph {et~al.}(2015)\citenamefont
  {Mirocha}, \citenamefont {Harker},\ and\ \citenamefont
  {Burns}}]{Mirocha:2015jra}%
  \BibitemOpen
  \bibfield  {author} {\bibinfo {author} {\bibfnamefont {J.}~\bibnamefont
  {Mirocha}}, \bibinfo {author} {\bibfnamefont {G.~J.~A.}\ \bibnamefont
  {Harker}}, \ and\ \bibinfo {author} {\bibfnamefont {J.~O.}\ \bibnamefont
  {Burns}},\ }\href {\doibase 10.1088/0004-637X/813/1/11} {\bibfield  {journal}
  {\bibinfo  {journal} {Astrophys. J.}\ }\textbf {\bibinfo {volume} {813}},\
  \bibinfo {pages} {11} (\bibinfo {year} {2015})},\ \Eprint
  {http://arxiv.org/abs/1509.07868} {arXiv:1509.07868 [astro-ph.CO]}
  \BibitemShut {NoStop}%
\bibitem [{\citenamefont {Eke}\ \emph {et~al.}(2006)\citenamefont {Eke},
  \citenamefont {Baugh}, \citenamefont {Cole}, \citenamefont {Frenk},\ and\
  \citenamefont {Navarro}}]{Eke:2005za}%
  \BibitemOpen
  \bibfield  {author} {\bibinfo {author} {\bibfnamefont {V.~R.}\ \bibnamefont
  {Eke}}, \bibinfo {author} {\bibfnamefont {C.}~\bibnamefont {Baugh}}, \bibinfo
  {author} {\bibfnamefont {S.}~\bibnamefont {Cole}}, \bibinfo {author}
  {\bibfnamefont {C.}~\bibnamefont {Frenk}}, \ and\ \bibinfo {author}
  {\bibfnamefont {J.}~\bibnamefont {Navarro}},\ }\href {\doibase
  10.1111/j.1365-2966.2006.10568.x} {\bibfield  {journal} {\bibinfo  {journal}
  {Mon.Not.Roy.Astron.Soc.}\ }\textbf {\bibinfo {volume} {370}},\ \bibinfo
  {pages} {1147} (\bibinfo {year} {2006})},\ \Eprint
  {http://arxiv.org/abs/astro-ph/0510643} {arXiv:astro-ph/0510643 [astro-ph]}
  \BibitemShut {NoStop}%
\bibitem [{\citenamefont {Rines}\ \emph {et~al.}(2008)\citenamefont {Rines},
  \citenamefont {Diaferio},\ and\ \citenamefont {Natarajan}}]{Rines:2008tr}%
  \BibitemOpen
  \bibfield  {author} {\bibinfo {author} {\bibfnamefont {K.}~\bibnamefont
  {Rines}}, \bibinfo {author} {\bibfnamefont {A.}~\bibnamefont {Diaferio}}, \
  and\ \bibinfo {author} {\bibfnamefont {P.}~\bibnamefont {Natarajan}},\ }\href
  {\doibase 10.1086/588783} {\bibfield  {journal} {\bibinfo  {journal}
  {Astrophys.J.}\ }\textbf {\bibinfo {volume} {679}},\ \bibinfo {pages} {L1}
  (\bibinfo {year} {2008})},\ \Eprint {http://arxiv.org/abs/0803.1843}
  {arXiv:0803.1843 [astro-ph]} \BibitemShut {NoStop}%
\bibitem [{\citenamefont {Vikhlinin}\ \emph {et~al.}(2009)\citenamefont
  {Vikhlinin} \emph {et~al.}}]{Vikhlinin:2008ym}%
  \BibitemOpen
  \bibfield  {author} {\bibinfo {author} {\bibfnamefont {A.}~\bibnamefont
  {Vikhlinin}} \emph {et~al.},\ }\href {\doibase 10.1088/0004-637X/692/2/1060}
  {\bibfield  {journal} {\bibinfo  {journal} {Astrophys.J.}\ }\textbf {\bibinfo
  {volume} {692}},\ \bibinfo {pages} {1060} (\bibinfo {year} {2009})},\ \Eprint
  {http://arxiv.org/abs/0812.2720} {arXiv:0812.2720 [astro-ph]} \BibitemShut
  {NoStop}%
\bibitem [{\citenamefont {Rozo}\ \emph {et~al.}(2010)\citenamefont {Rozo} \emph
  {et~al.}}]{Rozo:2009jj}%
  \BibitemOpen
  \bibfield  {author} {\bibinfo {author} {\bibfnamefont {E.}~\bibnamefont
  {Rozo}} \emph {et~al.} (\bibinfo {collaboration} {SDSS Collaboration}),\
  }\href {\doibase 10.1088/0004-637X/708/1/645} {\bibfield  {journal} {\bibinfo
   {journal} {Astrophys.J.}\ }\textbf {\bibinfo {volume} {708}},\ \bibinfo
  {pages} {645} (\bibinfo {year} {2010})},\ \Eprint
  {http://arxiv.org/abs/0902.3702} {arXiv:0902.3702 [astro-ph.CO]} \BibitemShut
  {NoStop}%
\bibitem [{\citenamefont {Barkana}\ and\ \citenamefont
  {Loeb}(2004)}]{Barkana:2003qk}%
  \BibitemOpen
  \bibfield  {author} {\bibinfo {author} {\bibfnamefont {R.}~\bibnamefont
  {Barkana}}\ and\ \bibinfo {author} {\bibfnamefont {A.}~\bibnamefont {Loeb}},\
  }\href {\doibase 10.1086/421079} {\bibfield  {journal} {\bibinfo  {journal}
  {Astrophys. J.}\ }\textbf {\bibinfo {volume} {609}},\ \bibinfo {pages} {474}
  (\bibinfo {year} {2004})},\ \Eprint {http://arxiv.org/abs/astro-ph/0310338}
  {arXiv:astro-ph/0310338 [astro-ph]} \BibitemShut {NoStop}%
\bibitem [{\citenamefont {Barkana}\ and\ \citenamefont
  {Loeb}(2008)}]{Barkana:2007xj}%
  \BibitemOpen
  \bibfield  {author} {\bibinfo {author} {\bibfnamefont {R.}~\bibnamefont
  {Barkana}}\ and\ \bibinfo {author} {\bibfnamefont {A.}~\bibnamefont {Loeb}},\
  }\href {\doibase 10.1111/j.1365-2966.2007.12729.x} {\bibfield  {journal}
  {\bibinfo  {journal} {Mon. Not. Roy. Astron. Soc.}\ }\textbf {\bibinfo
  {volume} {384}},\ \bibinfo {pages} {1069} (\bibinfo {year} {2008})},\ \Eprint
  {http://arxiv.org/abs/0705.3246} {arXiv:0705.3246 [astro-ph]} \BibitemShut
  {NoStop}%
\bibitem [{\citenamefont {Iliev}\ \emph {et~al.}(2006)\citenamefont {Iliev}
  \emph {et~al.}}]{Iliev:2005sz}%
  \BibitemOpen
  \bibfield  {author} {\bibinfo {author} {\bibfnamefont {I.~T.}\ \bibnamefont
  {Iliev}} \emph {et~al.},\ }\href {\doibase 10.1111/j.1365-2966.2006.10502.x}
  {\bibfield  {journal} {\bibinfo  {journal} {Mon. Not. Roy. Astron. Soc.}\
  }\textbf {\bibinfo {volume} {369}},\ \bibinfo {pages} {1625} (\bibinfo {year}
  {2006})},\ \Eprint {http://arxiv.org/abs/astro-ph/0512187}
  {arXiv:astro-ph/0512187 [astro-ph]} \BibitemShut {NoStop}%
\bibitem [{\citenamefont {Reed}\ \emph {et~al.}(2007)\citenamefont {Reed},
  \citenamefont {Bower}, \citenamefont {Frenk}, \citenamefont {Jenkins},\ and\
  \citenamefont {Theuns}}]{Reed:2006rw}%
  \BibitemOpen
  \bibfield  {author} {\bibinfo {author} {\bibfnamefont {D.}~\bibnamefont
  {Reed}}, \bibinfo {author} {\bibfnamefont {R.}~\bibnamefont {Bower}},
  \bibinfo {author} {\bibfnamefont {C.}~\bibnamefont {Frenk}}, \bibinfo
  {author} {\bibfnamefont {A.}~\bibnamefont {Jenkins}}, \ and\ \bibinfo
  {author} {\bibfnamefont {T.}~\bibnamefont {Theuns}},\ }\href {\doibase
  10.1111/j.1365-2966.2006.11204.x} {\bibfield  {journal} {\bibinfo  {journal}
  {Mon. Not. Roy. Astron. Soc.}\ }\textbf {\bibinfo {volume} {374}},\ \bibinfo
  {pages} {2} (\bibinfo {year} {2007})},\ \Eprint
  {http://arxiv.org/abs/astro-ph/0607150} {arXiv:astro-ph/0607150 [astro-ph]}
  \BibitemShut {NoStop}%
\bibitem [{\citenamefont {Lukic}\ \emph {et~al.}(2007)\citenamefont {Lukic},
  \citenamefont {Heitmann}, \citenamefont {Habib}, \citenamefont {Bashinsky},\
  and\ \citenamefont {Ricker}}]{Lukic:2007fc}%
  \BibitemOpen
  \bibfield  {author} {\bibinfo {author} {\bibfnamefont {Z.}~\bibnamefont
  {Lukic}}, \bibinfo {author} {\bibfnamefont {K.}~\bibnamefont {Heitmann}},
  \bibinfo {author} {\bibfnamefont {S.}~\bibnamefont {Habib}}, \bibinfo
  {author} {\bibfnamefont {S.}~\bibnamefont {Bashinsky}}, \ and\ \bibinfo
  {author} {\bibfnamefont {P.~M.}\ \bibnamefont {Ricker}},\ }\href {\doibase
  10.1086/523083} {\bibfield  {journal} {\bibinfo  {journal} {Astrophys.J.}\
  }\textbf {\bibinfo {volume} {671}},\ \bibinfo {pages} {1160} (\bibinfo {year}
  {2007})},\ \Eprint {http://arxiv.org/abs/astro-ph/0702360}
  {arXiv:astro-ph/0702360 [astro-ph]} \BibitemShut {NoStop}%
\bibitem [{\citenamefont {Eisenstein}\ and\ \citenamefont
  {Hu}(1997)}]{Eisenstein:1997jh}%
  \BibitemOpen
  \bibfield  {author} {\bibinfo {author} {\bibfnamefont {D.~J.}\ \bibnamefont
  {Eisenstein}}\ and\ \bibinfo {author} {\bibfnamefont {W.}~\bibnamefont
  {Hu}},\ }\href {\doibase 10.1086/306640} {\bibfield  {journal} {\bibinfo
  {journal} {Astrophys.J.}\ }\textbf {\bibinfo {volume} {511}},\ \bibinfo
  {pages} {5} (\bibinfo {year} {1997})},\ \Eprint
  {http://arxiv.org/abs/astro-ph/9710252} {arXiv:astro-ph/9710252 [astro-ph]}
  \BibitemShut {NoStop}%
\bibitem [{\citenamefont {Molin\'e}\ \emph {et~al.}(2015)\citenamefont
  {Molin\'e}, \citenamefont {Ibarra},\ and\ \citenamefont
  {Palomares-Ruiz}}]{Moline:2014xua}%
  \BibitemOpen
  \bibfield  {author} {\bibinfo {author} {\bibfnamefont {A.}~\bibnamefont
  {Molin\'e}}, \bibinfo {author} {\bibfnamefont {A.}~\bibnamefont {Ibarra}}, \
  and\ \bibinfo {author} {\bibfnamefont {S.}~\bibnamefont {Palomares-Ruiz}},\
  }\href {\doibase 10.1088/1475-7516/2015/06/005} {\bibfield  {journal}
  {\bibinfo  {journal} {JCAP}\ }\textbf {\bibinfo {volume} {1506}},\ \bibinfo
  {pages} {005} (\bibinfo {year} {2015})},\ \Eprint
  {http://arxiv.org/abs/1412.4308} {arXiv:1412.4308 [astro-ph.CO]} \BibitemShut
  {NoStop}%
\bibitem [{\citenamefont {Barkana}\ and\ \citenamefont
  {Loeb}(2001)}]{Barkana:2000fd}%
  \BibitemOpen
  \bibfield  {author} {\bibinfo {author} {\bibfnamefont {R.}~\bibnamefont
  {Barkana}}\ and\ \bibinfo {author} {\bibfnamefont {A.}~\bibnamefont {Loeb}},\
  }\href {\doibase 10.1016/S0370-1573(01)00019-9} {\bibfield  {journal}
  {\bibinfo  {journal} {Phys. Rept.}\ }\textbf {\bibinfo {volume} {349}},\
  \bibinfo {pages} {125} (\bibinfo {year} {2001})},\ \Eprint
  {http://arxiv.org/abs/astro-ph/0010468} {arXiv:astro-ph/0010468 [astro-ph]}
  \BibitemShut {NoStop}%
\bibitem [{\citenamefont {{Evrard}}(1990)}]{Evrard:1990}%
  \BibitemOpen
  \bibfield  {author} {\bibinfo {author} {\bibfnamefont {A.~E.}\ \bibnamefont
  {{Evrard}}},\ }\href {\doibase 10.1086/169350} {\bibfield  {journal}
  {\bibinfo  {journal} {Astrophys. J.}\ }\textbf {\bibinfo {volume} {363}},\
  \bibinfo {pages} {349} (\bibinfo {year} {1990})}\BibitemShut {NoStop}%
\bibitem [{\citenamefont {{Blanchard}}\ \emph {et~al.}(1992)\citenamefont
  {{Blanchard}}, \citenamefont {{Valls-Gabaud}},\ and\ \citenamefont
  {{Mamon}}}]{Blanchard:1992}%
  \BibitemOpen
  \bibfield  {author} {\bibinfo {author} {\bibfnamefont {A.}~\bibnamefont
  {{Blanchard}}}, \bibinfo {author} {\bibfnamefont {D.}~\bibnamefont
  {{Valls-Gabaud}}}, \ and\ \bibinfo {author} {\bibfnamefont {G.~A.}\
  \bibnamefont {{Mamon}}},\ }\href@noop {} {\bibfield  {journal} {\bibinfo
  {journal} {Astron. Astrophys.}\ }\textbf {\bibinfo {volume} {264}},\ \bibinfo
  {pages} {365} (\bibinfo {year} {1992})}\BibitemShut {NoStop}%
\bibitem [{\citenamefont {Tegmark}\ \emph {et~al.}(1997)\citenamefont {Tegmark}
  \emph {et~al.}}]{Tegmark:1996yt}%
  \BibitemOpen
  \bibfield  {author} {\bibinfo {author} {\bibfnamefont {M.}~\bibnamefont
  {Tegmark}} \emph {et~al.},\ }\href {\doibase 10.1086/303434} {\bibfield
  {journal} {\bibinfo  {journal} {Astrophys. J.}\ }\textbf {\bibinfo {volume}
  {474}},\ \bibinfo {pages} {1} (\bibinfo {year} {1997})},\ \Eprint
  {http://arxiv.org/abs/astro-ph/9603007} {arXiv:astro-ph/9603007 [astro-ph]}
  \BibitemShut {NoStop}%
\bibitem [{\citenamefont {Haiman}\ \emph {et~al.}(2000)\citenamefont {Haiman},
  \citenamefont {Abel},\ and\ \citenamefont {Rees}}]{Haiman:1999mn}%
  \BibitemOpen
  \bibfield  {author} {\bibinfo {author} {\bibfnamefont {Z.}~\bibnamefont
  {Haiman}}, \bibinfo {author} {\bibfnamefont {T.}~\bibnamefont {Abel}}, \ and\
  \bibinfo {author} {\bibfnamefont {M.~J.}\ \bibnamefont {Rees}},\ }\href
  {\doibase 10.1086/308723} {\bibfield  {journal} {\bibinfo  {journal}
  {Astrophys. J.}\ }\textbf {\bibinfo {volume} {534}},\ \bibinfo {pages} {11}
  (\bibinfo {year} {2000})},\ \Eprint {http://arxiv.org/abs/astro-ph/9903336}
  {arXiv:astro-ph/9903336 [astro-ph]} \BibitemShut {NoStop}%
\bibitem [{\citenamefont {Ciardi}\ \emph {et~al.}(2000)\citenamefont {Ciardi},
  \citenamefont {Ferrara}, \citenamefont {Governato},\ and\ \citenamefont
  {Jenkins}}]{Ciardi:1999mx}%
  \BibitemOpen
  \bibfield  {author} {\bibinfo {author} {\bibfnamefont {B.}~\bibnamefont
  {Ciardi}}, \bibinfo {author} {\bibfnamefont {A.}~\bibnamefont {Ferrara}},
  \bibinfo {author} {\bibfnamefont {F.}~\bibnamefont {Governato}}, \ and\
  \bibinfo {author} {\bibfnamefont {A.}~\bibnamefont {Jenkins}},\ }\href
  {\doibase 10.1046/j.1365-8711.2000.03365.x} {\bibfield  {journal} {\bibinfo
  {journal} {Mon. Not. Roy. Astron. Soc.}\ }\textbf {\bibinfo {volume} {314}},\
  \bibinfo {pages} {611} (\bibinfo {year} {2000})},\ \Eprint
  {http://arxiv.org/abs/astro-ph/9907189} {arXiv:astro-ph/9907189 [astro-ph]}
  \BibitemShut {NoStop}%
\bibitem [{\citenamefont {Greig}\ and\ \citenamefont
  {Mesinger}(2015)}]{Greig:2015qca}%
  \BibitemOpen
  \bibfield  {author} {\bibinfo {author} {\bibfnamefont {B.}~\bibnamefont
  {Greig}}\ and\ \bibinfo {author} {\bibfnamefont {A.}~\bibnamefont
  {Mesinger}},\ }\href {\doibase 10.1093/mnras/stv571} {\bibfield  {journal}
  {\bibinfo  {journal} {Mon. Not. Roy. Astron. Soc.}\ }\textbf {\bibinfo
  {volume} {449}},\ \bibinfo {pages} {4246} (\bibinfo {year} {2015})},\ \Eprint
  {http://arxiv.org/abs/1501.06576} {arXiv:1501.06576 [astro-ph.CO]}
  \BibitemShut {NoStop}%
\bibitem [{\citenamefont {Pacucci}\ \emph {et~al.}(2014)\citenamefont
  {Pacucci}, \citenamefont {Mesinger}, \citenamefont {Mineo},\ and\
  \citenamefont {Ferrara}}]{Pacucci:2014wwa}%
  \BibitemOpen
  \bibfield  {author} {\bibinfo {author} {\bibfnamefont {F.}~\bibnamefont
  {Pacucci}}, \bibinfo {author} {\bibfnamefont {A.}~\bibnamefont {Mesinger}},
  \bibinfo {author} {\bibfnamefont {S.}~\bibnamefont {Mineo}}, \ and\ \bibinfo
  {author} {\bibfnamefont {A.}~\bibnamefont {Ferrara}},\ }\href {\doibase
  10.1093/mnras/stu1240} {\bibfield  {journal} {\bibinfo  {journal} {Mon. Not.
  Roy. Astron. Soc.}\ }\textbf {\bibinfo {volume} {443}},\ \bibinfo {pages}
  {678} (\bibinfo {year} {2014})},\ \Eprint {http://arxiv.org/abs/1403.6125}
  {arXiv:1403.6125 [astro-ph.CO]} \BibitemShut {NoStop}%
\bibitem [{\citenamefont {Ciardi}\ and\ \citenamefont
  {Madau}(2003)}]{Ciardi:2003hg}%
  \BibitemOpen
  \bibfield  {author} {\bibinfo {author} {\bibfnamefont {B.}~\bibnamefont
  {Ciardi}}\ and\ \bibinfo {author} {\bibfnamefont {P.}~\bibnamefont {Madau}},\
  }\href {\doibase 10.1086/377634} {\bibfield  {journal} {\bibinfo  {journal}
  {Astrophys. J.}\ }\textbf {\bibinfo {volume} {596}},\ \bibinfo {pages} {1}
  (\bibinfo {year} {2003})},\ \Eprint {http://arxiv.org/abs/astro-ph/0303249}
  {arXiv:astro-ph/0303249 [astro-ph]} \BibitemShut {NoStop}%
\bibitem [{\citenamefont {Furlanetto}(2006)}]{Furlanetto:2006tf}%
  \BibitemOpen
  \bibfield  {author} {\bibinfo {author} {\bibfnamefont {S.}~\bibnamefont
  {Furlanetto}},\ }\href {\doibase 10.1111/j.1365-2966.2006.10725.x} {\bibfield
   {journal} {\bibinfo  {journal} {Mon. Not. Roy. Astron. Soc.}\ }\textbf
  {\bibinfo {volume} {371}},\ \bibinfo {pages} {867} (\bibinfo {year}
  {2006})},\ \Eprint {http://arxiv.org/abs/astro-ph/0604040}
  {arXiv:astro-ph/0604040 [astro-ph]} \BibitemShut {NoStop}%
\bibitem [{\citenamefont {Springel}\ \emph {et~al.}(2005)\citenamefont
  {Springel} \emph {et~al.}}]{Springel:2005nw}%
  \BibitemOpen
  \bibfield  {author} {\bibinfo {author} {\bibfnamefont {V.}~\bibnamefont
  {Springel}} \emph {et~al.},\ }\href {\doibase 10.1038/nature03597} {\bibfield
   {journal} {\bibinfo  {journal} {Nature}\ }\textbf {\bibinfo {volume}
  {435}},\ \bibinfo {pages} {629} (\bibinfo {year} {2005})},\ \Eprint
  {http://arxiv.org/abs/astro-ph/0504097} {arXiv:astro-ph/0504097 [astro-ph]}
  \BibitemShut {NoStop}%
\bibitem [{\citenamefont {Heitmann}\ \emph {et~al.}(2006)\citenamefont
  {Heitmann}, \citenamefont {Higdon}, \citenamefont {Nakhleh},\ and\
  \citenamefont {Habib}}]{Heitmann:2006hr}%
  \BibitemOpen
  \bibfield  {author} {\bibinfo {author} {\bibfnamefont {K.}~\bibnamefont
  {Heitmann}}, \bibinfo {author} {\bibfnamefont {D.}~\bibnamefont {Higdon}},
  \bibinfo {author} {\bibfnamefont {C.}~\bibnamefont {Nakhleh}}, \ and\
  \bibinfo {author} {\bibfnamefont {S.}~\bibnamefont {Habib}},\ }\href
  {\doibase 10.1086/506448} {\bibfield  {journal} {\bibinfo  {journal}
  {Astrophys.J.}\ }\textbf {\bibinfo {volume} {646}},\ \bibinfo {pages} {L1}
  (\bibinfo {year} {2006})},\ \Eprint {http://arxiv.org/abs/astro-ph/0606154}
  {arXiv:astro-ph/0606154 [astro-ph]} \BibitemShut {NoStop}%
\bibitem [{\citenamefont {Shull}\ and\ \citenamefont {van
  Steenberg}(1985)}]{Shull:1985}%
  \BibitemOpen
  \bibfield  {author} {\bibinfo {author} {\bibfnamefont {J.~M.}\ \bibnamefont
  {Shull}}\ and\ \bibinfo {author} {\bibfnamefont {M.~E.}\ \bibnamefont {van
  Steenberg}},\ }\href {\doibase 10.1086/163605} {\bibfield  {journal}
  {\bibinfo  {journal} {Astrophys. J.}\ }\textbf {\bibinfo {volume} {298}},\
  \bibinfo {pages} {268} (\bibinfo {year} {1985})}\BibitemShut {NoStop}%
\bibitem [{\citenamefont {Slatyer}(2015)}]{Slatyer:2015kla}%
  \BibitemOpen
  \bibfield  {author} {\bibinfo {author} {\bibfnamefont {T.~R.}\ \bibnamefont
  {Slatyer}},\ }\href@noop {} {\  (\bibinfo {year} {2015})},\ \Eprint
  {http://arxiv.org/abs/1506.03812} {arXiv:1506.03812 [astro-ph.CO]}
  \BibitemShut {NoStop}%
\bibitem [{\citenamefont {Schmid}\ \emph {et~al.}(1999)\citenamefont {Schmid},
  \citenamefont {Schwarz},\ and\ \citenamefont {Widerin}}]{Schmid:1998mx}%
  \BibitemOpen
  \bibfield  {author} {\bibinfo {author} {\bibfnamefont {C.}~\bibnamefont
  {Schmid}}, \bibinfo {author} {\bibfnamefont {D.~J.}\ \bibnamefont {Schwarz}},
  \ and\ \bibinfo {author} {\bibfnamefont {P.}~\bibnamefont {Widerin}},\ }\href
  {\doibase 10.1103/PhysRevD.59.043517} {\bibfield  {journal} {\bibinfo
  {journal} {Phys.Rev.}\ }\textbf {\bibinfo {volume} {D59}},\ \bibinfo {pages}
  {043517} (\bibinfo {year} {1999})},\ \Eprint
  {http://arxiv.org/abs/astro-ph/9807257} {arXiv:astro-ph/9807257 [astro-ph]}
  \BibitemShut {NoStop}%
\bibitem [{\citenamefont {Zybin}\ \emph {et~al.}(1999)\citenamefont {Zybin},
  \citenamefont {Vysotsky},\ and\ \citenamefont {Gurevich}}]{Zybin:1999ic}%
  \BibitemOpen
  \bibfield  {author} {\bibinfo {author} {\bibfnamefont {K.}~\bibnamefont
  {Zybin}}, \bibinfo {author} {\bibfnamefont {M.}~\bibnamefont {Vysotsky}}, \
  and\ \bibinfo {author} {\bibfnamefont {A.}~\bibnamefont {Gurevich}},\ }\href
  {\doibase 10.1016/S0375-9601(99)00434-X} {\bibfield  {journal} {\bibinfo
  {journal} {Phys.Lett.}\ }\textbf {\bibinfo {volume} {A260}},\ \bibinfo
  {pages} {262} (\bibinfo {year} {1999})}\BibitemShut {NoStop}%
\bibitem [{\citenamefont {Boehm}\ \emph {et~al.}(2001)\citenamefont {Boehm},
  \citenamefont {Fayet},\ and\ \citenamefont {Schaeffer}}]{Boehm:2000gq}%
  \BibitemOpen
  \bibfield  {author} {\bibinfo {author} {\bibfnamefont {C.}~\bibnamefont
  {Boehm}}, \bibinfo {author} {\bibfnamefont {P.}~\bibnamefont {Fayet}}, \ and\
  \bibinfo {author} {\bibfnamefont {R.}~\bibnamefont {Schaeffer}},\ }\href
  {\doibase 10.1016/S0370-2693(01)01060-7} {\bibfield  {journal} {\bibinfo
  {journal} {Phys. Lett.}\ }\textbf {\bibinfo {volume} {B518}},\ \bibinfo
  {pages} {8} (\bibinfo {year} {2001})},\ \Eprint
  {http://arxiv.org/abs/astro-ph/0012504} {arXiv:astro-ph/0012504 [astro-ph]}
  \BibitemShut {NoStop}%
\bibitem [{\citenamefont {Chen}\ \emph {et~al.}(2001)\citenamefont {Chen},
  \citenamefont {Kamionkowski},\ and\ \citenamefont {Zhang}}]{Chen:2001jz}%
  \BibitemOpen
  \bibfield  {author} {\bibinfo {author} {\bibfnamefont {X.-l.}\ \bibnamefont
  {Chen}}, \bibinfo {author} {\bibfnamefont {M.}~\bibnamefont {Kamionkowski}},
  \ and\ \bibinfo {author} {\bibfnamefont {X.-m.}\ \bibnamefont {Zhang}},\
  }\href {\doibase 10.1103/PhysRevD.64.021302} {\bibfield  {journal} {\bibinfo
  {journal} {Phys. Rev.}\ }\textbf {\bibinfo {volume} {D64}},\ \bibinfo {pages}
  {021302} (\bibinfo {year} {2001})},\ \Eprint
  {http://arxiv.org/abs/astro-ph/0103452} {arXiv:astro-ph/0103452 [astro-ph]}
  \BibitemShut {NoStop}%
\bibitem [{\citenamefont {Hofmann}\ \emph {et~al.}(2001)\citenamefont
  {Hofmann}, \citenamefont {Schwarz},\ and\ \citenamefont
  {Stoecker}}]{Hofmann:2001bi}%
  \BibitemOpen
  \bibfield  {author} {\bibinfo {author} {\bibfnamefont {S.}~\bibnamefont
  {Hofmann}}, \bibinfo {author} {\bibfnamefont {D.~J.}\ \bibnamefont
  {Schwarz}}, \ and\ \bibinfo {author} {\bibfnamefont {H.}~\bibnamefont
  {Stoecker}},\ }\href {\doibase 10.1103/PhysRevD.64.083507} {\bibfield
  {journal} {\bibinfo  {journal} {Phys. Rev.}\ }\textbf {\bibinfo {volume}
  {D64}},\ \bibinfo {pages} {083507} (\bibinfo {year} {2001})},\ \Eprint
  {http://arxiv.org/abs/astro-ph/0104173} {arXiv:astro-ph/0104173 [astro-ph]}
  \BibitemShut {NoStop}%
\bibitem [{\citenamefont {Berezinsky}\ \emph {et~al.}(2003)\citenamefont
  {Berezinsky}, \citenamefont {Dokuchaev},\ and\ \citenamefont
  {Eroshenko}}]{Berezinsky:2003vn}%
  \BibitemOpen
  \bibfield  {author} {\bibinfo {author} {\bibfnamefont {V.}~\bibnamefont
  {Berezinsky}}, \bibinfo {author} {\bibfnamefont {V.}~\bibnamefont
  {Dokuchaev}}, \ and\ \bibinfo {author} {\bibfnamefont {Y.}~\bibnamefont
  {Eroshenko}},\ }\href {\doibase 10.1103/PhysRevD.68.103003} {\bibfield
  {journal} {\bibinfo  {journal} {Phys.Rev.}\ }\textbf {\bibinfo {volume}
  {D68}},\ \bibinfo {pages} {103003} (\bibinfo {year} {2003})},\ \Eprint
  {http://arxiv.org/abs/astro-ph/0301551} {arXiv:astro-ph/0301551 [astro-ph]}
  \BibitemShut {NoStop}%
\bibitem [{\citenamefont {Boehm}\ \emph {et~al.}(2005)\citenamefont {Boehm},
  \citenamefont {Mathis}, \citenamefont {Devriendt},\ and\ \citenamefont
  {Silk}}]{Boehm:2003xr}%
  \BibitemOpen
  \bibfield  {author} {\bibinfo {author} {\bibfnamefont {C.}~\bibnamefont
  {Boehm}}, \bibinfo {author} {\bibfnamefont {H.}~\bibnamefont {Mathis}},
  \bibinfo {author} {\bibfnamefont {J.}~\bibnamefont {Devriendt}}, \ and\
  \bibinfo {author} {\bibfnamefont {J.}~\bibnamefont {Silk}},\ }\href {\doibase
  10.1111/j.1365-2966.2005.09032.x} {\bibfield  {journal} {\bibinfo  {journal}
  {Mon. Not. Roy. Astron. Soc.}\ }\textbf {\bibinfo {volume} {360}},\ \bibinfo
  {pages} {282} (\bibinfo {year} {2005})},\ \Eprint
  {http://arxiv.org/abs/astro-ph/0309652} {arXiv:astro-ph/0309652 [astro-ph]}
  \BibitemShut {NoStop}%
\bibitem [{\citenamefont {Boehm}\ and\ \citenamefont
  {Schaeffer}(2005)}]{Boehm:2004th}%
  \BibitemOpen
  \bibfield  {author} {\bibinfo {author} {\bibfnamefont {C.}~\bibnamefont
  {Boehm}}\ and\ \bibinfo {author} {\bibfnamefont {R.}~\bibnamefont
  {Schaeffer}},\ }\href {\doibase 10.1051/0004-6361:20042238} {\bibfield
  {journal} {\bibinfo  {journal} {Astron. Astrophys.}\ }\textbf {\bibinfo
  {volume} {438}},\ \bibinfo {pages} {419} (\bibinfo {year} {2005})},\ \Eprint
  {http://arxiv.org/abs/astro-ph/0410591} {arXiv:astro-ph/0410591 [astro-ph]}
  \BibitemShut {NoStop}%
\bibitem [{\citenamefont {Green}\ \emph {et~al.}(2005)\citenamefont {Green},
  \citenamefont {Hofmann},\ and\ \citenamefont {Schwarz}}]{Green:2005fa}%
  \BibitemOpen
  \bibfield  {author} {\bibinfo {author} {\bibfnamefont {A.~M.}\ \bibnamefont
  {Green}}, \bibinfo {author} {\bibfnamefont {S.}~\bibnamefont {Hofmann}}, \
  and\ \bibinfo {author} {\bibfnamefont {D.~J.}\ \bibnamefont {Schwarz}},\
  }\href {\doibase 10.1088/1475-7516/2005/08/003} {\bibfield  {journal}
  {\bibinfo  {journal} {JCAP}\ }\textbf {\bibinfo {volume} {0508}},\ \bibinfo
  {pages} {003} (\bibinfo {year} {2005})},\ \Eprint
  {http://arxiv.org/abs/astro-ph/0503387} {arXiv:astro-ph/0503387 [astro-ph]}
  \BibitemShut {NoStop}%
\bibitem [{\citenamefont {Loeb}\ and\ \citenamefont
  {Zaldarriaga}(2005)}]{Loeb:2005pm}%
  \BibitemOpen
  \bibfield  {author} {\bibinfo {author} {\bibfnamefont {A.}~\bibnamefont
  {Loeb}}\ and\ \bibinfo {author} {\bibfnamefont {M.}~\bibnamefont
  {Zaldarriaga}},\ }\href {\doibase 10.1103/PhysRevD.71.103520} {\bibfield
  {journal} {\bibinfo  {journal} {Phys. Rev.}\ }\textbf {\bibinfo {volume}
  {D71}},\ \bibinfo {pages} {103520} (\bibinfo {year} {2005})},\ \Eprint
  {http://arxiv.org/abs/astro-ph/0504112} {arXiv:astro-ph/0504112 [astro-ph]}
  \BibitemShut {NoStop}%
\bibitem [{\citenamefont {Profumo}\ \emph {et~al.}(2006)\citenamefont
  {Profumo}, \citenamefont {Sigurdson},\ and\ \citenamefont
  {Kamionkowski}}]{Profumo:2006bv}%
  \BibitemOpen
  \bibfield  {author} {\bibinfo {author} {\bibfnamefont {S.}~\bibnamefont
  {Profumo}}, \bibinfo {author} {\bibfnamefont {K.}~\bibnamefont {Sigurdson}},
  \ and\ \bibinfo {author} {\bibfnamefont {M.}~\bibnamefont {Kamionkowski}},\
  }\href {\doibase 10.1103/PhysRevLett.97.031301} {\bibfield  {journal}
  {\bibinfo  {journal} {Phys.Rev.Lett.}\ }\textbf {\bibinfo {volume} {97}},\
  \bibinfo {pages} {031301} (\bibinfo {year} {2006})},\ \Eprint
  {http://arxiv.org/abs/astro-ph/0603373} {arXiv:astro-ph/0603373 [astro-ph]}\BibitemShut {NoStop}%
\bibitem [{\citenamefont {Bertschinger}(2006)}]{Bertschinger:2006nq}%
  \BibitemOpen
  \bibfield  {author} {\bibinfo {author} {\bibfnamefont {E.}~\bibnamefont
  {Bertschinger}},\ }\href {\doibase 10.1103/PhysRevD.74.063509} {\bibfield
  {journal} {\bibinfo  {journal} {Phys. Rev.}\ }\textbf {\bibinfo {volume}
  {D74}},\ \bibinfo {pages} {063509} (\bibinfo {year} {2006})},\ \Eprint
  {http://arxiv.org/abs/astro-ph/0607319} {arXiv:astro-ph/0607319 [astro-ph]}
  \BibitemShut {NoStop}%
\bibitem [{\citenamefont {Bringmann}\ and\ \citenamefont
  {Hofmann}(2007)}]{Bringmann:2006mu}%
  \BibitemOpen
  \bibfield  {author} {\bibinfo {author} {\bibfnamefont {T.}~\bibnamefont
  {Bringmann}}\ and\ \bibinfo {author} {\bibfnamefont {S.}~\bibnamefont
  {Hofmann}},\ }\href {\doibase 10.1088/1475-7516/2007/04/016} {\bibfield
  {journal} {\bibinfo  {journal} {JCAP}\ }\textbf {\bibinfo {volume} {0407}},\
  \bibinfo {pages} {016} (\bibinfo {year} {2007})},\ \Eprint
  {http://arxiv.org/abs/hep-ph/0612238} {arXiv:hep-ph/0612238 [hep-ph]}
  \BibitemShut {NoStop}%
\bibitem [{\citenamefont {Bringmann}(2009)}]{Bringmann:2009vf}%
  \BibitemOpen
  \bibfield  {author} {\bibinfo {author} {\bibfnamefont {T.}~\bibnamefont
  {Bringmann}},\ }\href {\doibase 10.1088/1367-2630/11/10/105027} {\bibfield
  {journal} {\bibinfo  {journal} {New J. Phys.}\ }\textbf {\bibinfo {volume}
  {11}},\ \bibinfo {pages} {105027} (\bibinfo {year} {2009})},\ \Eprint
  {http://arxiv.org/abs/0903.0189} {arXiv:0903.0189 [astro-ph.CO]} \BibitemShut
  {NoStop}%
\bibitem [{\citenamefont {van~den Aarssen}\ \emph {et~al.}(2012)\citenamefont
  {van~den Aarssen}, \citenamefont {Bringmann},\ and\ \citenamefont
  {Goedecke}}]{vandenAarssen:2012ag}%
  \BibitemOpen
  \bibfield  {author} {\bibinfo {author} {\bibfnamefont {L.~G.}\ \bibnamefont
  {van~den Aarssen}}, \bibinfo {author} {\bibfnamefont {T.}~\bibnamefont
  {Bringmann}}, \ and\ \bibinfo {author} {\bibfnamefont {Y.~C.}\ \bibnamefont
  {Goedecke}},\ }\href {\doibase 10.1103/PhysRevD.85.123512} {\bibfield
  {journal} {\bibinfo  {journal} {Phys. Rev.}\ }\textbf {\bibinfo {volume}
  {D85}},\ \bibinfo {pages} {123512} (\bibinfo {year} {2012})},\ \Eprint
  {http://arxiv.org/abs/1202.5456} {arXiv:1202.5456 [hep-ph]} \BibitemShut
  {NoStop}%
\bibitem [{\citenamefont {Cornell}\ and\ \citenamefont
  {Profumo}(2012)}]{Cornell:2012tb}%
  \BibitemOpen
  \bibfield  {author} {\bibinfo {author} {\bibfnamefont {J.~M.}\ \bibnamefont
  {Cornell}}\ and\ \bibinfo {author} {\bibfnamefont {S.}~\bibnamefont
  {Profumo}},\ }\href {\doibase 10.1088/1475-7516/2012/06/011} {\bibfield
  {journal} {\bibinfo  {journal} {JCAP}\ }\textbf {\bibinfo {volume} {1206}},\
  \bibinfo {pages} {011} (\bibinfo {year} {2012})},\ \Eprint
  {http://arxiv.org/abs/1203.1100} {arXiv:1203.1100 [hep-ph]} \BibitemShut
  {NoStop}%
\bibitem [{\citenamefont {Gondolo}\ \emph {et~al.}(2012)\citenamefont
  {Gondolo}, \citenamefont {Hisano},\ and\ \citenamefont
  {Kadota}}]{Gondolo:2012vh}%
  \BibitemOpen
  \bibfield  {author} {\bibinfo {author} {\bibfnamefont {P.}~\bibnamefont
  {Gondolo}}, \bibinfo {author} {\bibfnamefont {J.}~\bibnamefont {Hisano}}, \
  and\ \bibinfo {author} {\bibfnamefont {K.}~\bibnamefont {Kadota}},\ }\href
  {\doibase 10.1103/PhysRevD.86.083523} {\bibfield  {journal} {\bibinfo
  {journal} {Phys.Rev.}\ }\textbf {\bibinfo {volume} {D86}},\ \bibinfo {pages}
  {083523} (\bibinfo {year} {2012})},\ \Eprint {http://arxiv.org/abs/1205.1914}
  {arXiv:1205.1914 [hep-ph]} \BibitemShut {NoStop}%
\bibitem [{\citenamefont {Cornell}\ \emph {et~al.}(2013)\citenamefont
  {Cornell}, \citenamefont {Profumo},\ and\ \citenamefont
  {Shepherd}}]{Cornell:2013rza}%
  \BibitemOpen
  \bibfield  {author} {\bibinfo {author} {\bibfnamefont {J.~M.}\ \bibnamefont
  {Cornell}}, \bibinfo {author} {\bibfnamefont {S.}~\bibnamefont {Profumo}}, \
  and\ \bibinfo {author} {\bibfnamefont {W.}~\bibnamefont {Shepherd}},\ }\href
  {\doibase 10.1103/PhysRevD.88.015027} {\bibfield  {journal} {\bibinfo
  {journal} {Phys. Rev. D 88,}\ }\textbf {\bibinfo {volume} {015027}} (\bibinfo
  {year} {2013}),\ 10.1103/PhysRevD.88.015027},\ \Eprint
  {http://arxiv.org/abs/1305.4676} {arXiv:1305.4676 [hep-ph]} \BibitemShut
  {NoStop}%
\bibitem [{\citenamefont {Shoemaker}(2013)}]{Shoemaker:2013tda}%
  \BibitemOpen
  \bibfield  {author} {\bibinfo {author} {\bibfnamefont {I.~M.}\ \bibnamefont
  {Shoemaker}},\ }\href {\doibase 10.1016/j.dark.2013.07.002} {\bibfield
  {journal} {\bibinfo  {journal} {Phys. Dark Univ.}\ }\textbf {\bibinfo
  {volume} {2}},\ \bibinfo {pages} {157} (\bibinfo {year} {2013})},\ \Eprint
  {http://arxiv.org/abs/1305.1936} {arXiv:1305.1936 [hep-ph]} \BibitemShut
  {NoStop}%
\bibitem [{\citenamefont {Diamanti}\ \emph {et~al.}(2015)\citenamefont
  {Diamanti}, \citenamefont {Catalan},\ and\ \citenamefont
  {Ando}}]{Diamanti:2015kma}%
  \BibitemOpen
  \bibfield  {author} {\bibinfo {author} {\bibfnamefont {R.}~\bibnamefont
  {Diamanti}}, \bibinfo {author} {\bibfnamefont {M.~E.~C.}\ \bibnamefont
  {Catalan}}, \ and\ \bibinfo {author} {\bibfnamefont {S.}~\bibnamefont
  {Ando}},\ }\href {\doibase 10.1103/PhysRevD.92.065029} {\bibfield  {journal}
  {\bibinfo  {journal} {Phys. Rev.}\ }\textbf {\bibinfo {volume} {D92}},\
  \bibinfo {pages} {065029} (\bibinfo {year} {2015})},\ \Eprint
  {http://arxiv.org/abs/1506.01529} {arXiv:1506.01529 [hep-ph]} \BibitemShut
  {NoStop}%
\bibitem [{\citenamefont {Navarro}\ \emph {et~al.}(1996)\citenamefont
  {Navarro}, \citenamefont {Frenk},\ and\ \citenamefont
  {White}}]{Navarro:1995iw}%
  \BibitemOpen
  \bibfield  {author} {\bibinfo {author} {\bibfnamefont {J.~F.}\ \bibnamefont
  {Navarro}}, \bibinfo {author} {\bibfnamefont {C.~S.}\ \bibnamefont {Frenk}},
  \ and\ \bibinfo {author} {\bibfnamefont {S.~D.}\ \bibnamefont {White}},\
  }\href {\doibase 10.1086/177173} {\bibfield  {journal} {\bibinfo  {journal}
  {Astrophys.J.}\ }\textbf {\bibinfo {volume} {462}},\ \bibinfo {pages} {563}
  (\bibinfo {year} {1996})},\ \Eprint {http://arxiv.org/abs/astro-ph/9508025}
  {arXiv:astro-ph/9508025 [astro-ph]} \BibitemShut {NoStop}%
\bibitem [{\citenamefont {Navarro}\ \emph {et~al.}(1997)\citenamefont
  {Navarro}, \citenamefont {Frenk},\ and\ \citenamefont
  {White}}]{Navarro:1996gj}%
  \BibitemOpen
  \bibfield  {author} {\bibinfo {author} {\bibfnamefont {J.~F.}\ \bibnamefont
  {Navarro}}, \bibinfo {author} {\bibfnamefont {C.~S.}\ \bibnamefont {Frenk}},
  \ and\ \bibinfo {author} {\bibfnamefont {S.~D.}\ \bibnamefont {White}},\
  }\href {\doibase 10.1086/304888} {\bibfield  {journal} {\bibinfo  {journal}
  {Astrophys.J.}\ }\textbf {\bibinfo {volume} {490}},\ \bibinfo {pages} {493}
  (\bibinfo {year} {1997})},\ \Eprint {http://arxiv.org/abs/astro-ph/9611107}
  {arXiv:astro-ph/9611107 [astro-ph]} \BibitemShut {NoStop}%
\bibitem [{\citenamefont {Prada}\ \emph {et~al.}(2012)\citenamefont {Prada},
  \citenamefont {Klypin}, \citenamefont {Cuesta}, \citenamefont
  {Betancort-Rijo},\ and\ \citenamefont {Primack}}]{Prada:2011jf}%
  \BibitemOpen
  \bibfield  {author} {\bibinfo {author} {\bibfnamefont {F.}~\bibnamefont
  {Prada}}, \bibinfo {author} {\bibfnamefont {A.~A.}\ \bibnamefont {Klypin}},
  \bibinfo {author} {\bibfnamefont {A.~J.}\ \bibnamefont {Cuesta}}, \bibinfo
  {author} {\bibfnamefont {J.~E.}\ \bibnamefont {Betancort-Rijo}}, \ and\
  \bibinfo {author} {\bibfnamefont {J.}~\bibnamefont {Primack}},\ }\href
  {\doibase 10.1111/j.1365-2966.2012.21007.x} {\bibfield  {journal} {\bibinfo
  {journal} {Mon.Not.Roy.Astron.Soc.}\ }\textbf {\bibinfo {volume} {428}},\
  \bibinfo {pages} {3018} (\bibinfo {year} {2012})},\ \Eprint
  {http://arxiv.org/abs/1104.5130} {arXiv:1104.5130 [astro-ph.CO]} \BibitemShut
  {NoStop}%
\bibitem [{\citenamefont {Strigari}\ \emph {et~al.}(2007)\citenamefont
  {Strigari}, \citenamefont {Koushiappas}, \citenamefont {Bullock},\ and\
  \citenamefont {Kaplinghat}}]{Strigari:2006rd}%
  \BibitemOpen
  \bibfield  {author} {\bibinfo {author} {\bibfnamefont {L.~E.}\ \bibnamefont
  {Strigari}}, \bibinfo {author} {\bibfnamefont {S.~M.}\ \bibnamefont
  {Koushiappas}}, \bibinfo {author} {\bibfnamefont {J.~S.}\ \bibnamefont
  {Bullock}}, \ and\ \bibinfo {author} {\bibfnamefont {M.}~\bibnamefont
  {Kaplinghat}},\ }\href {\doibase 10.1103/PhysRevD.75.083526} {\bibfield
  {journal} {\bibinfo  {journal} {Phys. Rev.}\ }\textbf {\bibinfo {volume}
  {D75}},\ \bibinfo {pages} {083526} (\bibinfo {year} {2007})},\ \Eprint
  {http://arxiv.org/abs/astro-ph/0611925} {arXiv:astro-ph/0611925 [astro-ph]}
  \BibitemShut {NoStop}%
\bibitem [{\citenamefont {Kuhlen}\ \emph {et~al.}(2008)\citenamefont {Kuhlen},
  \citenamefont {Diemand},\ and\ \citenamefont {Madau}}]{Kuhlen:2008aw}%
  \BibitemOpen
  \bibfield  {author} {\bibinfo {author} {\bibfnamefont {M.}~\bibnamefont
  {Kuhlen}}, \bibinfo {author} {\bibfnamefont {J.}~\bibnamefont {Diemand}}, \
  and\ \bibinfo {author} {\bibfnamefont {P.}~\bibnamefont {Madau}},\
  }\href@noop {} {\bibfield  {journal} {\bibinfo  {journal} {Astrophys.J.}\
  }\textbf {\bibinfo {volume} {686}},\ \bibinfo {pages} {262} (\bibinfo {year}
  {2008})},\ \Eprint {http://arxiv.org/abs/0805.4416} {arXiv:0805.4416
  [astro-ph]} \BibitemShut {NoStop}%
\bibitem [{\citenamefont {Diemand}\ \emph
  {et~al.}(2007{\natexlab{a}})\citenamefont {Diemand}, \citenamefont {Kuhlen},\
  and\ \citenamefont {Madau}}]{Diemand:2006ik}%
  \BibitemOpen
  \bibfield  {author} {\bibinfo {author} {\bibfnamefont {J.}~\bibnamefont
  {Diemand}}, \bibinfo {author} {\bibfnamefont {M.}~\bibnamefont {Kuhlen}}, \
  and\ \bibinfo {author} {\bibfnamefont {P.}~\bibnamefont {Madau}},\ }\href
  {\doibase 10.1086/510736} {\bibfield  {journal} {\bibinfo  {journal}
  {Astrophys. J.}\ }\textbf {\bibinfo {volume} {657}},\ \bibinfo {pages} {262}
  (\bibinfo {year} {2007}{\natexlab{a}})},\ \Eprint
  {http://arxiv.org/abs/astro-ph/0611370} {arXiv:astro-ph/0611370 [astro-ph]}
  \BibitemShut {NoStop}%
\bibitem [{\citenamefont {Madau}\ \emph {et~al.}(2008)\citenamefont {Madau},
  \citenamefont {Diemand},\ and\ \citenamefont {Kuhlen}}]{Madau:2008fr}%
  \BibitemOpen
  \bibfield  {author} {\bibinfo {author} {\bibfnamefont {P.}~\bibnamefont
  {Madau}}, \bibinfo {author} {\bibfnamefont {J.}~\bibnamefont {Diemand}}, \
  and\ \bibinfo {author} {\bibfnamefont {M.}~\bibnamefont {Kuhlen}},\ }\href
  {\doibase 10.1086/587545} {\bibfield  {journal} {\bibinfo  {journal}
  {Astrophys. J.}\ }\textbf {\bibinfo {volume} {679}},\ \bibinfo {pages} {1260}
  (\bibinfo {year} {2008})},\ \Eprint {http://arxiv.org/abs/0802.2265}
  {arXiv:0802.2265 [astro-ph]} \BibitemShut {NoStop}%
\bibitem [{\citenamefont {Springel}\ \emph {et~al.}(2008)\citenamefont
  {Springel} \emph {et~al.}}]{Springel:2008cc}%
  \BibitemOpen
  \bibfield  {author} {\bibinfo {author} {\bibfnamefont {V.}~\bibnamefont
  {Springel}} \emph {et~al.},\ }\href {\doibase
  10.1111/j.1365-2966.2008.14066.x} {\bibfield  {journal} {\bibinfo  {journal}
  {Mon. Not. Roy. Astron. Soc.}\ }\textbf {\bibinfo {volume} {391}},\ \bibinfo
  {pages} {1685} (\bibinfo {year} {2008})},\ \Eprint
  {http://arxiv.org/abs/0809.0898} {arXiv:0809.0898 [astro-ph]} \BibitemShut
  {NoStop}%
\bibitem [{\citenamefont {S\'anchez-Conde}\ and\ \citenamefont
  {Prada}(2014)}]{Sanchez-Conde:2013yxa}%
  \BibitemOpen
  \bibfield  {author} {\bibinfo {author} {\bibfnamefont {M.~A.}\ \bibnamefont
  {S\'anchez-Conde}}\ and\ \bibinfo {author} {\bibfnamefont {F.}~\bibnamefont
  {Prada}},\ }\href {\doibase 10.1093/mnras/stu1014} {\bibfield  {journal}
  {\bibinfo  {journal} {Mon.Not.Roy.Astron.Soc.}\ }\textbf {\bibinfo {volume}
  {442}},\ \bibinfo {pages} {2271} (\bibinfo {year} {2014})},\ \Eprint
  {http://arxiv.org/abs/1312.1729} {arXiv:1312.1729 [astro-ph.CO]} \BibitemShut
  {NoStop}%
\bibitem [{\citenamefont {Martinez}\ \emph {et~al.}(2009)\citenamefont
  {Martinez}, \citenamefont {Bullock}, \citenamefont {Kaplinghat},
  \citenamefont {Strigari},\ and\ \citenamefont {Trotta}}]{Martinez:2009jh}%
  \BibitemOpen
  \bibfield  {author} {\bibinfo {author} {\bibfnamefont {G.~D.}\ \bibnamefont
  {Martinez}}, \bibinfo {author} {\bibfnamefont {J.~S.}\ \bibnamefont
  {Bullock}}, \bibinfo {author} {\bibfnamefont {M.}~\bibnamefont {Kaplinghat}},
  \bibinfo {author} {\bibfnamefont {L.~E.}\ \bibnamefont {Strigari}}, \ and\
  \bibinfo {author} {\bibfnamefont {R.}~\bibnamefont {Trotta}},\ }\href
  {\doibase 10.1088/1475-7516/2009/06/014} {\bibfield  {journal} {\bibinfo
  {journal} {JCAP}\ }\textbf {\bibinfo {volume} {0906}},\ \bibinfo {pages}
  {014} (\bibinfo {year} {2009})},\ \Eprint {http://arxiv.org/abs/0902.4715}
  {arXiv:0902.4715 [astro-ph.HE]} \BibitemShut {NoStop}%
\bibitem [{\citenamefont {Ishiyama}(2014)}]{Ishiyama:2014uoa}%
  \BibitemOpen
  \bibfield  {author} {\bibinfo {author} {\bibfnamefont {T.}~\bibnamefont
  {Ishiyama}},\ }\href {\doibase 10.1088/0004-637X/788/1/27} {\bibfield
  {journal} {\bibinfo  {journal} {Astrophys. J.}\ }\textbf {\bibinfo {volume}
  {788}},\ \bibinfo {pages} {27} (\bibinfo {year} {2014})},\ \Eprint
  {http://arxiv.org/abs/1404.1650} {arXiv:1404.1650 [astro-ph.CO]} \BibitemShut
  {NoStop}%
\bibitem [{\citenamefont {Molin\'e}\ \emph {et~al.}(2016)\citenamefont
  {Molin\'e}, \citenamefont {S\'anchez-Conde}, \citenamefont {Palomares-Ruiz},\
  and\ \citenamefont {Prada}}]{Moline:2016pbm}%
  \BibitemOpen
  \bibfield  {author} {\bibinfo {author} {\bibfnamefont {A.}~\bibnamefont
  {Molin\'e}}, \bibinfo {author} {\bibfnamefont {M.~A.}\ \bibnamefont
  {S\'anchez-Conde}}, \bibinfo {author} {\bibfnamefont {S.}~\bibnamefont
  {Palomares-Ruiz}}, \ and\ \bibinfo {author} {\bibfnamefont {F.}~\bibnamefont
  {Prada}},\ }\href@noop {} {\  (\bibinfo {year} {2016})},\ \Eprint
  {http://arxiv.org/abs/1603.04057} {arXiv:1603.04057 [astro-ph.CO]}
  \BibitemShut {NoStop}%
\bibitem [{\citenamefont {Ghigna}\ \emph {et~al.}(2000)\citenamefont {Ghigna}
  \emph {et~al.}}]{Ghigna:1999sn}%
  \BibitemOpen
  \bibfield  {author} {\bibinfo {author} {\bibfnamefont {S.}~\bibnamefont
  {Ghigna}} \emph {et~al.},\ }\href {\doibase 10.1086/317221} {\bibfield
  {journal} {\bibinfo  {journal} {Astrophys.J.}\ }\textbf {\bibinfo {volume}
  {544}},\ \bibinfo {pages} {616} (\bibinfo {year} {2000})},\ \Eprint
  {http://arxiv.org/abs/astro-ph/9910166} {arXiv:astro-ph/9910166 [astro-ph]}
  \BibitemShut {NoStop}%
\bibitem [{\citenamefont {Bullock}\ \emph {et~al.}(2001)\citenamefont {Bullock}
  \emph {et~al.}}]{Bullock:1999he}%
  \BibitemOpen
  \bibfield  {author} {\bibinfo {author} {\bibfnamefont {J.~S.}\ \bibnamefont
  {Bullock}} \emph {et~al.},\ }\href {\doibase
  10.1046/j.1365-8711.2001.04068.x} {\bibfield  {journal} {\bibinfo  {journal}
  {Mon.Not.Roy.Astron.Soc.}\ }\textbf {\bibinfo {volume} {321}},\ \bibinfo
  {pages} {559} (\bibinfo {year} {2001})},\ \Eprint
  {http://arxiv.org/abs/astro-ph/9908159} {arXiv:astro-ph/9908159 [astro-ph]}
  \BibitemShut {NoStop}%
\bibitem [{\citenamefont {Ullio}\ \emph {et~al.}(2002)\citenamefont {Ullio},
  \citenamefont {Bergstr{\"o}m}, \citenamefont {Edsj{\"o}},\ and\ \citenamefont
  {Lacey}}]{Ullio:2002pj}%
  \BibitemOpen
  \bibfield  {author} {\bibinfo {author} {\bibfnamefont {P.}~\bibnamefont
  {Ullio}}, \bibinfo {author} {\bibfnamefont {L.}~\bibnamefont
  {Bergstr{\"o}m}}, \bibinfo {author} {\bibfnamefont {J.}~\bibnamefont
  {Edsj{\"o}}}, \ and\ \bibinfo {author} {\bibfnamefont {C.~G.}\ \bibnamefont
  {Lacey}},\ }\href {\doibase 10.1103/PhysRevD.66.123502} {\bibfield  {journal}
  {\bibinfo  {journal} {Phys.Rev.}\ }\textbf {\bibinfo {volume} {D66}},\
  \bibinfo {pages} {123502} (\bibinfo {year} {2002})},\ \Eprint
  {http://arxiv.org/abs/astro-ph/0207125} {arXiv:astro-ph/0207125 [astro-ph]}
  \BibitemShut {NoStop}%
\bibitem [{\citenamefont {Diemand}\ \emph
  {et~al.}(2007{\natexlab{b}})\citenamefont {Diemand}, \citenamefont {Kuhlen},\
  and\ \citenamefont {Madau}}]{Diemand:2007qr}%
  \BibitemOpen
  \bibfield  {author} {\bibinfo {author} {\bibfnamefont {J.}~\bibnamefont
  {Diemand}}, \bibinfo {author} {\bibfnamefont {M.}~\bibnamefont {Kuhlen}}, \
  and\ \bibinfo {author} {\bibfnamefont {P.}~\bibnamefont {Madau}},\ }\href
  {\doibase 10.1086/520573} {\bibfield  {journal} {\bibinfo  {journal}
  {Astrophys.J.}\ }\textbf {\bibinfo {volume} {667}},\ \bibinfo {pages} {859}
  (\bibinfo {year} {2007}{\natexlab{b}})},\ \Eprint
  {http://arxiv.org/abs/astro-ph/0703337} {arXiv:astro-ph/0703337 [astro-ph]}
  \BibitemShut {NoStop}%
\bibitem [{\citenamefont {Diemand}\ \emph {et~al.}(2008)\citenamefont {Diemand}
  \emph {et~al.}}]{Diemand:2008in}%
  \BibitemOpen
  \bibfield  {author} {\bibinfo {author} {\bibfnamefont {J.}~\bibnamefont
  {Diemand}} \emph {et~al.},\ }\href {\doibase 10.1038/nature07153} {\bibfield
  {journal} {\bibinfo  {journal} {Nature}\ }\textbf {\bibinfo {volume} {454}},\
  \bibinfo {pages} {735} (\bibinfo {year} {2008})},\ \Eprint
  {http://arxiv.org/abs/0805.1244} {arXiv:0805.1244 [astro-ph]} \BibitemShut
  {NoStop}%
\bibitem [{\citenamefont {Diemand}\ and\ \citenamefont
  {Moore}(2011)}]{Diemand:2009bm}%
  \BibitemOpen
  \bibfield  {author} {\bibinfo {author} {\bibfnamefont {J.}~\bibnamefont
  {Diemand}}\ and\ \bibinfo {author} {\bibfnamefont {B.}~\bibnamefont
  {Moore}},\ }\href {\doibase 10.1166/asl.2011.1211} {\bibfield  {journal}
  {\bibinfo  {journal} {Adv. Sci. Lett.}\ }\textbf {\bibinfo {volume} {4}},\
  \bibinfo {pages} {297} (\bibinfo {year} {2011})},\ \Eprint
  {http://arxiv.org/abs/0906.4340} {arXiv:0906.4340 [astro-ph.CO]} \BibitemShut
  {NoStop}%
\bibitem [{\citenamefont {Valdes}\ \emph {et~al.}(2010)\citenamefont {Valdes},
  \citenamefont {Evoli},\ and\ \citenamefont {Ferrara}}]{Valdes:2009cq}%
  \BibitemOpen
  \bibfield  {author} {\bibinfo {author} {\bibfnamefont {M.}~\bibnamefont
  {Valdes}}, \bibinfo {author} {\bibfnamefont {C.}~\bibnamefont {Evoli}}, \
  and\ \bibinfo {author} {\bibfnamefont {A.}~\bibnamefont {Ferrara}},\
  }\href@noop {} {\bibfield  {journal} {\bibinfo  {journal} {Mon. Not. Roy.
  Astron. Soc.}\ }\textbf {\bibinfo {volume} {404}},\ \bibinfo {pages} {1569}
  (\bibinfo {year} {2010})},\ \Eprint {http://arxiv.org/abs/0911.1125}
  {arXiv:0911.1125 [astro-ph.CO]} \BibitemShut {NoStop}%
\bibitem [{\citenamefont {Chluba}(2010)}]{Chluba:2009uv}%
  \BibitemOpen
  \bibfield  {author} {\bibinfo {author} {\bibfnamefont {J.}~\bibnamefont
  {Chluba}},\ }\href {\doibase 10.1111/j.1365-2966.2009.15957.x} {\bibfield
  {journal} {\bibinfo  {journal} {Mon.Not.Roy.Astron.Soc.}\ }\textbf {\bibinfo
  {volume} {402}},\ \bibinfo {pages} {1195} (\bibinfo {year} {2010})},\ \Eprint
  {http://arxiv.org/abs/0910.3663} {arXiv:0910.3663 [astro-ph.CO]} \BibitemShut
  {NoStop}%
\bibitem [{\citenamefont {Seager}\ \emph {et~al.}(1999)\citenamefont {Seager},
  \citenamefont {Sasselov},\ and\ \citenamefont {Scott}}]{Seager:1999bc}%
  \BibitemOpen
  \bibfield  {author} {\bibinfo {author} {\bibfnamefont {S.}~\bibnamefont
  {Seager}}, \bibinfo {author} {\bibfnamefont {D.~D.}\ \bibnamefont
  {Sasselov}}, \ and\ \bibinfo {author} {\bibfnamefont {D.}~\bibnamefont
  {Scott}},\ }\href {\doibase 10.1086/312250} {\bibfield  {journal} {\bibinfo
  {journal} {Astrophys. J.}\ }\textbf {\bibinfo {volume} {523}},\ \bibinfo
  {pages} {L1} (\bibinfo {year} {1999})},\ \Eprint
  {http://arxiv.org/abs/astro-ph/9909275} {arXiv:astro-ph/9909275 [astro-ph]}
  \BibitemShut {NoStop}%
\bibitem [{\citenamefont {Poulin}\ \emph {et~al.}(2015)\citenamefont {Poulin},
  \citenamefont {Serpico},\ and\ \citenamefont {Lesgourgues}}]{Poulin:2015pna}%
  \BibitemOpen
  \bibfield  {author} {\bibinfo {author} {\bibfnamefont {V.}~\bibnamefont
  {Poulin}}, \bibinfo {author} {\bibfnamefont {P.~D.}\ \bibnamefont {Serpico}},
  \ and\ \bibinfo {author} {\bibfnamefont {J.}~\bibnamefont {Lesgourgues}},\
  }\href {\doibase 10.1088/1475-7516/2015/12/041} {\bibfield  {journal}
  {\bibinfo  {journal} {JCAP}\ }\textbf {\bibinfo {volume} {1512}},\ \bibinfo
  {pages} {041} (\bibinfo {year} {2015})},\ \Eprint
  {http://arxiv.org/abs/1508.01370} {arXiv:1508.01370 [astro-ph.CO]}
  \BibitemShut {NoStop}%
\bibitem [{\citenamefont {Pober}\ \emph {et~al.}(2014)\citenamefont {Pober}
  \emph {et~al.}}]{Pober:2013jna}%
  \BibitemOpen
  \bibfield  {author} {\bibinfo {author} {\bibfnamefont {J.~C.}\ \bibnamefont
  {Pober}} \emph {et~al.},\ }\href {\doibase 10.1088/0004-637X/782/2/66,
  10.1088/0004-637X/788/1/96} {\bibfield  {journal} {\bibinfo  {journal}
  {Astrophys. J.}\ }\textbf {\bibinfo {volume} {782}},\ \bibinfo {pages} {66}
  (\bibinfo {year} {2014})},\ \Eprint {http://arxiv.org/abs/1310.7031}
  {arXiv:1310.7031 [astro-ph.CO]} \BibitemShut {NoStop}%
\bibitem [{\citenamefont {Pober}\ \emph
  {et~al.}(2013{\natexlab{b}})\citenamefont {Pober} \emph
  {et~al.}}]{Pober:2012zz}%
  \BibitemOpen
  \bibfield  {author} {\bibinfo {author} {\bibfnamefont {J.~C.}\ \bibnamefont
  {Pober}} \emph {et~al.},\ }\href {\doibase 10.1088/0004-6256/145/3/65}
  {\bibfield  {journal} {\bibinfo  {journal} {Astron. J.}\ }\textbf {\bibinfo
  {volume} {145}},\ \bibinfo {pages} {65} (\bibinfo {year}
  {2013}{\natexlab{b}})},\ \Eprint {http://arxiv.org/abs/1210.2413}
  {arXiv:1210.2413 [astro-ph.CO]} \BibitemShut {NoStop}%
\bibitem [{\citenamefont {Parsons}\ \emph {et~al.}(2012)\citenamefont
  {Parsons}, \citenamefont {McQuinn}, \citenamefont {Jacobs}, \citenamefont
  {Aguirre},\ and\ \citenamefont {Pober}}]{Parsons:2011ew}%
  \BibitemOpen
  \bibfield  {author} {\bibinfo {author} {\bibfnamefont {A.}~\bibnamefont
  {Parsons}}, \bibinfo {author} {\bibfnamefont {M.}~\bibnamefont {McQuinn}},
  \bibinfo {author} {\bibfnamefont {D.}~\bibnamefont {Jacobs}}, \bibinfo
  {author} {\bibfnamefont {J.}~\bibnamefont {Aguirre}}, \ and\ \bibinfo
  {author} {\bibfnamefont {J.}~\bibnamefont {Pober}},\ }\href {\doibase
  10.1088/0004-637X/753/1/81} {\bibfield  {journal} {\bibinfo  {journal}
  {Astrophys. J.}\ }\textbf {\bibinfo {volume} {753}},\ \bibinfo {pages} {81}
  (\bibinfo {year} {2012})},\ \Eprint {http://arxiv.org/abs/1103.2135}
  {arXiv:1103.2135 [astro-ph.IM]}\BibitemShut {NoStop}%
\bibitem [{\citenamefont {Beardsley}\ \emph {et~al.}(2015)\citenamefont
  {Beardsley}, \citenamefont {Morales}, \citenamefont {Lidz}, \citenamefont
  {Malloy},\ and\ \citenamefont {Sutter}}]{Beardsley:2014bea}%
  \BibitemOpen
  \bibfield  {author} {\bibinfo {author} {\bibfnamefont {A.~P.}\ \bibnamefont
  {Beardsley}}, \bibinfo {author} {\bibfnamefont {M.~F.}\ \bibnamefont
  {Morales}}, \bibinfo {author} {\bibfnamefont {A.}~\bibnamefont {Lidz}},
  \bibinfo {author} {\bibfnamefont {M.}~\bibnamefont {Malloy}}, \ and\ \bibinfo
  {author} {\bibfnamefont {P.~M.}\ \bibnamefont {Sutter}},\ }\href {\doibase
  10.1088/0004-637X/800/2/128} {\bibfield  {journal} {\bibinfo  {journal}
  {Astrophys. J.}\ }\textbf {\bibinfo {volume} {800}},\ \bibinfo {pages} {128}
  (\bibinfo {year} {2015})},\ \Eprint {http://arxiv.org/abs/1410.5427}
  {arXiv:1410.5427 [astro-ph.CO]} \BibitemShut {NoStop}%
\bibitem [{\citenamefont {Ewall-Wice}\ \emph {et~al.}(2015)\citenamefont
  {Ewall-Wice} \emph {et~al.}}]{Ewall-Wice:2015uul}%
  \BibitemOpen
  \bibfield  {author} {\bibinfo {author} {\bibfnamefont {A.}~\bibnamefont
  {Ewall-Wice}} \emph {et~al.},\ }\href@noop {} {\  (\bibinfo {year} {2015})},\
  \Eprint {http://arxiv.org/abs/1511.04101} {arXiv:1511.04101 [astro-ph.CO]}
  \BibitemShut {NoStop}%
\end{thebibliography}%

\end{document}